\documentclass[english]{article}
\usepackage[T1]{fontenc}
\usepackage[latin9]{inputenc}
\usepackage{amsbsy}
\usepackage{amssymb}
\usepackage{graphicx}
\usepackage{caption}
\usepackage{subcaption}
\usepackage{fancybox}
\usepackage{fullpage}
\usepackage{hyperref}
\usepackage{color}
\usepackage{url,textcomp}
\usepackage{amsmath}
\usepackage{amsfonts}
\usepackage{amsthm}
\usepackage{babel}
\usepackage{enumerate}
\usepackage[section]{placeins}
\usepackage{bbm}

\definecolor{bugs}{rgb}{0.5,0.1,0.1}
\definecolor{matlab}{rgb}{0.1,0.1,0.5}
\definecolor{grc}{rgb}{0.0,0.55,0.0}
\definecolor{olive}{rgb}{0.4,0.7,0.2}



\title{An Empirical approach to Survival Density Estimation for randomly-censored data using Wavelets}

\author{German A. Schnaidt Grez  \\ email \href{mailto:gschnaidt@gatech.edu}{gschnaidt@gatech.edu} \and Brani Vidakovic \\ email
\href{mailto:brani@gatech.edu}{brani@gatech.edu}}

\date{Georgia Institute of Technology}

\begin{document}
\maketitle

\begin{abstract}
Density estimation is a classical problem in statistics and has received considerable attention when both the data has been fully observed and in the case of partially observed (censored) samples. In survival analysis or clinical trials, a typical problem encountered in the data collection stage is that the samples may be censored from the right. The variable of interest could be observed partially due to the presence of a set of events that occur at random and potentially censor the data. Consequently, developing a methodology that enables robust estimation of the lifetimes in such setting is of high interest for researchers.

\medskip

In this paper, we propose a non-parametric linear density estimator using empirical wavelet coefficients that are fully data driven. We derive an asymptotically unbiased estimator constructed from the complete sample based on an inductive bias correction procedure. Also, we provide upper bounds for the bias and analyze the large sample behavior of the expected $\mathbb{L}_{2}$ estimation error based on the approach used by Stute (1995), showing that the estimates are asymptotically normal and possess global mean square consistency.

\medskip

In addition, we evaluate the proposed approach via a theoretical simulation study using different exemplary baseline distributions with different sample sizes. In this study, we choose a censoring scheme that produces a censoring proportion of 40\% on average. Finally, we apply the proposed estimator to real data-sets previously published, showing that the proposed wavelet estimator provides a robust and useful tool for the non-parametric estimation of the survival time density function.
\end{abstract}

\newpage

\section{Introduction}\label{Intro}

Density estimation is a classical problem in statistics and has received considerable attention when both the data has been fully observed and also in the case of partially observed (censored) samples. See \cite{Deroye1985,Parzen1962,Rossenblat1956} for thorough discussions about this topic. In areas such as survival analysis, the estimate of the lifetime density function is of a major importance. In fact, the knowledge of how the lifetimes behave in medical follow-up research or reliability analysis is paramount to get insights, draw conclusions, derive results, make comparisons and/or characterize the underlying death/failure process.

\medskip

In general, the density estimation problem can be approached from either a parametric or non-parametric perspective. In the first case, an assumption is made about the particular distribution or family of distributions to which the density of interest belongs. As can immediately be observed, that approach causes the estimated function to be completely dependant on the such assumption which may prove of high benefit in the case when it is correct or close-to correct. However, if the elicited family for the target density is not correct, the parametric approach may lead to unsatisfactory results.

\medskip

Because of the uncertainty about parametric family, the non-parametric approach for density estimation has become a popular topic of research in statistics. In particular, popular methods for density estimation include kernel and nearest neighbors methods \cite{Antoniadis1997}. Another approach for the aforementioned problem consists of the use of orthogonal series (see \cite{Efromovich1999,Antoniadis1999}). In this approach. wavelets can be utilized since they can generate orthonormal bases for functions belonging to $\mathbb{L}_{2}(\mathbb{R})$.

\medskip

One of the first uses of wavelets in density estimation could be traced back to papers by Doukhan and Leon (1990), Antoniadis and Carmona (1991) Kerkyacharian and Picard (1992) and Walter (1992). Moreover, due to their locality in both time and frequency and their exceptional approximation properties, wavelets provide a good choice for density estimation. See e.g. Meyer (1992), Daubechies (1992)\cite{Daubechies1992}, Donoho and Johnstone (1994, 1995, 1998) for detailed discussions about the properties of wavelets in this context. Also, in Vidakovic (1999)\cite{Vidakovic1999} an extensive and thorough discussion of wavelets and their application in statistical modeling can be found.

\medskip
Even though wavelets offer major advantages for curve estimation, there is a potential problem associated with their use in density estimation: there is no guarantee that the estimates are positive or integrate to 1 when using general scaling functions $\phi$. As described in \cite{Antoniadis1997}, the negative values may appear often in the tails of the target distribution. Nonetheless, that can be addressed; a possible remedial approach is the estimation of the square root of the density which allows then to square back to get a non-negative estimate integrating to 1 (as can be see in Pinheiro and Vidakovic (1997) \cite{Pinheiro1997}).

\medskip

In survival analysis or clinical trials, a typical problem encountered in the data collection stage is that the samples may be censored from the right. The variable of interest may be prevented to be fully observed due to the presence of random events (typically assumed to be independent of the variable of interest) and potentially censor the data. A common example of right censoring in clinical trials is the situation in which a patient leaves the study before its termination or was still alive by the end of the observation period. In these cases, only a subset of the observations are fully observed lifetimes; the others are partially observed and it is only known that the actual lifetime was greater than equal to the time at which the subject ceased to be observed (i.e. the censored time).

\medskip

Let $X_{1},...,X_{N}$ be i.i.d. survival times with a common unknown density function $f$. Also, let $T_{1},...,T_{N}$ be i.i.d. censoring times with a common unknown density $g$. Typically (and in the sequel) it is assumed that for $i=1,...,N$ $X_{i} \perp T_{i}$ (here, $\perp$ stands for statistical independence). In the context of partially observed data, instead of fully observing $X_{1},...,X_{N}$, we observed an i.i.d. sequence $\left\{Y_{i}\, , \delta_{i}\right\}_{i=1}^{N}$, where $Y_{i}=\min \left({X_{i}\, , T_{i}}\right)$ and $\delta_{i}=\mathbf{1}_{\left(X_{i}\leq T_{i}\right)}$. The function $\mathbb{1}_{(\cdot)}$ stands for the indicator function.

\medskip

In this paper, we propose a linear estimator based on an orthogonal projection onto a defined multiresolution space $V_{J}$ using empirical wavelet coefficients that are fully data driven. We derive an asymptotically unbiased estimator constructed from the complete sample based on a an inductive bias correction. Also, we provide estimates for the bias and large sample behavior of the expected $\mathbb{L}_{2}$ error based on the approach used by Stute (1995). In addition, we evaluate the performance of the proposed estimator via a simulation study using different exemplary unimodal and multimodal baseline distributions under different sample sizes. For this purpose, we chose an exponential censoring scheme that produces a censoring proportion of 40\% on average. Finally, we apply the proposed estimator to real data-sets previously used in other published results in the field of non-parametric density estimation.

\medskip

Our results are based on wavelets periodic on the interval $[0,1]$ and are derived under the assumption that both densities $f$ and $g$ are continuous and the survival function of the censoring random variable $T$ is bounded from below by an exponentially decaying function. Also, we assume that the scaling function $\phi$ is absolutely integrable and the multiresolution space index $J$ used for the projection is chosen as a function of the sample size $N$ as $J=\lfloor \log_{2}(N)-\log_{2}(\log(N)) \rfloor$. The only assumption that we impose on the target density $f$ is that it belongs to the s-sobolev space $H^{s}$.

\subsection{Overview of previous and current work in the area}

In the context of wavelets applied to density estimation with complete data, Donoho, et al. (1992) \cite{Donoho1996} proposed a wavelet estimator based on thresholded empirical wavelet coefficients and investigate the minimax rates of convergence over a wide range of Besov function classes $B_{\sigma p q}$. They choose the resolution of projection spaces such that the estimator achieves the proper convergence rates. As it can be seen in recent literature, their work is fundamental for subsequent research in the field.

\medskip

A work by Vanucci (1998) \cite{Vanucci1998} provides overview of different wavelet-based density estimators, emphasizing their properties and comparison with classical estimators. In her paper, the author provides a general description of an orthonormal wavelet basis, focusing on the properties that are essential for the construction of wavelet density estimators. Also, a description of linear and thresholded density estimators is provided.  This works constitutes a comprehensive reference for density estimation in the context of complete data.

\medskip

Following the available results in the context of complete-data density estimation (i.e. no censoring), Pinheiro and Vidakovic (1997) \cite{Pinheiro1997} propose estimators of the square root of a density based on compactly supported wavelets. Their estimator is a bona-fide density with $\mathbb{L}_{1}$ norm equal to 1, taking care of possible negative values resulting from the usual estimation of the density $f$.

\medskip

Now in the context of density estimation with censored data, Antoniadis et al. (1999) \cite{Antoniadis1999} proposed a wavelet method based on dividing the time axis into a dyadic number of intervals and counting the number of occurrences within each one. Then, they use wavelets smoothers based on wavelets on the interval (see \cite{Daubechies1992}) to get the survival function of the observations. Also, they obtain the best possible asymptotic mean integrated square error (MISE) convergence rate under the assumption that the target density $f$ is $r-$times continuously differentiable and the censoring density $g$ is continuous.

\medskip

Later on, Li (2003)\cite{Li2003} provides a non-linear wavelet-based density estimator under random censorship that uses a thresholded series expansion of the sub-density $f_{1}(x)=f(x)\mathbbm{1}_{\left\{x\leq T\right\}}$ where $T<\tau_H$ and $\tau_{H}=\inf \left\{x\,:\,F_{Y}(x)=1 \right\}$. This approach is based on compactly supported $\phi$ and $\psi$ (father and mother wavelet, respectively) and detail coefficients $d_{jk}$ are thresholded according to $\tilde{d}_{jk}=\hat{d}_{jk}\mathbbm{1}_{\left\{|\hat{d}_{jk} |>\delta \right\}}$ for a suitable defined threshold $\delta$ and parameter $j=q$ for the wavelet expansion. In his work, Li provides and asymptotic expansion for the MISE and calculate the convergence rates under smoothness and regularity assumptions on the target density $f$. This work is then further extended in Li (2007) \cite{Li2007}, where the minimax optimality of the thresholded wavelet-based estimator is investigated over a large range of Besov function classes.

\medskip

One of the most recent works in the context of censored data was developed by Zou and Liang (2017) \cite{Zou2017}. They define a non-linear wavelet estimator for the right censoring model in the case when the censoring indicator $\delta$ is missing at random. They develop an asymptotic expression for the MISE which is robust under the presence of discontinuities in $f$. Their estimator reduces to the one proposed by Li (2003) when the censoring indicator missing at random does not happen and a bandwidth in non-parametric estimation is close to zero.

\subsection{About Periodic Wavelets}\label{wavelets}

For the implementation of the functional estimator, we choose periodic wavelets as an orthonormal basis. Even though this kind of wavelets exhibit poor behaviour near the boundaries (when the analyzed function is not periodic, high amplitude wavelet coefficients are generated in the neighborhood of the boundaries) they are typically used due to the relatively simple numerical implementation and compact support. Also, as was suggested by Johnstone (1994), this simplification affects only a small number of wavelet coefficients at each resolution level.

\medskip

Periodic wavelets in $[0,1]$ are defined by a modification of the standard scaling and wavelet functions:
\begin{eqnarray}
 & \phi^{per}_{j,k}(x)=\sum_{l \in \mathbb{Z}}\phi_{j,k}(x-l) \,,\\
 & \psi^{per}_{j,k}(x)=\sum_{l \in \mathbb{Z}}\psi_{j,k}(x-l)\,.
\end{eqnarray}

It is possible to show, as in \cite{Restrepo1996}, that $\left\{ \phi^{per}_{j,k}(x), 0\leq k \leq 2^{j}-1 , j\geq 0\right\}$ constitutes an orthonormal basis
for $\mathbb{L}_{2}[0,1]$. Consequently, $\cup_{j=0}^{\infty} V_{j}^{per}=\mathbb{L}_{2}[0,1]$, where $V_{j}^{per}$ is the space spanned by $\left\{ \phi^{per}_{j,k}(x), 0\leq k \leq 2^{j}-1 \right\}$. This allows to represent a function $f$ with support in $[0,1]$ as:

\begin{equation}\label{eq:1b}
f(x)=\langle f(x),\phi^{per}_{0,0}(x) \rangle \phi^{per}_{0,0}(x) + \sum_{j\geq 0}\sum_{k=0}^{2^{j}-1}\langle f(x),\psi^{per}_{j,k}(x) \rangle \psi^{per}_{j,k}(x)\,.
\end{equation}

\medskip
Also, for a fixed $j=J$, we can obtain an orthogonal projection of $f(x)$ onto $V_{J}$ denoted as $\textbf{P}_{J}(f(x))$ given by:

\begin{equation}\label{eq:1c}
\textbf{P}_{J}(f(x))=\sum_{k=0}^{2^{J}-1}\langle f(x),\phi^{per}_{J,k}(x) \rangle \phi^{per}_{J,k}(x)
\end{equation}
Since periodized wavelets provide a basis for $\mathbb{L}^{2}([0,1])$, we have that $\parallel f(x) - \textbf{P}_{J}(f(x)) \parallel_{2} \rightarrow 0$ as $ J \rightarrow  \infty $. Also, it can be shown that $\parallel f(x) - \textbf{P}_{J}(f(x)) \parallel_{\infty} \rightarrow 0$ as $ J \rightarrow  \infty $. Therefore, we can see that $\textbf{P}_{J}(f(x))$ uniformly converges to $f$ as $J \rightarrow \infty$.
\medskip
Similarly, as discussed in \cite{Daubechies1992} it is possible to assess the approximation error for a certain density of interest $f$ using a truncated projection (i.e. for a certain chosen detail space $J$). For example, using the $s$-th Sobolev norm of a function defined as:

\begin{equation}
\parallel f(x) \parallel_{H^{s}}=\sqrt{\int(1+|x|^{2})^{s}|f(x)|^{2}dx}\,,
\end{equation}

one defines the $H^{s}$ sobolev space, as the space that consists of all functions $f$ whose s-Sobolev norm exists and is finite. As it is shown in \cite{Daubechies1992}:

\begin{equation}\label{eq:1d}
\parallel f(x) - \textbf{P}_{J}(f(x)) \parallel_{2} \leq 2^{-J\cdot s}\cdot \parallel f \parallel _{H^{s}[0,1]}\,.
\end{equation}

From (\ref{eq:1d}), for a pre-specified $\epsilon>0$ one can choose $J$ such that $\parallel f(x) - \textbf{P}_{J}(f(x)) \parallel_{2} \leq \epsilon$. In fact, a possible choice of
J could be:

\begin{equation}\label{eq:1e}
J \geq -\left\lceil \frac{1}{s} \log_{2}\left(\frac{\epsilon}{\parallel f \parallel _{H^{s}[0,1]}}\right) \right\rceil\,.
\end{equation}

Therefore, it is possible to approximate a desired function to arbitrary precision using the MRA generated by a wavelet basis. In this context, extensive work has been done regarding the proper choice of the scale level $J$ for the estimator in the MRA. In fact, \cite{Donoho1993} suggests that the choice $J=\lfloor \log_{2}(N)-\log_{2}(\log(N)) \rfloor$ can guarantee consistency of the estimator, under the proper regularity conditions on the scaling functions and underlying density $f$.

\section{Survival Density Estimation for right-censored data using Periodized Wavelets}\label{method}

\subsection{Problem statement, assumptions and derivation of the estimator for a density $f(x)$.}\label{problemstatement}

Consider a sample of iid lifetimes (non-negative) of the form $\tilde{X}_{1},...,\tilde{X}_{N}$ drawn from a random variable $\tilde{X}\sim \tilde{f}(\cdot)$, with unknown density $\tilde{f}\in\mathbb{L}_{2}(\mathbb{R})$.  Furthermore, let $\tau_{\tilde{X}}=\inf{\left\{\tilde{x}:\tilde{F}_{\tilde{X}}(\tilde{x})=1 \right\}}$, where $\tilde{F}_{\tilde{X}}(\tilde{x})$ corresponds to the cumulative density function (cdf) of the random variable $\tilde{X}$.

\medskip

Define the target density (i.e. the density to be estimated) as $\tilde{f}_{c}(\tilde{x})=\tilde{f}(\tilde{x})\mathbbm{1}_{\left\{\tilde{x}\leq \tau_{\tilde{X}} \right\}}$, which corresponds to $\tilde{f}(\cdot)$ constrained to the interval $[0,\tau_{\tilde{X}}]$. This definition implies that $\tilde{f}_{c}(\tilde{x})=\tilde{f}(\tilde{x})$, for $\tilde{x}\leq \tau_{\tilde{X}}$.

\medskip

From the observed sample $\tilde{X}_{1},...,\tilde{X}_{N}$, and a pre-specified $\tau>0$, define the normalized random variable $X=\frac{1}{\tau}\tilde{X}$. Then, it follows:
\begin{equation}
f_{X}(x)=\tau\,f_{\tilde{X}}(\tau x)\mathbbm{1}_{\left\{x\leq \frac{\tau_{\tilde{X}}}{\tau}\right\}}\,,\label{eq:1f}
\end{equation}
for the domain-restricted density $\tilde{f}_{c}(\tilde{x})$.

\paragraph*{Remarks}
\begin{enumerate}[(i)]
\item If $\tau = \tau_{\tilde{X}}$ the normalized random variable $X$ has support in [0,1] with density given by $f(x)=f_{X}(x)$.
\item In practice, since $\tilde{f}$ is not known, it is possible to select $\tau=\max\left\{\tilde{X}_{1},...,\tilde{X}_{N} \right\}$; this, since in general $\tilde{X}_{(N)} \mathop{\rightarrow}\limits^{\mathbb{P}} \tau_{\tilde{X}}$ where the operator $\mathop{\rightarrow}\limits^{\mathbb{P}}$ denotes convergence in probability.
\item Note that the definition $\tilde{f}_{c}(\tilde{x})=\tilde{f}(\tilde{x})\mathbbm{1}_{\left\{\tilde{x}\leq \tau_{\tilde{X}} \right\}}$ corresponds exactly to the conditional density $\tilde{f}_{\tilde{X}|\tilde{X}\leq \tau_{\tilde{X}}}(\tilde{x})$.
\end{enumerate}

\medskip
\medskip

In the sequel, it will be assumed that the random variable $X$ was obtained presented above, with a probability density of the form (\ref{eq:1f}).

\subsubsection*{Representing $f(x)$ using Wavelets}
Using a multiresolution analysis (MRA) based on periodized wavelets in $[0,1]$, the density $f(\cdot)$ can be expressed as:
\begin{equation}\label{eq:1}
  f(x)=\sum_{j\in\mathbb{Z}}\sum_{k\geq 0}d_{jk}\cdot\psi_{jk}^{per}(x)\,.
\end{equation}
Using the hierarchical structure of the MRA, for a pre-specified multiresolution scale $J=J_{0}$, (\ref{eq:1}) can be expressed as:
\begin{equation}\label{eq:2}
f(x)=\sum_{k\in\mathbb{Z}}c_{J_{0},k}\cdot\phi_{J_{0},k}^{per}(x)+\sum_{j\geq J_{0}}\sum_{k\in\mathbb{Z}}d_{jk}\cdot\psi_{jk}^{per}(x)\,,
\end{equation}
for $\phi_{jk}^{per}(x)=2^{\frac{j}{2}}\phi^{per}(2^{j}x-k)$, and $\psi_{jk}^{per}(x)=2^{\frac{j}{2}}\psi^{per}(2^{j}x-k)$ for $j,k\in\mathbb{Z}$.

\medskip

Because periodic extensions of wavelets in $[0,1]$ are used, the support of the scaling function $\phi_{jk}^{per}(x)$ and the wavelet function $\psi_{jk}^{per}(x)$ is
 $[k\cdot 2^{-j},(k+1)\cdot 2^{-j}]$ where $k=0,...,2^{j-1}$, and by the Strang-fix condition $j\geq 0$.

\medskip

From (\ref{eq:2}), the summation over the MRA scale index $j$ goes from $J_{0}$ to $\infty$. This implies that it is possible to approximate $f(\cdot)$
by truncating the summation up to scale index $J^{*}$. Therefore, it follows:
\begin{equation}\label{eq:3}
\hat{f}_{J^{*}}(x)=\sum_{k\in\mathbf{K}(J_{0})}c_{J_{0},k}\cdot\phi_{J_{0},k}^{per}(x)+\sum_{j\geq J_{0}}^{J^{*}}\sum_{k\in\mathbf{K}(j)}d_{jk}\cdot\psi_{jk}^{per}(x)\,,
\end{equation}
where $\mathbf{K}(J_{0})=\left\{k\in\mathbb{N}\,|\,0\leq k\leq2^{J_{0}-1}\right\}$ and $\mathbf{K}(j)=\left\{k\in\mathbb{N}\,|\,0\leq k\leq2^{j-1}\right\}$.
\medskip
In the sequel, the value of $J^{*}$ will be assumed to be selected as a function of the sample size $N$.

\medskip

In the wavelet series approximation of $f(\cdot)$ defined by (\ref{eq:3}), the coefficients $c_{J_{0},k}$ and $d_{jk}$ are given by the orthogonal projection of $f(\cdot)$ onto each
subspace $V_{J_{0}}^{per}$ and $W_{j}^{per}$ in the MRA\footnote{In fact, from the MRA approach we have that $V_{J^{*}}^{per}=V_{J_{0}}^{per}\oplus\cup_{j=J_{0}}^{J^{*}}W_{j}^{per}$.}. Here, $V_{J_{0}}^{per}$ and $W_{j}^{per}$ correspond to the functional spaces spanned by $\left\{\phi_{J_{0},k}^{per}\,,0\leq k \leq 2^{J_{0}}-1 \right\} $, and $\left\{\psi_{j,k}^{per}\,,0\leq k \leq 2^{j}-1\,, J_{0}\leq j \leq J^{*} \right\} $ respectively. Using this definitions, it follows:
\begin{eqnarray}
c_{J_{0},k}&=&\int_{0}^{1}f(x)\cdot\phi^{per}_{J_{0},k}(x)dx=\langle f(x),\phi^{per}_{J_{0},k}(x) \rangle \,,\label{eq:4}\\
d_{jk}&=&\int_{0}^{1}f(x)\cdot\psi^{per}_{j,k}(x)dx=\langle f(x),\psi^{per}_{j,k}(x) \rangle\,.\label{eq:5}
\end{eqnarray}
Clearly, since $f$ is a probability density, (\ref{eq:4}) and (\ref{eq:5}) can be represented as:
\begin{eqnarray}
c_{J_{0},k}&=&\mathbb{E}_{f}[\phi^{per}_{J_{0},k}(X)]\,,\label{eq:6} \\
d_{jk}&=&\mathbb{E}_{f}[\psi^{per}_{j,k}(X)] \,.\label{eq:7}
\end{eqnarray}
\medskip
Substituting (\ref{eq:6}) and (\ref{eq:7}) in (\ref{eq:3}), $\hat{f}_{J^{*}}(x)$ takes the form:
\begin{equation}\label{eq:8}
\hat{f}_{J^{*}}(x)=\sum_{k\in\mathbf{K}(J_{0})}\mathbb{E}_{f}[\phi^{per}_{J_{0},k}(X)]\cdot\phi^{per}_{J_{0},k}(x)+\sum_{j\geq J_{0}}^{J^{*}}\sum_{k\in\mathbf{K}(j)}\mathbb{E}_{f}[\psi^{per}_{j,k}(X)]\cdot\psi^{per}_{jk}(x)\,.
\end{equation}
\medskip
Using (\ref{eq:8}) and assuming $X_{1},...,X_{N}\sim f(\cdot)$ are iid, for $f(\cdot)$ unknown, it is possible to estimate the coefficients $c_{J_{0},k}$ and $d_{jk}$ from the sample as follows:
\begin{eqnarray}
\tilde{c}_{J_{0},k}&=&\frac{1}{N}\sum_{i=1}^{N}\phi^{per}_{J_{0},k}(X_{i})\,, \label{eq:9}\\
\tilde{d}_{j,k}&=&\frac{1}{N}\sum_{i=1}^{N}\psi^{per}_{j,k}(X_{i})\,. \label{eq:10}
\end{eqnarray}
Therefore, the data-driven estimated density $\hat{f}_{J^{*}}(x)$ can be expressed as:
\begin{equation}\label{eq:11}
\hat{f}_{J^{*}}(x)=\sum_{k\in\mathbf{K}(J_{0})}\left( \frac{1}{N}\sum_{i=1}^{N}\phi^{per}_{J_{0},k}(X_{i}) \right) \cdot\phi^{per}_{J_{0},k}(x)+\sum_{j\geq J_{0}}^{J^{*}}\sum_{k\in\mathbf{K}(j)}\left( \frac{1}{N}\sum_{i=1}^{N}\psi^{per}_{j,k}(X_{i}) \right) \cdot\psi^{per}_{jk}(x)\,.
\end{equation}

\medskip

From (\ref{eq:11}), it follows that $\hat{f}_{J^{*}}(x)$ was constructed based on fully observed realizations of the lifetime random variable $X$. Therefore, a natural extension is the modification of
(\ref{eq:11}) to allow the introduction of partially observed (censored) samples; in particular, we will focus on the case of right-censored data.

\subsection{Estimating $\hat{f}_{J^{*}}(x)$  in the case of partially observed data.}

Consider a random variable $X$ that is distributed with an unknown density $f(x)$. Furthermore, suppose an observed sample $\left\{Y_{i},\delta_{i}\right\}_{i=1}^{N}$ that is composed on both fully, and partially observed realizations of $X$. In the sample, $Y_{i}$ is defined as:
\begin{equation}\label{eq:12}
Y_{i}=\min\left(X_{i},T_{i}\right) \quad i=1,...,N\,,
\end{equation}
for $T_{1},...,T_{N}$ being iid random variables from an unknown distribution $T\sim g(t)$, which is
the right-censoring sequence that causes some realizations from $X$ to be partially observed, and is assumed to be independent of $X$. Also $\delta_{i}$, representing the censoring indicator, is defined as:
\begin{equation}\label{eq:13}
\delta_{i}=\mathbbm{1}_{(X_{i}\leq T_{i})} \quad i=1,...,N\,,
\end{equation}
where $\mathbbm{1}_{(X_{i}\leq T_{i})}=1$ if and only if $(X_{i}\leq T_{i})$ and $0$ otherwise. Therefore, $\delta_{i}=0$ represents a life-time $X_{i}$ that was observed only up to time $T_{i}$, for which we can only conclude that $X_{i}>T_{i}$.

\medskip
Since the observed data is $\left\{Y_{i},\delta_{i}\right\}_{i=1}^{N}$, from (\ref{eq:12}) and (\ref{eq:13}), the joint distribution of the pair $\left( Y,\delta \right)$ can be obtained as follows:
\begin{eqnarray}\label{eq:14}
\nonumber
\mathbb{P}(Y\leq y,\delta=1)&=&\mathbb{P}\left( \min(X,T)\leq y, X\leq T \right) \\
\nonumber
&=& \int_{-\infty}^{y}\mathbb{P}\left( T\geq x \right)f(x)dx \\
&=& \int_{-\infty}^{y}\left( 1-G(x) \right)f(x)dx \,,\label{eq:15}
\end{eqnarray}
where $G(x)=\mathbb{P}\left( T\leq x \right)$. Similarly, for $\mathbb{P}(Y\leq y,\delta=0)$ and a fixed $y$, it follows:

\begin{eqnarray}\label{eq:16}
\nonumber
\mathbb{P}(Y\leq y,\delta=0)&=&\mathbb{P}\left( \min(X,T)\leq y, X>T \right) \\
\nonumber
&=&\int_{-\infty}^{+\infty}\mathbb{P}\left( T\leq \min(x,y) \right)f(x)dx \\
\nonumber
&=&\int_{-\infty}^{y}\mathbb{P}\left( T\leq x \right)f(x)dx + \int_{y}^{+\infty}\mathbb{P}\left( T\leq y \right)f(x)dx \\
\nonumber
&=&\int_{-\infty}^{y}G(x)f(x)dx+G(y)\int_{y}^{+\infty}f(x)dx \\
&=& \int_{-\infty}^{y}G(x)f(x)dx+G(y)(1-F(y)) \,.\label{eq:17}
\end{eqnarray}
From (\ref{eq:15}) and (\ref{eq:17}) it follows:
\begin{equation}\label{eq:19}
f_{Y,\delta}(y,\delta)=f(y)^{\delta}(1-G(y))^{\delta}g(y)^{1-\delta}(1-F(y))^{1-\delta}\,.
\end{equation}
\medskip
Similarly, from (\ref{eq:19}), the marginal density of the complete-data sample $Y$ can be expressed as:
\begin{equation}\label{eq:21}
f_{Y}(y)=f_{X}(y)(1-G_{T}(y))+g_{T}(y)(1-F_{X}(y))\,,
\end{equation}
\medskip
where the subscripts $X$ and $T$ are placed to emphasize the relation between each density function and its corresponding random variable.
\medskip

Assuming $0<G_{T}(y)<1$, $f(x)$, from (\ref{eq:21}) it follows that $f(x)$ can be expressed as:
\begin{equation}\label{eq:22}
f_{X}(y)=\frac{f_{Y}(y)}{1-G_{T}(y)}-\frac{(1-F_{X}(y))g_{T}(y)}{1-G_{T}(y)}\,.
\end{equation}
\medskip
As was mentioned in \ref{problemstatement}, the next sections assume that the observed data has been normalized according to $\tau = \max \left\{Y_{1},...,Y_{N} \right\}$, to restrict the support of the random variable $X$ to the interval $[0,1]$.

\subsubsection{Complete Data Estimator}
\medskip
From (\ref{eq:9}) and (\ref{eq:10}), (\ref{eq:21}) and (\ref{eq:22}), the wavelet coefficients $c_{J_{0},k}$ in the orthogonal wavelet expansion can be expressed as:
\begin{eqnarray*}
c_{J_{0},k}&=&\int_{0}^{1}f(x)\cdot\phi^{per}_{J_{0},k}(x)dx \\
&=&\int_{0}^{1}\left(\frac{f_{Y}(y)}{1-G_{T}(y)}-\frac{(1-F_{X}(y))g_{T}(y)}{1-G_{T}(y)}\right) \cdot \phi^{per}_{J_{0},k}(x)dx\,.
\end{eqnarray*}
Therefore:
\begin{equation}\label{eq:23}
c_{J_{0},k}=\mathbb{E}_{Y}\left[ \frac{\phi^{per}_{J_{0},k}(Y)}{(1-G(Y))} \right]-\mathbb{E}_{T}\left[ \frac{(1-F(Y))\phi^{per}_{J_{0},k}(Y)}{(1-G(Y))}\right]\,.
\end{equation}
Similarly, for the coefficients $d_{j,k}$, it follows:
\begin{equation}\label{eq:24}
d_{j,k}=\mathbb{E}_{Y}\left[ \frac{\psi^{per}_{j,k}(Y)}{(1-G(Y))} \right]-\mathbb{E}_{T}\left[ \frac{(1-F(Y))\psi^{per}_{j,k}(Y)}{(1-G(Y))}\right]\,.
\end{equation}
\medskip
\paragraph*{Remarks:}
\begin{enumerate}[(i)]
\item Expressions (\ref{eq:23}) and (\ref{eq:24}) are valid assuming $0<G(y)<1$ for $y\in [0,1]$. 
\item In the case of non-censored data, $G=\delta_{\infty}$ (i.e. Dirac at $\infty$) and, for $i=1,...,N$\, $\delta_{i}=1$. Therefore, $f_{Y,\delta}=f(x)$. Thus, (\ref{eq:23}) and (\ref{eq:24}) collapse into $\frac{1}{N}\sum_{i=1}^{N}\phi_{Jk}^{per}(Y_{i})$
and $\frac{1}{N}\sum_{i=1}^{N}\psi_{Jk}^{per}(Y_{i})$ respectively, which is the usual orthogonal-series density estimator scheme.
\end{enumerate}
\medskip
\medskip
\hfill \break
Using an empirical approach as in (\ref{eq:9}) and (\ref{eq:10}), it follows:

\begin{equation}\label{eq:25}
\tilde{c}_{J_{0},k}=\frac{1}{N}\sum_{i=1}^{N}\frac{\phi^{per}_{J_{0},k}(Y_{i})}{1-G(Y_{i})}-\frac{1}{N}\sum_{i=1}^{N}\frac{\mathbbm{1}_{(\delta_{i}=0)}(1-F(Y_{i}))\phi^{per}_{J_{0},k}(Y_{i})}{(1-G(Y_{i}))}\,,
\end{equation}
provided $0<G(Y_{i})<1$, for $i=1,...,N$. 
\medskip

Finally, the data-driven estimated density $\hat{f}_{J^{*}}(x)$ can be expressed as:
\begin{equation}\label{eq:26}
\hat{f}_{J^{*}}(x)=\sum_{k\in\mathbf{K}(J_{0})}\left(\frac{1}{N} \sum_{i=1}^{N} \alpha^{\phi}_{i}\cdot \phi^{per}_{J_{0},k}(Y_{i}) \right) \cdot\phi^{per}_{J_{0},k}(x)+\sum_{j\geq J_{0}}^{J^{*}}\sum_{k\in\mathbf{K}(j)}\left( \frac{1}{N} \sum_{i=1}^{N} \alpha^{\psi}_{i}\cdot \psi^{per}_{j,k}(Y_{i}) \right) \cdot\psi^{per}_{jk}(x)\,,
\end{equation}
\medskip
where:
\begin{equation}
\alpha^{\phi}_{i}=\alpha^{\psi}_{i}=\frac{1}{1-G(Y_{i})}-\frac{\mathbbm{1}_{(\delta_{i}=0)}(1-F(Y_{i}))}{1-G(Y_{i})}\,, \label{eq:27}
\end{equation}
for $i=1,...,N$. 
\medskip

As can be seen from (\ref{eq:26}) and (\ref{eq:27}), the computation of (\ref{eq:26}) implies addressing the following issues:
\begin{enumerate}[(i)]
    \item Estimation of $G(Y_{i})$ and $F(Y_{i})$ for $i=1,...,N$.
    \item Computation of $\alpha^{\phi}_{i}$, for $i=1,...,N$.
    \item Computation of $\phi^{per}_{J_{0},k}(Y_{i})$ and $\psi^{per}_{j,k}(Y_{i})$ for $i=1,...,N$, $j=J_{0},...,J^{*}$ and $0 \leq k \leq 2^{j-1}$.
\end{enumerate}

\medskip
Naturally, $G(Y_{i})$ and $F(Y_{i})$  can be obtained using the Kaplan-Meier estimator, which is well known for its robustness in the presence of censored data. Similarly, $\phi^{per}_{J_{0},k}(Y_{i})$ and $\psi^{per}_{j,k}(Y_{i})$ we can computed using Daubechies-Lagarias algorithm.
\medskip

Denote $\left\{ (Y_{(i)},\tilde{\delta}_{(i)}) \right\}_{i=1}^{N}$ as the ranked sample $\left\{(Y_{i},\delta_{i})\right\}_{i=1}^{N}$ with respect to $Y_{i}$,
where $\tilde{\delta}_{(i)}=1-\delta_{(i)}$. Using Kaplan-Meier, it follows:
\begin{eqnarray}
\hat{G}_{N}(Y_{(i)})&=&\hat{G}(Y_{(i)})=\sum_{k=1}^{i}\left( \frac{\tilde{\delta}_{(k)}}{N-k+1} \prod_{j=1}^{k-1}( 1-\frac{\tilde{\delta}_{(j)}}{N-j+1} )\right)\,,\label{eq:29} \\
\hat{F}_{N}(Y_{(i)})&=&\hat{F}(Y_{(i)})=\sum_{k=1}^{i}\left( \frac{\delta_{(k)}}{N-k+1} \prod_{j=1}^{k-1}( 1-\frac{\delta_{(j)}}{N-j+1} )\right)\,,\label{eq:30}
\end{eqnarray}
for $i=1,...,N$. Thus, the estimated density $\hat{f}_{J^{*}}(x)$ can be expressed as:
\begin{eqnarray}
\nonumber
\hat{f}_{J^{*}}(x)&=& \sum_{k\in\mathbf{K}(J_{0})}\left(\frac{1}{N} \sum_{i=1}^{N} \alpha^{\phi}_{(i)}\cdot \phi^{per}_{J_{0},k}(Y_{(i)}) \right) \cdot\phi^{per}_{J_{0},k}(x) \\
\label{eq:31}
& & +\sum_{j\geq J_{0}}^{J^{*}}\sum_{k\in\mathbf{K}(j)}\left( \frac{1}{N} \sum_{i=1}^{N} \alpha^{\psi}_{(i)}\cdot \psi^{per}_{j,k}(Y_{(i)}) \right) \cdot\psi^{per}_{jk}(x)\,,
\end{eqnarray}
where:
\begin{equation} \label{eq:32}
\alpha^{\phi}_{(i)}=\alpha^{\psi}_{(i)}=\frac{1}{1-\hat{G}(Y_{(i)})} - \frac{\mathbbm{1}_{(\delta_{i}=0)}(1-\hat{F}(Y_{(i)}))}{1-\hat{G}(Y_{(i)})}\,,
\end{equation}
for $0<\hat{G}(Y_{(i)})<1$, $i\in \subset \left \{ 1,...,N \right \}$, $\mathbf{K}(J_{0})=\left\{0,1,...,2^{J_{0}}-1 \right\}$, and $\mathbf{K}(j)=\left\{0,1,...,2^{j}-1;\,j\geq J_{0} \right\}$.
\medskip

From section \ref{wavelets}, for a properly chosen multiresolution index $J$, the estimated density $\hat{f}_{J}(x)$ can be approximated by a truncated projection $\textbf{P}_{J}(f(x))$ onto a multiresolution space $V_{J}$ spanned by the functions $\left\{\phi_{Jk}^{per},\,0\leq k \leq 2^{J}-1\right\} $. Under this setting, $\hat{f}_{J^{*}}(x)$ takes the form:
\begin{equation}\label{eq:33}
\hat{f}_{J}(x)=\sum_{k=0}^{2^{J}-1}\tilde{c_{Jk}}\cdot \phi^{per}_{J,k}(x)\,,
\end{equation}
where:
\begin{equation}\label{eq:34}
\tilde{c_{Jk}}=\frac{1}{N}\sum_{i=1}^{N}\alpha^{\phi}_{(i)}\cdot \phi^{per}_{J,k}(Y_{(i)})\,.
\end{equation}

\subsubsection{Partial-Data Estimator assuming $G(y)$ is known.} \label{partial}

From definition (\ref{eq:33}), using an iterative bias-correction procedure it is possible to obtain an unbiased estimator for (\ref{eq:33}), which is given by:
\begin{equation}\label{eq:34c}
\hat{f}^{PD}(x)=\sum_{k=0}^{2^{J}-1}\tilde{c}_{Jk}\cdot \phi^{per}_{J,k}(x)\,,
\end{equation}
where:
\begin{eqnarray}
\tilde{c}_{Jk} &=& \frac{1}{N}\sum_{i=1}^{N}\frac{\mathbbm{1}_{(\delta_{i}=1)}}{1-\hat{G}(Y_{i})}\phi^{per}_{Jk}(Y_{i}) \label{eq:34d},\,\,\text{and}\\
\mathbb{E}\left[\tilde{c}_{Jk} \right] &=& c_{Jk}\,. \label{eq:34e}
\end{eqnarray}
The corresponding derivation can be found in section \ref{der:unbiased} of the appendix.
\medskip
\paragraph*{Remark}
From (\ref{eq:34d}), it is possible to observe that the "partial data" definition comes from the fact that the estimator uses only the samples corresponding to actual observations
of the survival time $X$, as opposed to (\ref{eq:33}) which uses the complete sample $Y_{1},...,Y_{N}$. A similar estimator is proposed by Efromovich in \cite{Efromovich1999} using a fourier basis instead of wavelets.

\subsection{Statistical properties of the Estimator assuming $G(y)$ is known .}

\subsubsection{Mean Square Consistency.}

Now we investigate the mean-square convergence of the estimator $\hat{f}^{PD}(x)$. 

\subsubsection*{Proposition 1} \label{prop1}
Define:
\begin{eqnarray}
  \mu_{J}(x)&=& \mathbb{E}\left[\hat{f}^{PD}(x) \right]=f_{J}(x) \,, \label{eq:68}\\
  \sigma^{2}_{J}(x)&=& Var\left[ \hat{f}^{PD}(x) \right]\,. \label{eq:69}
\end{eqnarray}
Assume the following conditions are satisfied:
\begin{enumerate}[(i)]
\item The scaling function $\phi$ that generates the orthonormal set $\left\{\phi_{Jk}^{per}, 0\leq k \leq 2^{J}\right\} $ has compact support and satisfies $||\theta_{\phi}(x)||_{\infty}=C<\infty$, for $\theta_{\phi}(x):=\sum_{r\in\mathbb{Z}}|\phi(x-r)|$.
\item $\exists$ $F\in \mathbb{L}_{2}(\mathbb{R})$ such that $|K(x,y)|\leq F(x-y)$, for all $x,y\, \in \mathbb{R}$, where $K(x,y)=\sum_{k\in\mathbb{Z}}\phi(x-k)\phi(y-k)$.
\item For $s=m+1$, $m\geq1$, integer, $\int |x|^{s}F(x)dx<\infty$.
\item $\int (y-x)^{l}K(x,y)dy=\delta_{0,l}$ for $l=0,...,s$.
\item The density $f$ belongs to the $s$-sobolev space $W_{2}^{s}([0,1])$, defined as:
\begin{equation}
\nonumber
W_{2}^{s}([0,1])=\left\{f\,|\,f\in\mathbb{L}_{2}([0,1]),\, \exists \, f^{(1)},...,f^{(s)}\,\text{s.t.}\,f^{(l)}\in \mathbb{L}_{2}([0,1]),\,l=1,...,s \right\}.
\end{equation}
\end{enumerate}

\medskip

Then, it follows:
\begin{equation}\label{eq:72a}\mathop{\sup}\limits_{f\in W_{2}^{s}([0,1])}\mathbb{E}\left[||\hat{f}^{PD}(x)-f(x) ||_{2}^{2} \right]\leq C_{1}\frac{2^{J}}{N}+C_{2}2^{-2sJ}\,,\text{and} \end{equation}
for $J=\lfloor \log_{2}(N)-\log_{2}(\log(N)) \rfloor$:
\begin{equation}\label{eq:72}
\sigma^{2}_{J}(x)=\mathcal{O}(\log(N)^{-1})\,,
\end{equation}
\begin{equation}\label{eq:46}
\mathbb{E}\left[\parallel f(x)-\hat{f}^{PD}(x) \parallel_{2}^{2}\right] \, \leq \, \mathcal{O}(N^{-s}\log(N)^{s})
\end{equation}
\medskip
for $C_{1}>0\,,\,C_{2}>0$ independent of $J$ and $N$, provided $\exists$ $ \alpha_{1}$ $\mid$ $0<\alpha_{1}<\infty$, $C_{T} \in (0,1) $ such that $(1-G(y)) \geq C_{T}e^{-\alpha_{1}y}$ for $y \in [0,1)$,
and $0 \leq F(y) \leq 1$ $\forall y \in [0,1]$.

\medskip

The proof can be found in section \ref{proof:prop1} of the appendix.

\hfill \break
\medskip
Based on (\ref{eq:72}), it is possible to observe that $\sigma^{2}_{J}(x) \rightarrow 0$ as $N \rightarrow \infty$,  which implies that $\hat{f}^{PD}(x)$ is consistent for $f(x)$, for all $x\in[0,1]$ and $f\in W_{2}^{s}([0,1])$.

\subsubsection*{Remarks}
Note that from (\ref{eq:46}), it is possible to choose the multiresolution level $J$ such that the upper bound for the $\mathbb{L}_{2}$ risk is minimized. In this context, it is possible to show that $J^{*}(N)=\frac{1}{2s+1}\log_{2}\left(\frac{2s\,C_{2}}{C_{1}} \right)+\frac{1}{2s+1}\log_{2}(N)$ achieves that result. Moreover, under this choice of $J$, it follows:
\begin{equation}\nonumber
\mathop{\sup}\limits_{f\in W_{2}^{s}([0,1])}\mathbb{E}\left[||\hat{f}^{PD}(x)-f(x) ||_{2}^{2} \right]\leq \tilde{C}N^{-\frac{2s}{2s+1}}\,.
\end{equation}

\subsection{Statistical properties for Partial Data Estimator assuming $G(y)$ unknown.}\label{properties2}

In the previous section, we showed that $f^{PD}(x)$ is unbiased for $f_{J}(x)$ and mean square consistent for $f(x)\in W_{2}^{s}([0,1])$, assuming $G$ known and the multiresolution index $J$ for the orthogonal projection onto the space $V_{J}$ was chosen as $J=\lfloor \log_{2}(N)-\log_{2}(\log(N)) \rfloor$.
\medskip

Naturally, assuming $G$ is known may be questionable because of both the nature of the non-parametric density estimation approach, and its practical application. In most of real life cases neither the target density $f$, nor the censoring density $g$ are known, so making assumptions about them could undermine the robustness and quality of the estimated functions.

\medskip

In this section we approach the problem of deriving the partial-data estimator using the data driven wavelet coefficients proposed in (\ref{eq:34d}). In particular, we investigate the statistical properties of the partial data estimator through the application the methodology proposed by Stute (1995) \cite{Stute1995} that approximates Kaplan-Meier integrals by the average of i.i.d. random variables plus a remainder that decays to zero at a certain rate.

\subsubsection{Asymptotic unbiasedness.}\label{StuteBias}

As was proposed in (\ref{eq:34d}), $\tilde{c}_{Jk} =\frac{1}{N}\sum_{i=1}^{N}\frac{\mathbbm{1}_{(\delta_{(i)}=1)}}{1-\hat{G}(Y_{i})}\phi^{per}_{Jk}(Y_{i})$. Using the methodology and results proposed by Stute in \cite{Stute1995}, and assumptions defined in \ref{prop1}, it follows:

\begin{equation}\label{eq:113}
\sum_{i=1}^{N}W_{(i)}\phi_{Jk}^{per}(Y_{(i)}) = \frac{1}{N}\sum_{i=1}^{N}\delta_{i}\phi_{Jk}^{per}(Y_{i})\gamma_{0}(Y_{i})+\frac{1}{N}\sum_{i=1}^{N}U_{i}+R_{N}\,,
\end{equation}
where $W_{(i)}=d\hat{F}_{N}(x)$ is the Kaplan-Meier probability mass function of the random variable $X$ based on the sample, $\gamma_{0}(Y_{i})=\frac{1}{1-G_{T}(Y_{i})}$ and $U_{i}=(1-\delta_{i})\gamma_{1}(Y_{i})-\gamma_{2}(Y_{i})$ for $i=1,...,N$.

\medskip

Similarly, $\gamma_{1}(x)=\gamma_{1,Jk}(x)$ and $\gamma_{2}(x)=\gamma_{2,Jk}(x)$ are given by the following expressions:

\begin{eqnarray}
 \nonumber 
  \gamma_{1,Jk}(x)& = &\frac{1}{1-F_{Y}(x)}\int_{x}^{\tau_{H}}\phi_{Jk}^{per}(u)f_{X}(u)du\,, \\
 \nonumber
 \gamma_{2,Jk}(x)& = &\int_{-\infty}^{\tau_{H}}C(\min{\left\{x,u\right\}})\phi_{Jk}^{per}(u)f_{X}(u)du \,,\text{where}\\
 \nonumber
 C(x)& = &\int_{-\infty}^{x^{-}}\frac{g_{T}(u)du}{(1-F_{Y}(u))(1-G_{T}(u))}\,.
\end{eqnarray}
In addition, assume the following conditions are satisfied (from Stute \cite{Stute1995}):

\begin{eqnarray}
\int \phi^{2}(x)\gamma_{0}^{2}(x)f_{Y,\delta=1}(x)dx &<& \infty \label{eq:140}\,, \\
\int |\phi(x)|\sqrt{C(x)}f_{X}(x)dx &<& \infty\,. \label{eq:141}
\end{eqnarray}
Condition (\ref{eq:140}) corresponds to the requirement of finite second moment (modified) on the scaling function $\phi(x)$, while condition (\ref{eq:141}) incorporates a modification on the first moment of $\phi(x)$ with respect to $f_{X}$ that allows to control de bias in $\int \phi_{Jk}^{per}(u)\hat{f}_{N}(u)du$. For further details, see \cite{Stute1995} and \cite{Stute1994}.

\medskip

From the definitions above, it follows:
\begin{equation}\label{eq:114}
\mathbb{E}\left[ \phi_{Jk}^{per}(Y)\delta \gamma_{0}(Y)\right]=c_{Jk}\,,
\end{equation}
assuming $x<\tau_{H}$ for $\tau_{H}=\inf{\left\{x:F_{Y}(x)=1\right\}}$.

\medskip

Also, from (\ref{eq:29}) and (\ref{eq:30}), it follows that $d\hat{F}_{N}(x)=\hat{f}_{N}(x)$; indeed:
$$
d\hat{F}_{N}(x)=
\begin{cases}
0 & \mbox{if } x \notin \left\{Y_{(1)},...,Y_{(N)} \right\} \\
\frac{\delta_{(i)}}{N-i+1} \prod_{j=1}^{i-1}\left(1-\frac{\delta_{(j)}}{n-j+1} \right) & \mbox{if } x = Y_{(i)}\,, i=1,...,N
\end{cases}
$$
After some algebra, it follows:
\begin{equation}\label{eq:115}
d\hat{F}_{N}(x)=\frac{\delta_{(i)}}{N-i+1} \prod_{j=1}^{i-1}\left(\frac{n-j}{n-j+1} \right)^{\delta_{(j)}}\,.
\end{equation}
Moreover, $\frac{1}{1-\hat{G}_{N}(Y_{(i)})}$ can be expressed as:
\begin{equation}\label{eq:116}
\frac{1}{1-\hat{G}_{N}(Y_{(i)})}=\frac{N}{N-i+1}\prod_{j=1}^{i-1}\left(\frac{n-j}{n-j+1} \right)^{\delta_{(j)}}\,.
\end{equation}
Therefore, putting together (\ref{eq:115}) and (\ref{eq:116}), it follows:
\begin{equation}\label{eq:117}
\frac{\delta_{(i)}}{N(1-\hat{G}_{N}(Y_{(i)}))}=\frac{\delta_{(i)}}{N-i+1} \prod_{j=1}^{i-1}\left(\frac{n-j}{n-j+1} \right)^{\delta_{(j)}}=d\hat{F}_{N}(x)\,.
\end{equation}
These results altogether imply:
\begin{equation}\label{eq:117b}
\int \phi_{Jk}^{per}(u)\hat{f}_{N}(u)du =\tilde{c}_{Jk}\,.
\end{equation}
\medskip
From Stute (1995), results (\ref{eq:140})-(\ref{eq:117b}) imply that (\ref{eq:113}) can be expressed as:
\begin{equation}\label{eq:118}
\int \phi_{Jk}^{per}(u)\hat{f}_{N}(u)du =\frac{1}{N}\sum_{i=1}^{N}\delta_{i}\phi_{Jk}^{per}(Y_{i})\gamma_{0}(Y_{i})+\frac{1}{N}\sum_{i=1}^{N}U_{i}+R_{N}\,,
\end{equation}
where $U_{i}$ i.i.d. for $i=1,...,N$ with $\mathbb{E}[U_{1}]=0$\,, $\mathbb{E}[U_{1}^{2}]=\sigma^{2}<\infty$ and $|R_{N}|=\mathcal{O}(N^{-1}\log(N))$.

\medskip

Therefore:
\begin{eqnarray}
\nonumber
\mathbb{E}\left[ \int \phi_{Jk}^{per}(u)\hat{f}_{N}(u)du \right] &=&\mathbb{E}\left[ \frac{1}{N}\sum_{i=1}^{N}\delta_{i}\phi_{Jk}^{per}(Y_{i})\gamma_{0}(Y_{i}) \right]+\mathbb{E}\left[ \frac{1}{N}\sum_{i=1}^{N}U_{i} \right]+\mathcal{O}(N^{-1}\log(N))\,, \\
&=& c_{Jk}+\mathcal{O}(N^{-1}\log(N)) \,.\label{eq:121}
\end{eqnarray}
Thus, $bias(\tilde{c}_{Jk})= \mathcal{O}(N^{-1}\log(N))$, which implies that the partial data approach is asymptotically unbiased. The exact bias can be obtained by following the details presented in \cite{Stute1995}.

\subsubsection{$\mathbb{L}_{2}$ Risk Analysis.}\label{StuteError}

Following the same methodology and assumptions used in the previous section, we investigate the estimation error for the partial data approach, in the case where $G$ is unknown.

\subsubsection*{Proposition 2} \label{prop2}
Under the assumptions and definitions stated in \ref{prop1} and \ref{StuteBias}, by choosing $J=\lfloor \log_{2}(N)-\log_{2}(\log(N)) \rfloor$, it follows:

\begin{eqnarray}
\mathop{\sup}\limits_{f\in W_{2}^{s}([0,1])}\mathbb{E}\left[\parallel f(x)-\hat{f}^{PD}(x) \parallel_{2}^{2}\right] &=& \mathcal{O}(N^{-s}\log(N)^{s})\,. \label{eq:StuteError2}\\
\end{eqnarray}
The corresponding proofs can be found in section \ref{proof:prop2} of the appendix.

\medskip

\subsubsection*{Remarks}
\begin{enumerate}[(i)]
\item Observe that by following the same methodology as in \ref{proof:prop1}, it is possible to obtain:
\begin{equation}\nonumber
\mathop{\sup}\limits_{f\in W_{2}^{s}([0,1])}\mathbb{E}\left[||\hat{f}^{PD}(x)-f(x) ||_{2}^{2} \right]\leq C_{1}\frac{2^{J}}{N}+C_{2}2^{-2sJ}\,,
\end{equation}
for $C_{1}=\frac{||F||_{2}^{2}e^{2\gamma}}{C^{2}}$ and $C_{2}>0$, independent of $N$ and $J$.
\item The last result implies that by choosing $J^{*}(N)=\frac{1}{2s+1}\log_{2}\left(\frac{2s\,C_{2}}{C_{1}} \right)+\frac{1}{2s+1}\log_{2}(N)$, the $\mathbb{L}_{2}$ risk of the estimator $\hat{f}^{PD}(x)$ (when $G$ is unknown) is also mean square consistent, and achieves a convergence rate of the order $\sim N^{-\frac{2s}{2s+1}}$. This implies that as long as the empirical survival function of the censoring random variable obtained from the Kaplan-Meier estimator is bounded from below by an exponentially decaying function, the knowledge of the its cdf does not affects the statistical properties of the estimator.
\end{enumerate}

\medskip

\subsubsection{Limiting Distribution.}\label{limitDist}

In this section, we investigate the limiting distribution of the partial data estimator $\hat{f}^{PD}(x)$. Similarly as in sections \ref{StuteBias} and \ref{StuteError}, we will use results proposed in \cite{Stute1995} as framework for our analysis.

\medskip

As seen in (\ref{eq:118}), (\ref{eq:121}), Theorem 1.1 of \cite{Stute1995} and the SLLN (Strong Law of Large Numbers), the following results hold:
\begin{eqnarray}
 \frac{1}{N}\sum_{i=1}^{N}\frac{\delta_{i}\phi_{Jk}^{per}(Y_{i})}{1-G(Y_{i})} &\mathop{\rightarrow}\limits^{\mathbb{P}}& c_{Jk} \label{eq:119}\,, \\
 R_{N} &\mathop{\rightarrow}\limits^{\mathbb{P}}& 0 \,, \label{eq:120}
\end{eqnarray}
where (\ref{eq:119}) follows from the SLLN (assuming the expectation is finite), and (\ref{eq:120}) from the fact that $|R_{N}|=\mathcal{O}_{\mathbb{P}}(\frac{1}{\sqrt{N}})$, as shown in \cite{Stute1995}. Using Slutzky's theorem (see \cite{DasGupta2008}), it follows:
\begin{equation} \label{eq:136}
\tilde{c}_{Jk}-\frac{1}{N}\sum_{i=1}^{N}\frac{\delta_{i}\phi_{Jk}^{per}(Y_{i})}{1-G(Y_{i})}-R_{N} \mathop{=}\limits^{\mathbb{D}} \frac{1}{N}\sum_{i=1}^{N}U_{i}\,,
\end{equation}
\medskip
where $U_{i}=(1-\delta_{i})\gamma_{1}(Y_{i})-\gamma_{2}(Y_{i})$, $i=1,...,N$ are i.i.d. zero-mean and finite variance random variables with $\mathbb{E}\left[ U_{1}^{2}\right]=\sigma^{2}$. Also, from the definitions of $\gamma_{1}(x)$ and $\gamma_{2}(x)$, it follows that $\sigma^{2}=\sigma^{2}_{Jk}$ since it depends on the scaling function $\phi_{Jk}^{per}(x)$.  Now, by the CLT (Central Limit Theorem) it follows:

\begin{equation}\label{eq:137}
\frac{1}{\sqrt{N}}\sum_{i=1}^{N}U_{i} \mathop{\rightarrow}\limits^{\mathbb{D}} N(0,\sigma_{Jk}^{2})\,.
\end{equation}
Combining results (\ref{eq:136}), (\ref{eq:137}), Slutzky's theorem implies:
\begin{equation}\label{eq:138}
\sqrt{N}(\tilde{c}_{Jk}-c_{Jk}) \mathop{\rightarrow}\limits^{\mathbb{D}} N(0,\sigma_{Jk}^{2})\,.
\end{equation}
Similarly, it follows:
\begin{equation}\label{eq:139}
\sqrt{N}\left(\hat{f}^{PD}(x)-f(x) \right) =\sum_{k=0}^{2^{J}-1}\sqrt{N}(\tilde{c}_{Jk}-c_{Jk})\phi_{Jk}^{per}(x)\,.
\end{equation}

\subsubsection*{Proposition 3} \label{prop3}

For $c>0$, $\beta >1$ and $x$ in a neighborhood of 1, assume the following conditions hold:

\begin{enumerate}[(i)]
\item $(1-F_{X})  \sim  c(1-G_{T})^{\beta}$ \label{eq:142}
\item $C(x)  \leq  \frac{1}{(1-F_{X}(x))(1-G_{T}(x))}$ \label{eq:143}
\end{enumerate}
Then, it follows:
\begin{equation}\label{eq:144}
\sqrt{N}\left(\hat{f}^{PD}(x)-f(x) \right) \mathop{\rightarrow}\limits^{\mathbb{D}} N\left(0 \, , \sum_{k=0}^{2^{J}-1}\sigma_{Jk}^{2}(\phi_{Jk}^{per}(x))^{2}+2\,\sum_{k<l}\sigma_{J,kl}\phi_{Jk}^{per}(x)\phi_{Jl}^{per}(x) \right)\,,
\end{equation}
for $k,l \,=0,...,2^{J}-1$, $\sigma_{Jk}^{2}=\mathbb{E}\left[\left((1-\delta)\gamma_{1,Jk}(Y)-\gamma_{2,Jk}(Y) \right)^{2} \right]$ and $\sigma_{J,kl}=\mathbb{E}\left[ \frac{\delta^{2}\phi_{Jk}^{per}(Y)\phi_{Jl}^{per}(Y)}{(1-G(Y))^{2}}-c_{Jk}c_{Jl}\right]$, provided assumptions detailed in \ref{prop1}, (\ref{eq:140}), (\ref{eq:141}) are satisfied and $J=\lfloor \log_{2}(N)-\log_{2}(\log(N)) \rfloor$. The corresponding proof can be found in section \ref{proof:prop3} of the appendix.

\subsubsection*{Remarks}

\begin{enumerate}[(a)]
  \item Note that condition (\ref{eq:143}) indicates that there is enough information about the tails of the target density $f$; also, the larger the values of $\beta$, the heavier the tails of the censoring distribution, compared to the tails of the survival time distribution.
  \item As described in \cite{Stute1994} and \cite{Stute1995}, the condition of $\beta>1$ is required so that the bias of $\tilde{c}_{Jk}-c_{Jk} $ achieves a convergence rate better that $a\,N^{-\frac{1}{2}}$ for some non-vanishing $a$ which may cause that (\ref{eq:113}) is no longer valid.
  \item As it can be seen in (\ref{eq:158}), the fact that $\hat{f}^{PD}(x)$ presents asymptotic normality brings to discussion the possibility that the estimates may be negative, as was previously mentioned in \ref{limitDist} and discussed in \cite{Antoniadis1997}.
\end{enumerate}

\section{Simulation Study}\label{Simulations}

In this section, we investigate the estimation performance of $\hat{f}^{PD}(x)$ and evaluate it with respect to the AMSE (Average Mean Squared Error) via a simulation study. For this purpose, we choose a set of exemplary baseline functions that resemble important features that continuous survival times that can be encountered in practice could posses. To simplify the simulations, we chose functions that are supported in an interval close to [0,1]. A brief description of each chosen function follows:

\begin{enumerate}
  \item \textbf{Delta.}
  This corresponds to a R.V. $X \sim N(0.5\,,0.02^{2})$. The idea is to have an extreme spatially heterogeneous curve that has support over a small region. The goal is to represent situations when a short but abrupt deviation from a process may happen.
  \item \textbf{Normal.}
  This corresponds to the usual Normal distribution with parameters $\mu=0.5$ and $\sigma=0.15$.
  \item \textbf{Bimodal.}
  This corresponds to a mixture of 2 Normal distributions and has the form $f(x)=0.5 \,X_{1} +0.5 \,X_{2}$ where $ X_{1} \sim N(0.4\,,0.12^{2})$ and $X_{2} \sim N(0.7\,,0.08^{2})$.
  \item \textbf{Strata.}
  This corresponds to a mixture of 2 Normal distributions and has the form $f(x)=0.5 \,X_{1} +0.5 \,X_{2}$ where $ X_{1} \sim N(0.2\,,0.06^{2})$ and $X_{2} \sim N(0.7\,,0.08^{2})$. The idea is to represent a function that is supported over 2 separate subintervals.
  \item \textbf{Multimodal.}
  This functions corresponds to a mixture of 3 Normal distributions and has the form $f(x)=\frac{1}{3} \,X_{1} +\frac{1}{3} \,X_{2} +\frac{1}{3} \,X_{3}$ where $ X_{1} \sim N(0.2\,,0.06^{2})$,  $X_{2} \sim N(0.5\,,0.05^{2})$ and  $X_{3} \sim N(0.7\,,0.05^{2})$. The idea of this function is to represent multimodal survival times which are expected to occur in heterogeneous populations.
\end{enumerate}

An advantage of using simulated data in the case of censored data is that the values for both $X$ and $T$ are known for all samples; also, the controlled-environment approach allows the investigation of the estimator's performance for different sample sizes and censoring schemes. For testing purposes, we choose a censoring random variable $T \sim Exp(\lambda)$ with $\lambda=0.8$, which produces approximately 45\% censored samples at each generated datasets. Also, we use samples sizes $N=100,200,500,1000$ and measure the global error given by:

\begin{equation}\label{eq:135}
\hat{MSE}=\frac{1}{B}\sum_{b=1}^{B}\frac{1}{N}\sum_{i=1}^{N}\left(f(x_{i})-\hat{f}_{N,b}(x_{i}) \right)^{2}\,,
\end{equation}
where $B$ is the number of replications of the experiment and $N$ is the number of samples. For all experiments we choose $B=1000$ and the wavelet filter Symmlet5. To implement simulations, we generate 2 independent random samples $\left\{X_{i}\right\}_{i=1}^{N}$ and $\left\{T_{i}\right\}_{i=1}^{N}$. $X_{i}$ random variables were drawn from each one of the aforementioned distributions, while $T_{i}\mathop{\sim}\limits^{i.i.d.}Exp(\lambda)$. Also, we included in the simulation study the complete data estimator as we found of interest to observe its performance and compare it to the partial data approach.

\subsection{Simulation Results.}

\begin{figure}[!htb]
   \centering
   \begin{subfigure}[b]{0.25\textwidth}
       \includegraphics[width=\textwidth]{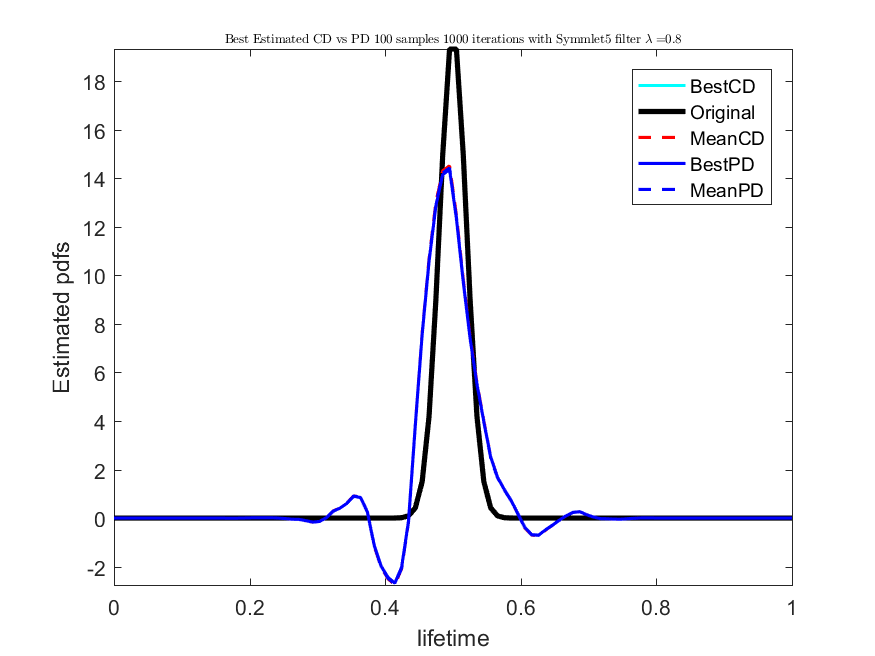}
       \caption{}
       \label{fig:delta100}
   \end{subfigure}\hfill
   \begin{subfigure}[b]{0.25\textwidth}
       \includegraphics[width=\textwidth]{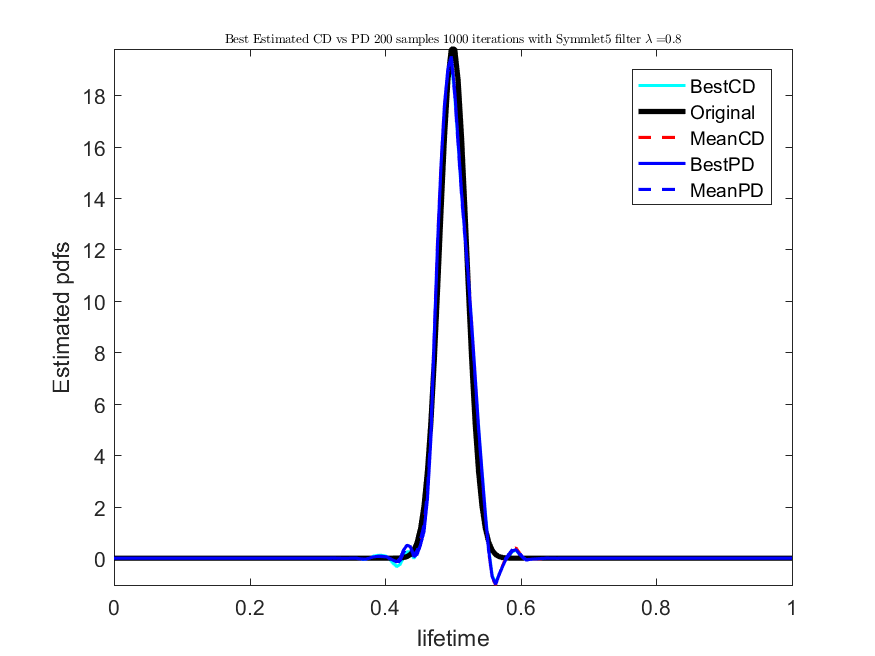}
       \caption{}
       \label{fig:delta200}
    \end{subfigure}\hfill
    \begin{subfigure}[b]{0.25\textwidth}
       \includegraphics[width=\textwidth]{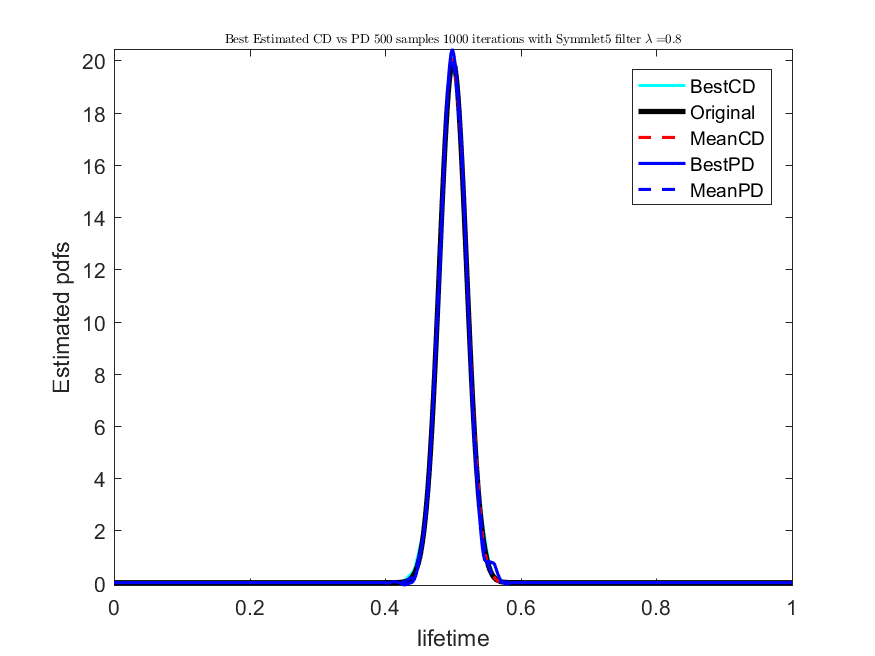}
       \caption{}
       \label{fig:delta500}
    \end{subfigure}\hfill
    \begin{subfigure}[b]{0.25\textwidth}
       \includegraphics[width=\textwidth]{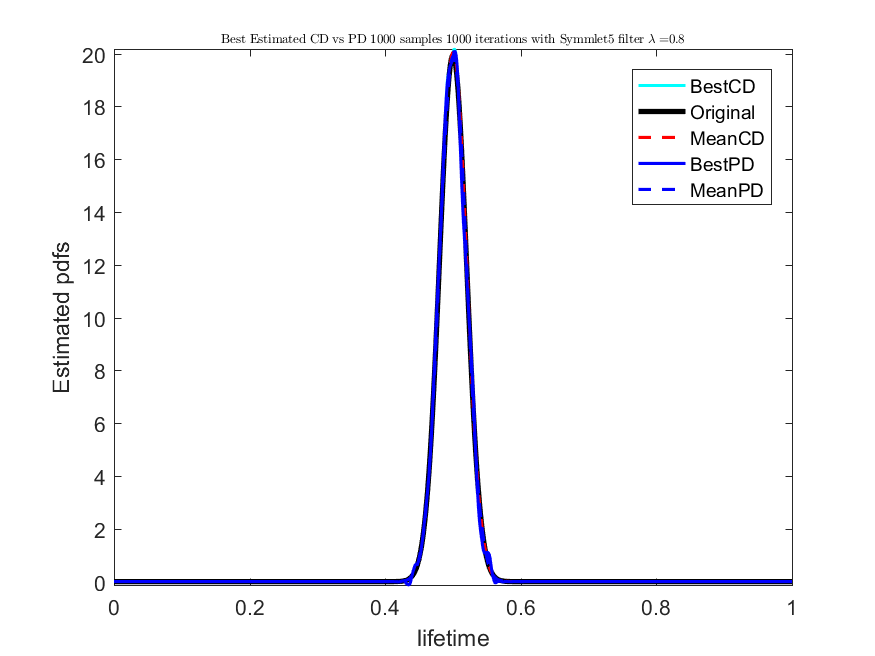}
       \caption{}
       \label{fig:delta1000}
    \end{subfigure}
    \caption{Estimate results for Delta distribution, $N=100,200,500,1000$ using Symmlet5.}\label{fig:deltaAll}
\end{figure}

\begin{figure}[!htb]
   \centering
   \begin{subfigure}[b]{0.25\textwidth}
       \includegraphics[width=\textwidth]{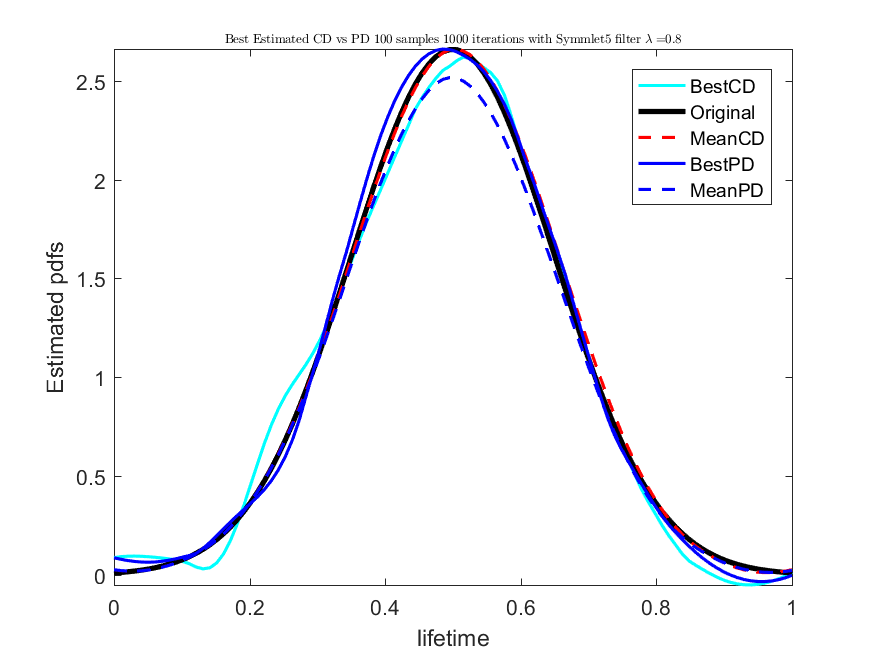}
       \caption{}
       \label{fig:normal100}
   \end{subfigure}\hfill
   \begin{subfigure}[b]{0.25\textwidth}
       \includegraphics[width=\textwidth]{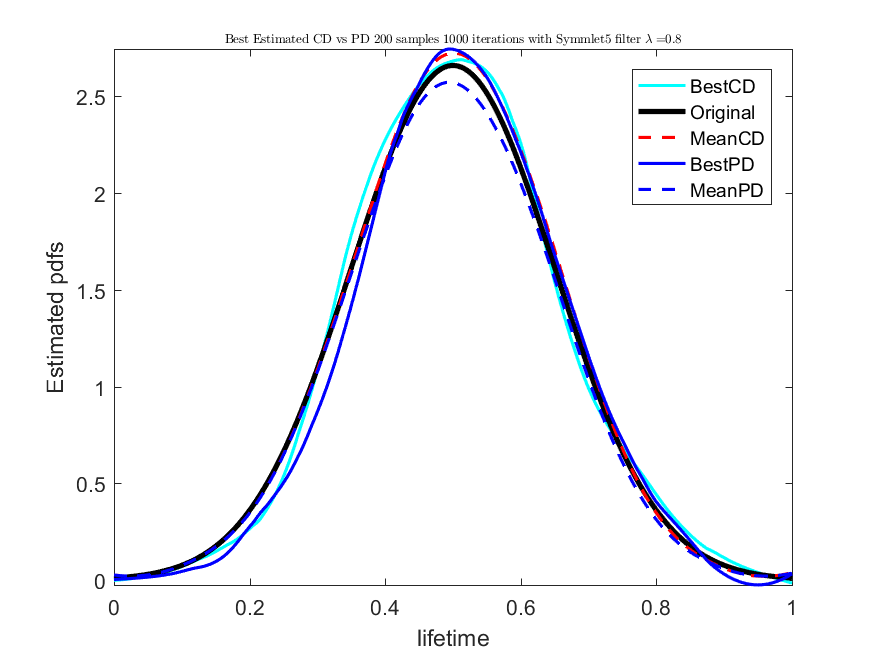}
       \caption{}
       \label{fig:normal200}
    \end{subfigure}\hfill
    \begin{subfigure}[b]{0.25\textwidth}
       \includegraphics[width=\textwidth]{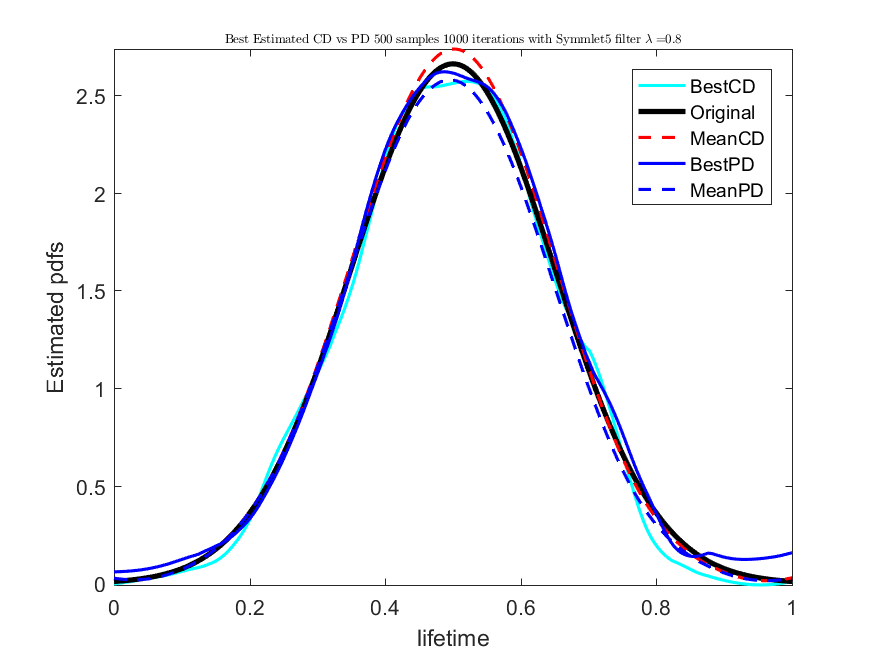}
       \caption{}
       \label{fig:normal500}
    \end{subfigure}\hfill
    \begin{subfigure}[b]{0.25\textwidth}
       \includegraphics[width=\textwidth]{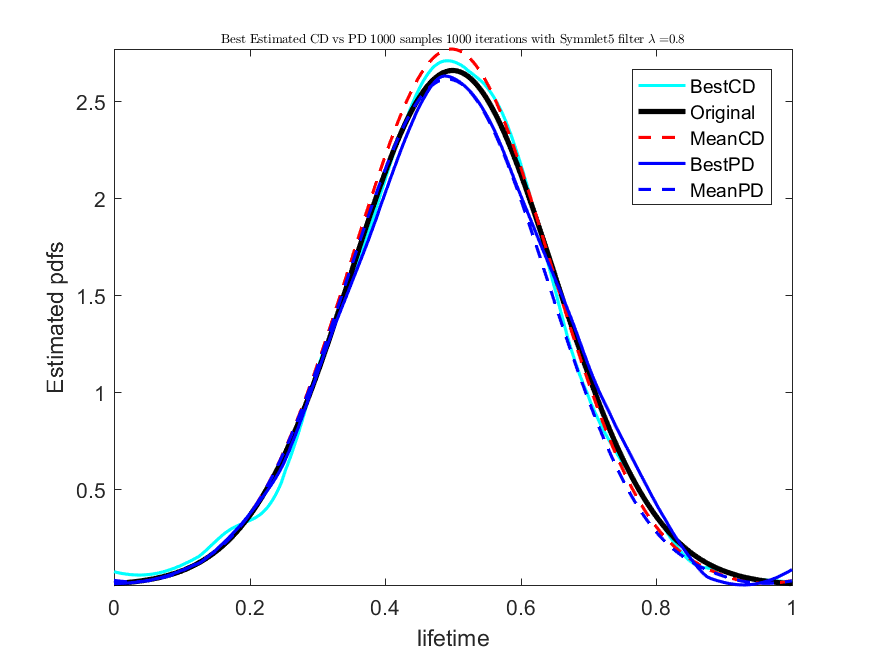}
       \caption{}
       \label{fig:normal1000}
    \end{subfigure}
    \caption{Estimate results for Normal distribution, $N=100,200,500,1000$ using Symmlet5.}\label{fig:normalAll}
\end{figure}

\begin{figure}[!htb]
   \centering
   \begin{subfigure}[b]{0.25\textwidth}
       \includegraphics[width=\textwidth]{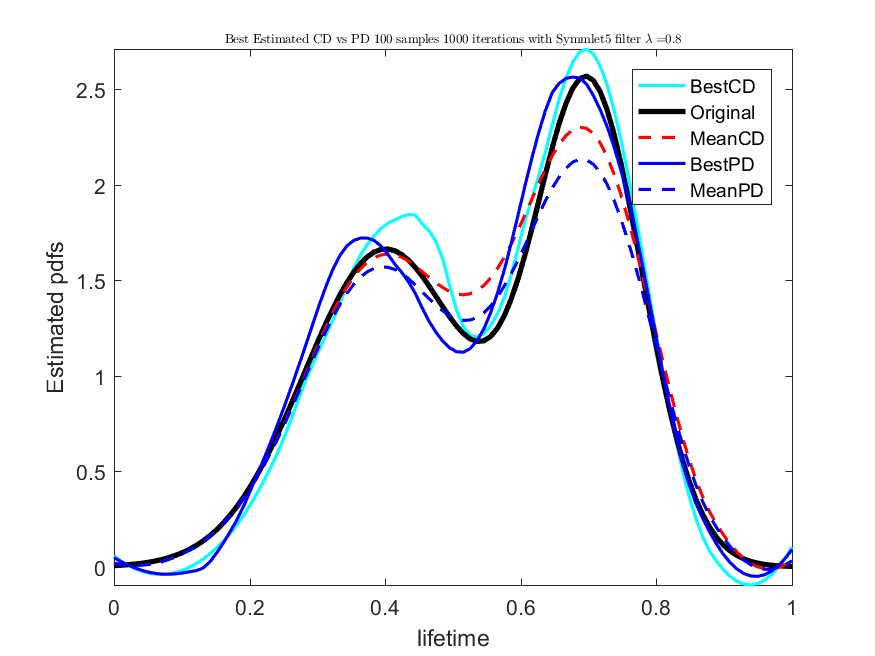}
       \caption{}
       \label{fig:bimodal100}
   \end{subfigure}\hfill
   \begin{subfigure}[b]{0.25\textwidth}
       \includegraphics[width=\textwidth]{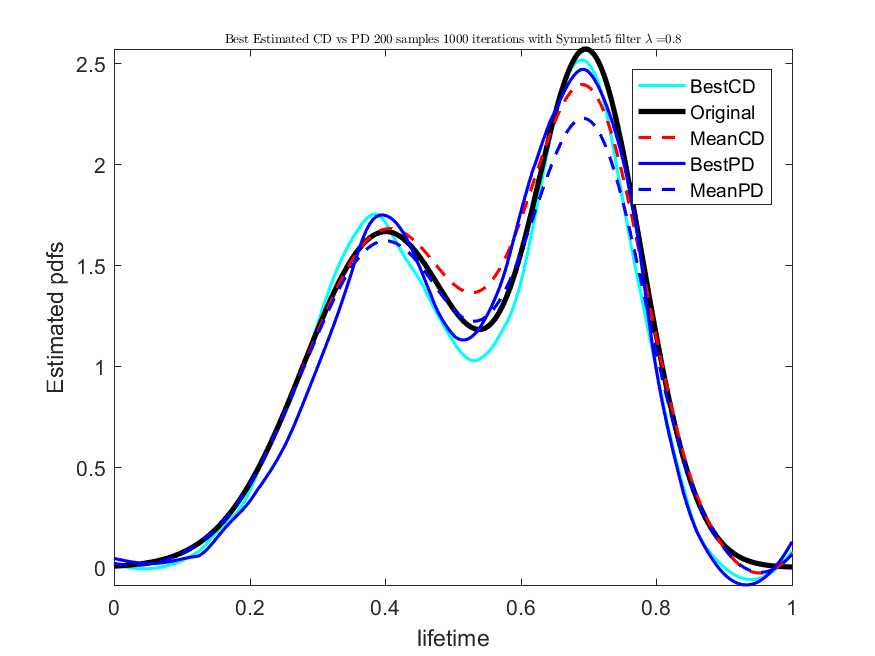}
       \caption{}
       \label{fig:bimodal200}
    \end{subfigure}\hfill
    \begin{subfigure}[b]{0.25\textwidth}
       \includegraphics[width=\textwidth]{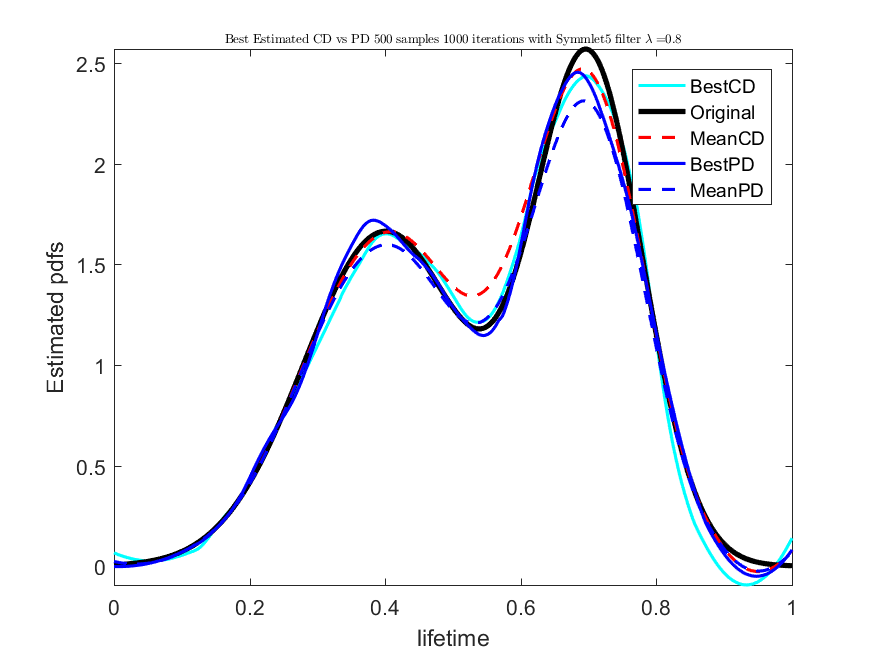}
       \caption{}
       \label{fig:bimodal500}
    \end{subfigure}\hfill
    \begin{subfigure}[b]{0.25\textwidth}
       \includegraphics[width=\textwidth]{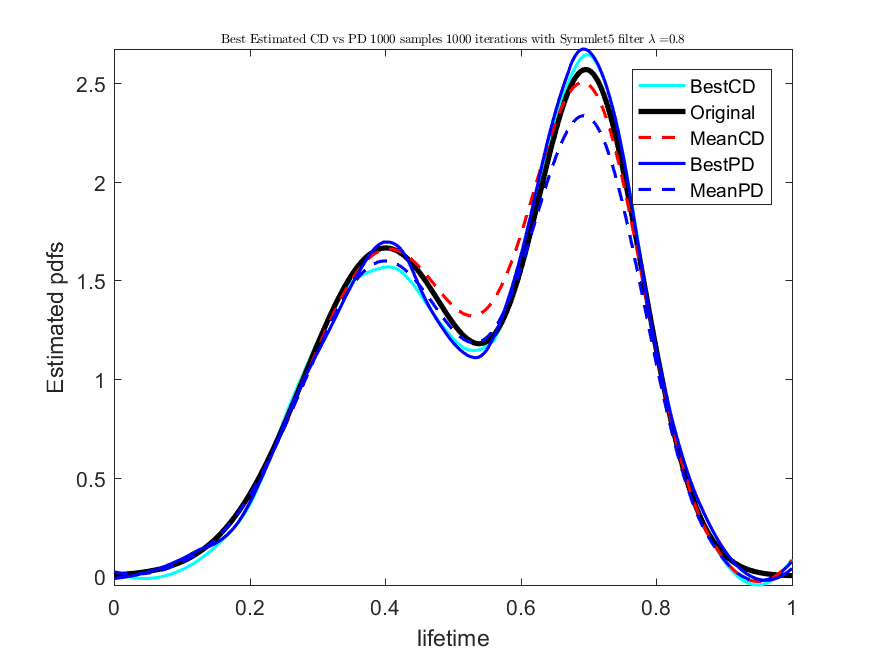}
       \caption{}
       \label{fig:bimodal1000}
    \end{subfigure}
    \caption{Estimate results for Bimodal distribution, $N=100,200,500,1000$ using Symmlet5.}\label{fig:bimodalAll}
\end{figure}

\begin{figure}[!htb]
   \centering
   \begin{subfigure}[b]{0.25\textwidth}
       \includegraphics[width=\textwidth]{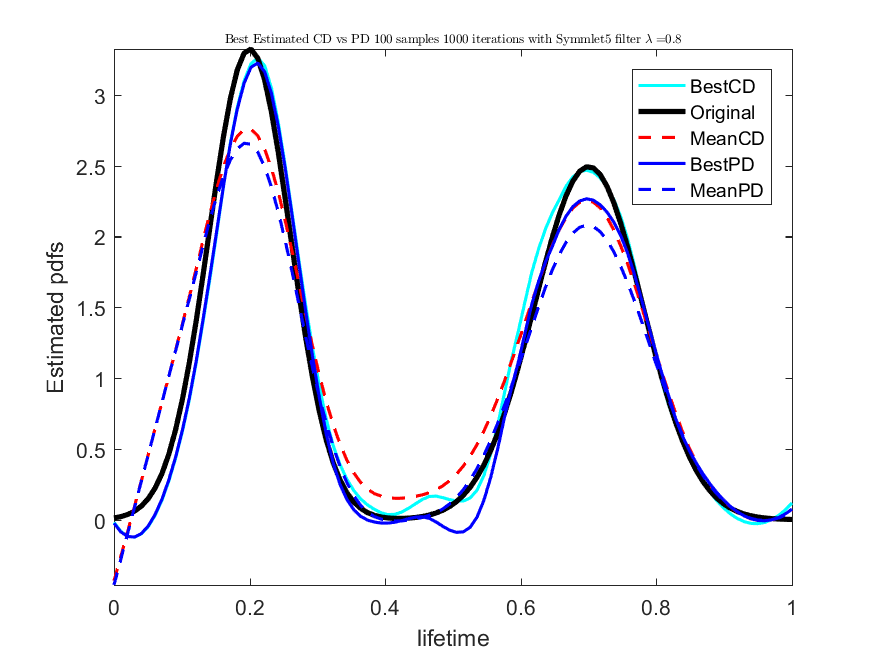}
       \caption{}
       \label{fig:strata100}
   \end{subfigure}\hfill
   \begin{subfigure}[b]{0.25\textwidth}
       \includegraphics[width=\textwidth]{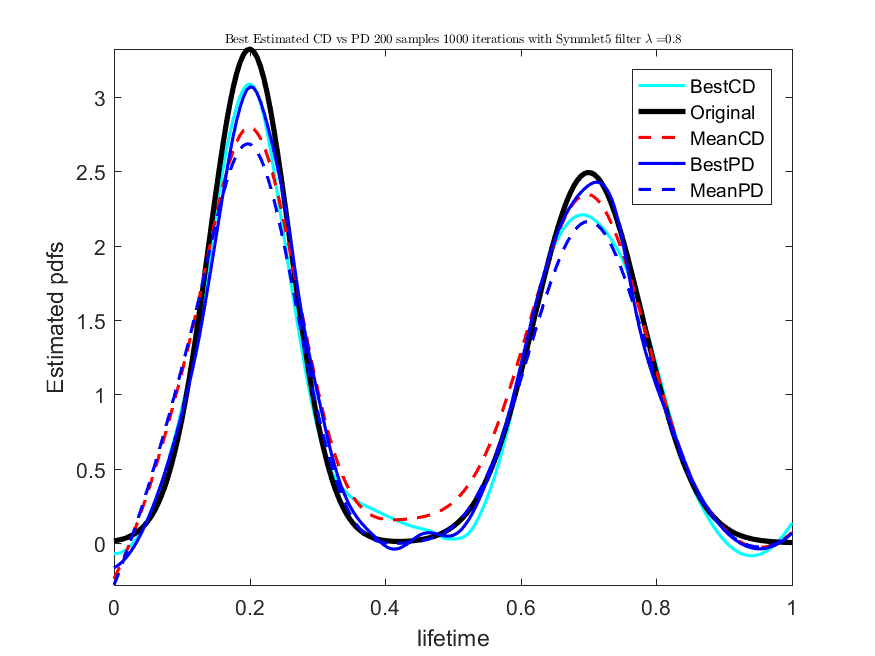}
       \caption{}
       \label{fig:strata200}
    \end{subfigure}\hfill
    \begin{subfigure}[b]{0.25\textwidth}
       \includegraphics[width=\textwidth]{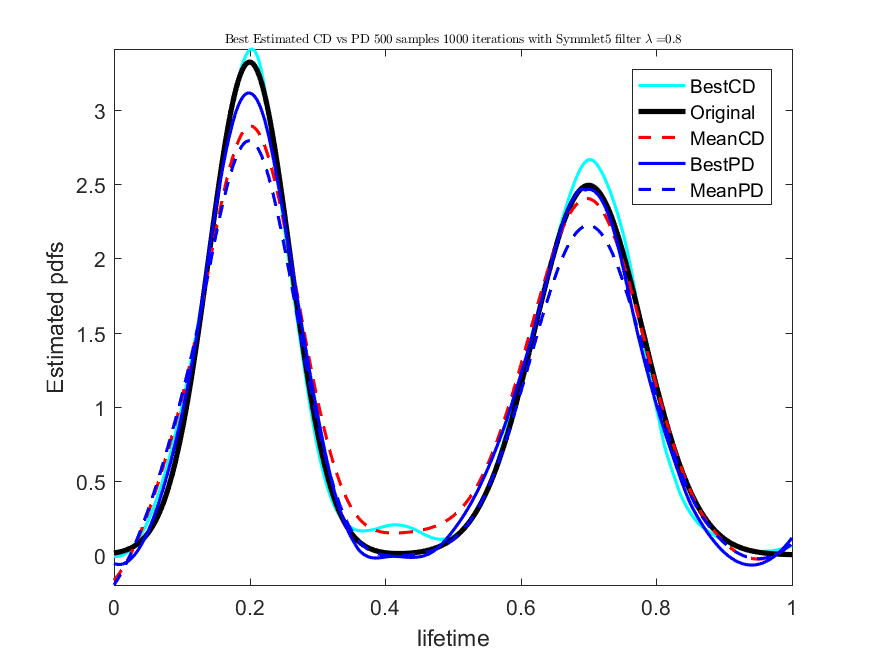}
       \caption{}
       \label{fig:strata500}
    \end{subfigure}\hfill
    \begin{subfigure}[b]{0.25\textwidth}
       \includegraphics[width=\textwidth]{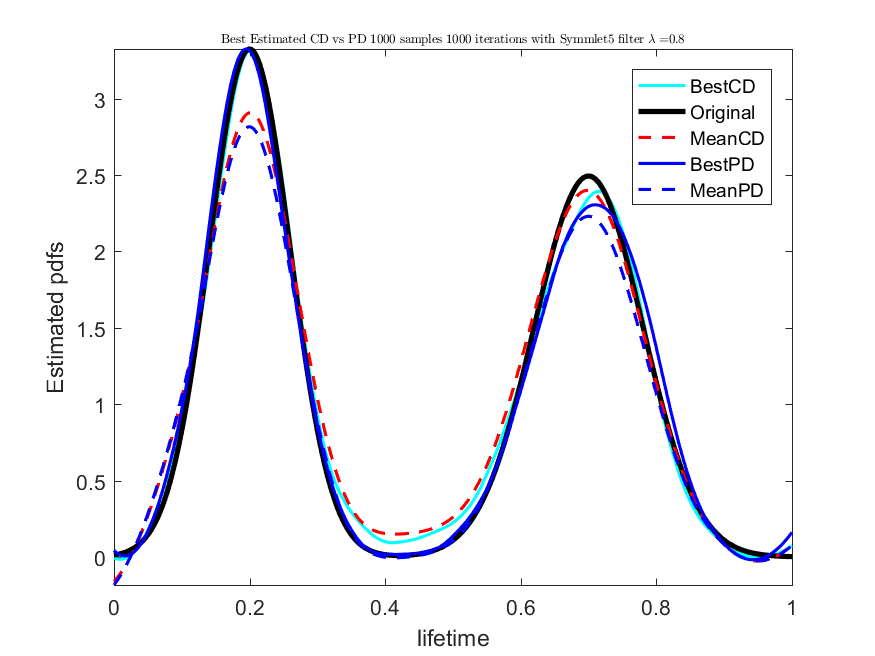}
       \caption{}
       \label{fig:strata1000}
    \end{subfigure}
    \caption{Estimate results for Strata distribution, $N=100,200,500,1000$ using Symmlet5.}\label{fig:stratalAll}
\end{figure}

\begin{figure}[!htb]
   \centering
   \begin{subfigure}[b]{0.25\textwidth}
       \includegraphics[width=\textwidth]{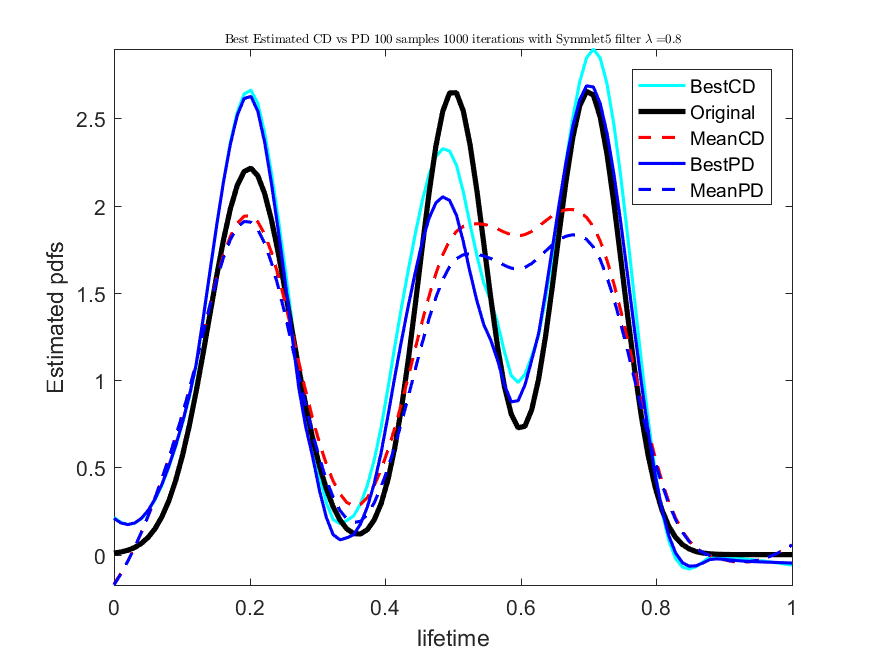}
       \caption{}
       \label{fig:multi100}
   \end{subfigure}\hfill
   \begin{subfigure}[b]{0.25\textwidth}
       \includegraphics[width=\textwidth]{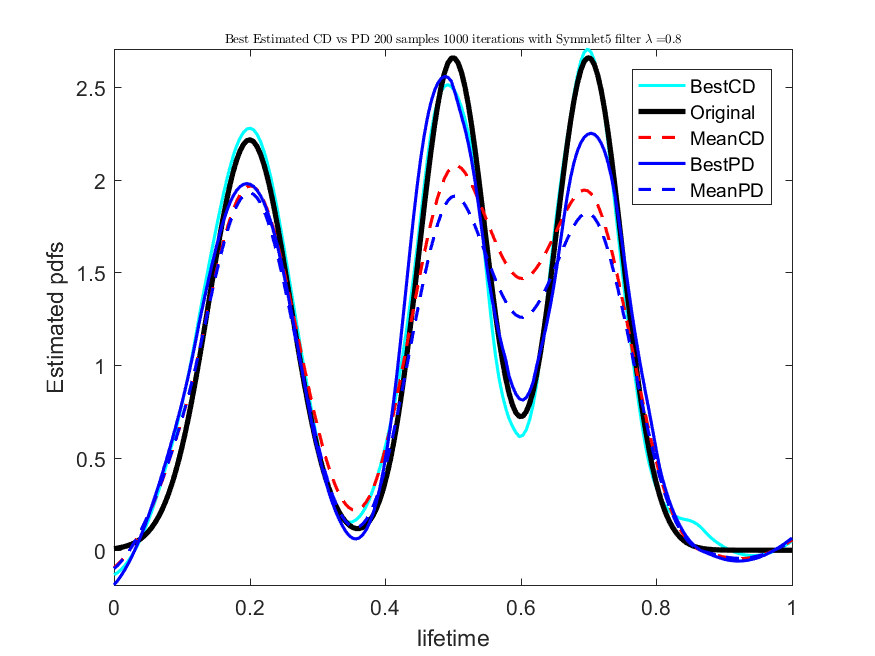}
       \caption{}
       \label{fig:multi200}
    \end{subfigure}\hfill
    \begin{subfigure}[b]{0.25\textwidth}
       \includegraphics[width=\textwidth]{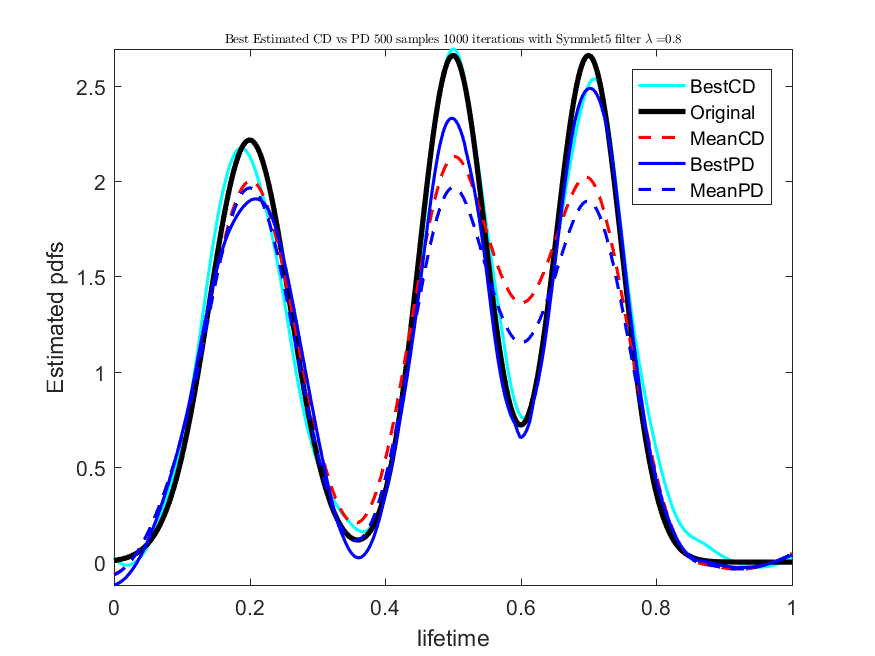}
       \caption{}
       \label{fig:multi500}
    \end{subfigure}\hfill
    \begin{subfigure}[b]{0.25\textwidth}
       \includegraphics[width=\textwidth]{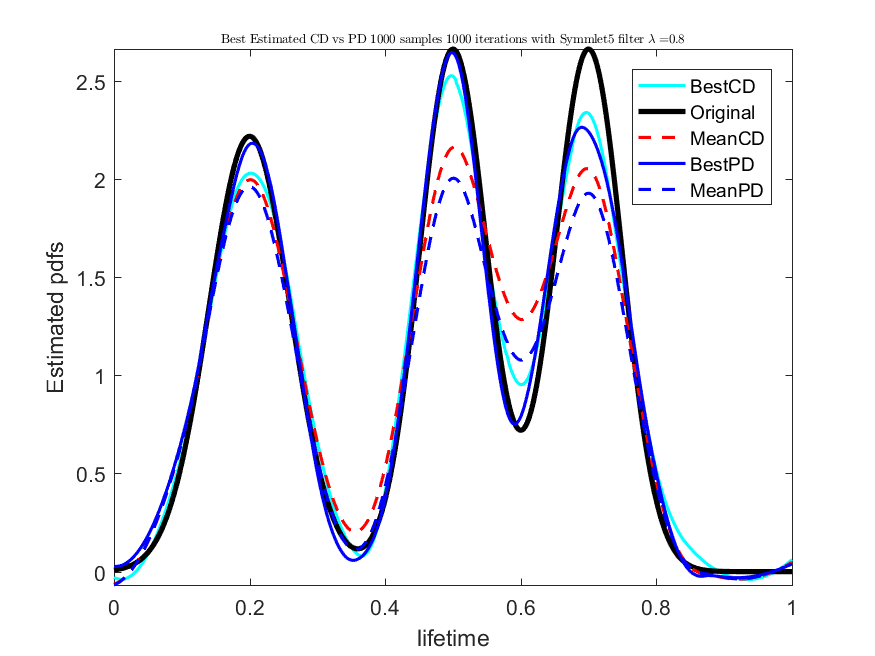}
       \caption{}
       \label{fig:multi1000}
    \end{subfigure}
    \caption{Estimate results for Multimodal distribution, $N=100,200,500,1000$ using Symmlet5.}\label{fig:multilAll}
\end{figure}

\begin{figure}[!htb]
   \centering
   \begin{subfigure}[b]{0.25\textwidth}
       \includegraphics[width=\textwidth]{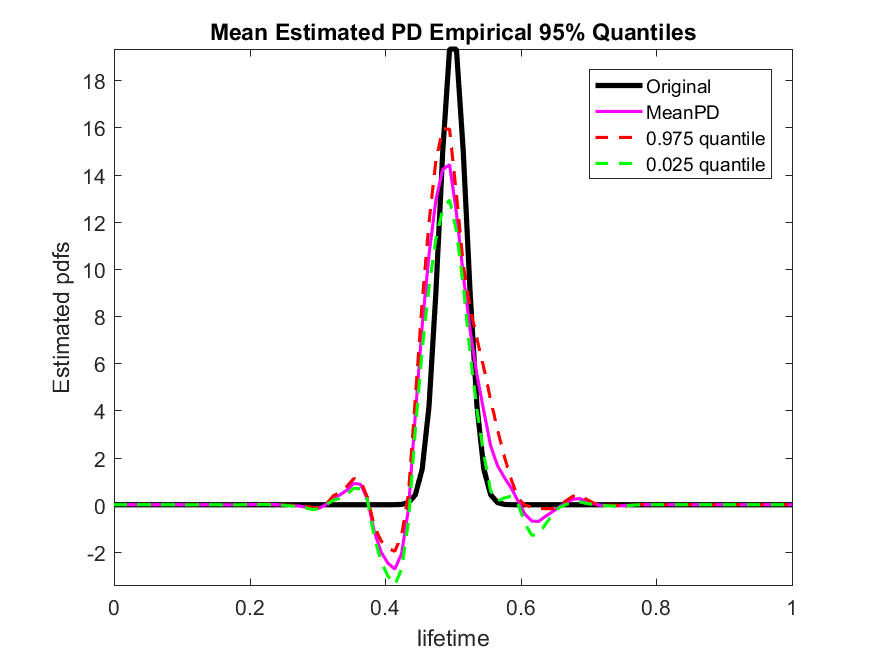}
       \caption{}
       \label{fig:deltaQ100Efro}
   \end{subfigure}\hfill
   \begin{subfigure}[b]{0.25\textwidth}
       \includegraphics[width=\textwidth]{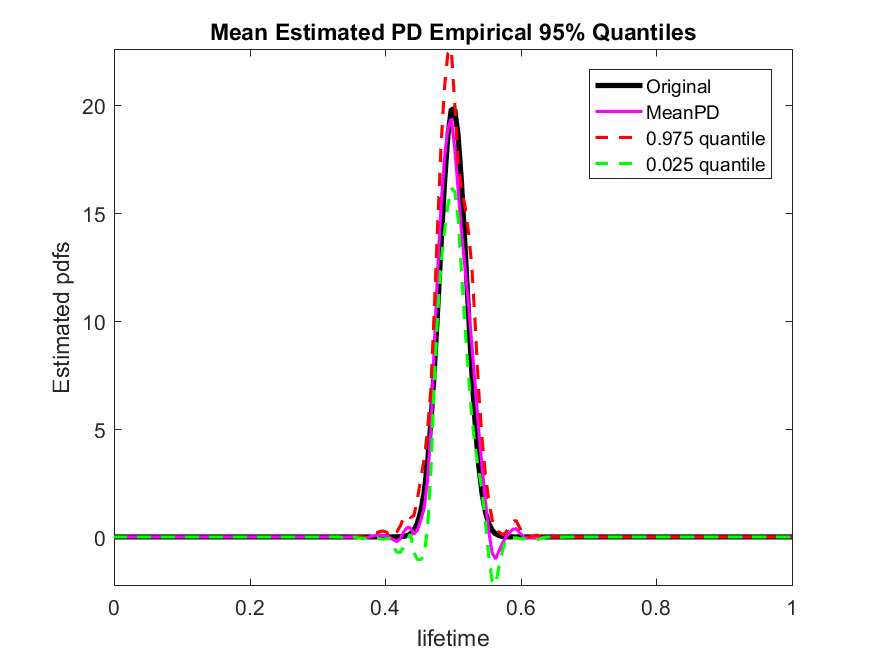}
       \caption{}
       \label{fig:deltaQ200Efro}
    \end{subfigure}\hfill
    \begin{subfigure}[b]{0.25\textwidth}
       \includegraphics[width=\textwidth]{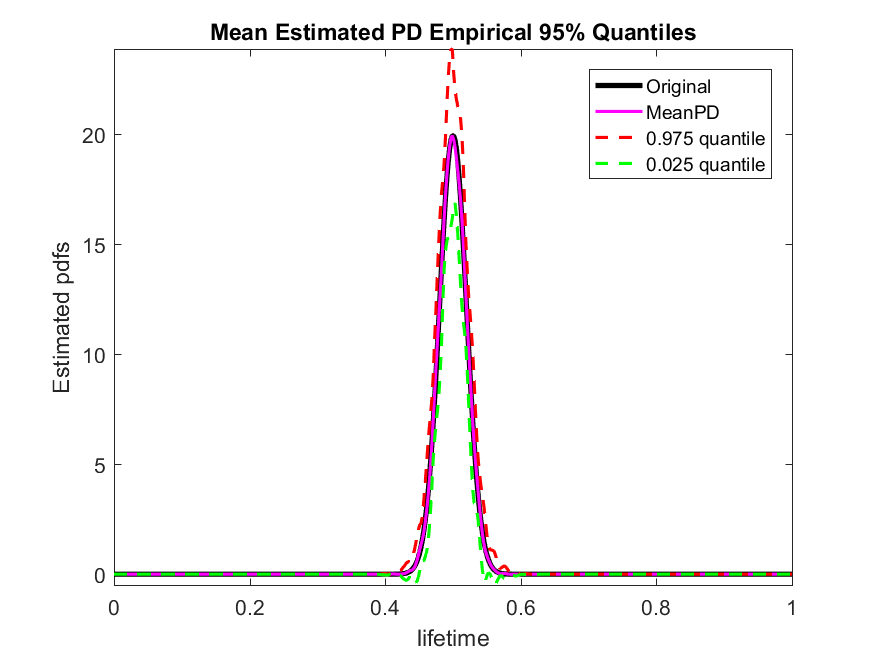}
       \caption{}
       \label{fig:deltaQ500Efro}
    \end{subfigure}\hfill
    \begin{subfigure}[b]{0.25\textwidth}
       \includegraphics[width=\textwidth]{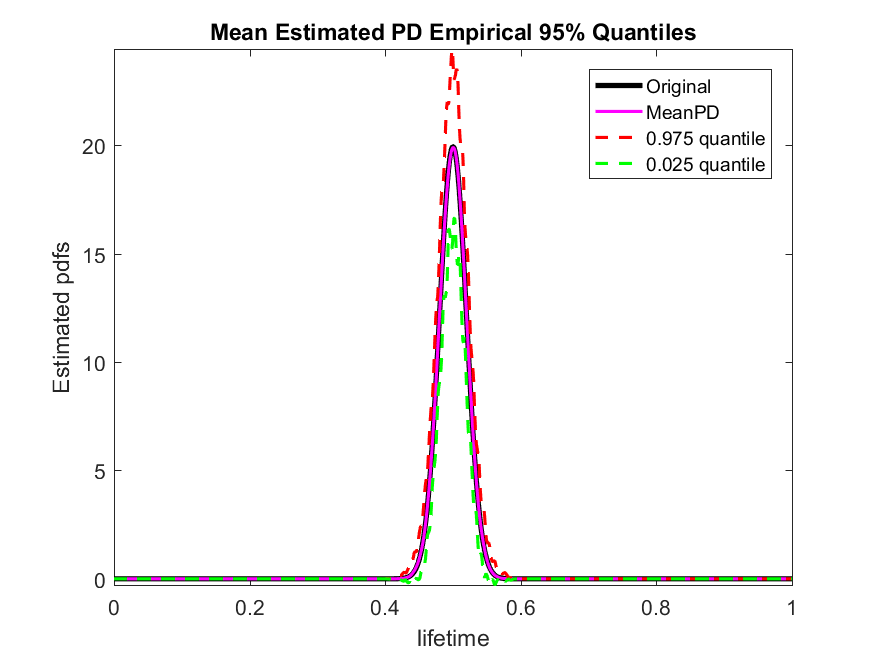}
       \caption{}
       \label{fig:deltaQ1000Efro}
    \end{subfigure}
\caption{Results for 95\% empirical quantiles and average estimate for Delta distribution using Symmlet5.(a)-(d) correspond to the partial data approach (for $N=100,200,500,1000$, respectively).}
\label{fig:qdelta}
\end{figure}

\begin{figure}[!htb]
   \centering
   \begin{subfigure}[b]{0.25\textwidth}
       \includegraphics[width=\textwidth]{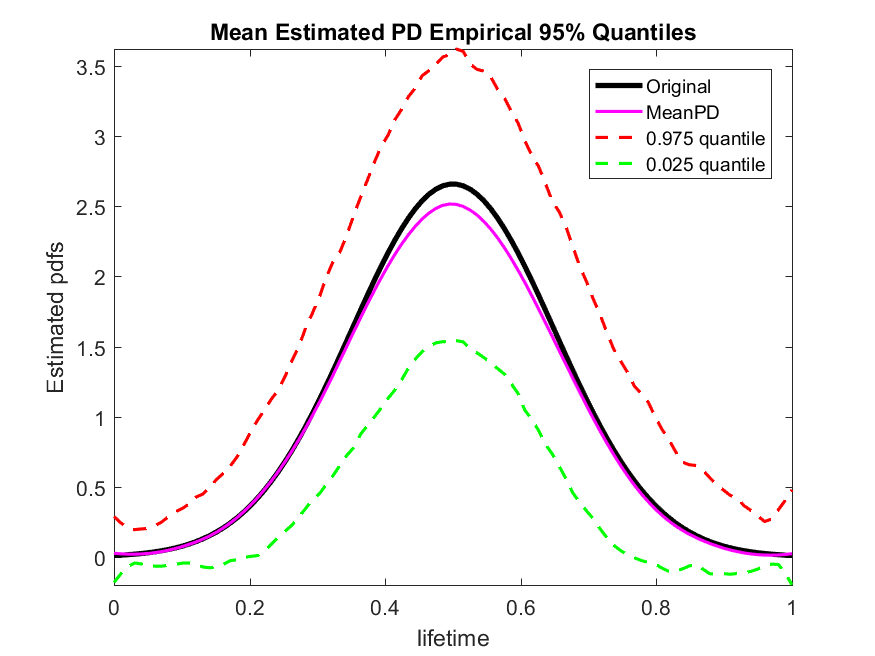}
       \caption{}
       \label{fig:normalQ100Efro}
   \end{subfigure}\hfill
   \begin{subfigure}[b]{0.25\textwidth}
       \includegraphics[width=\textwidth]{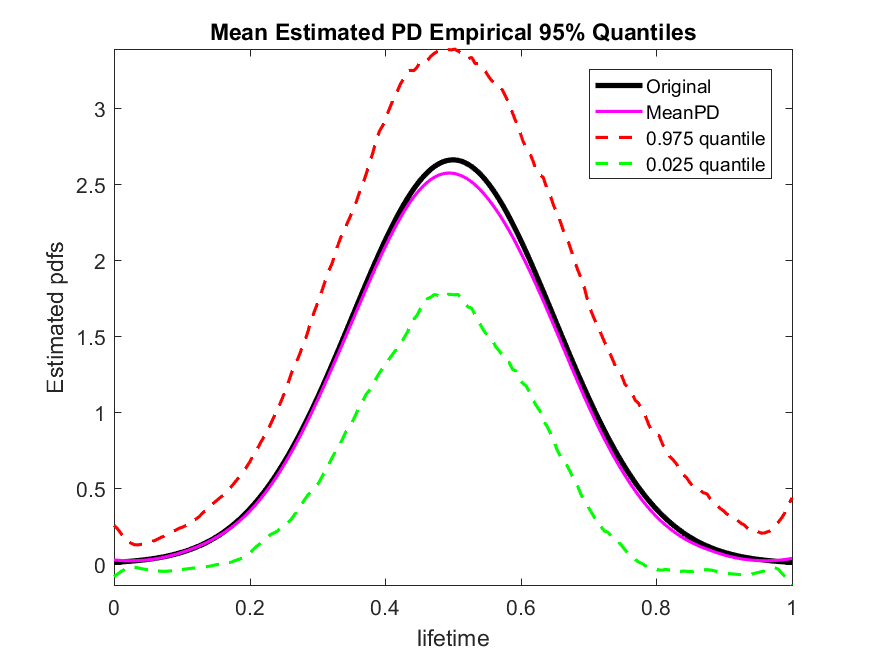}
       \caption{}
       \label{fig:normalQ200Efro}
    \end{subfigure}\hfill
    \begin{subfigure}[b]{0.25\textwidth}
       \includegraphics[width=\textwidth]{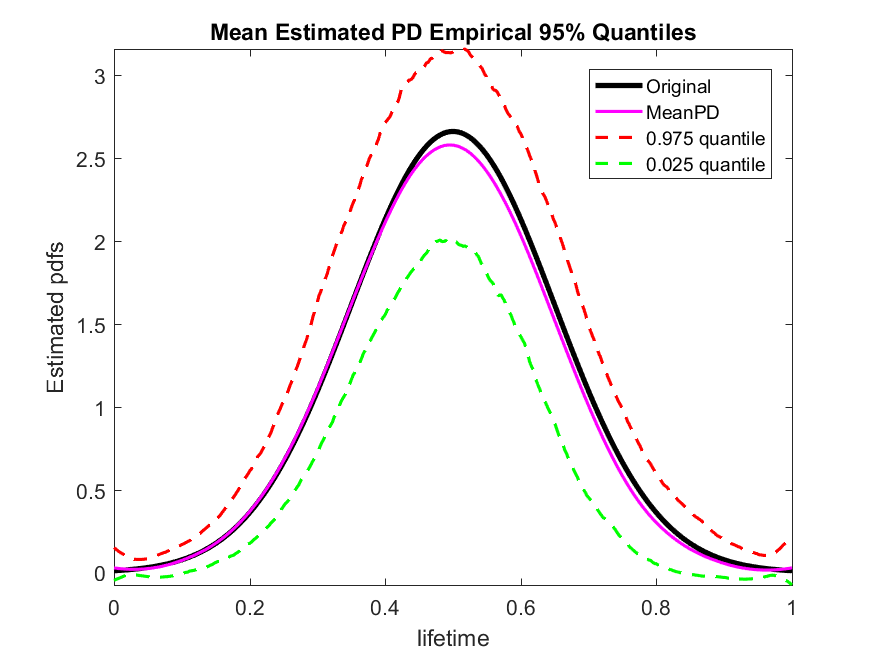}
       \caption{}
       \label{fig:normalQ500Efro}
    \end{subfigure}\hfill
    \begin{subfigure}[b]{0.25\textwidth}
       \includegraphics[width=\textwidth]{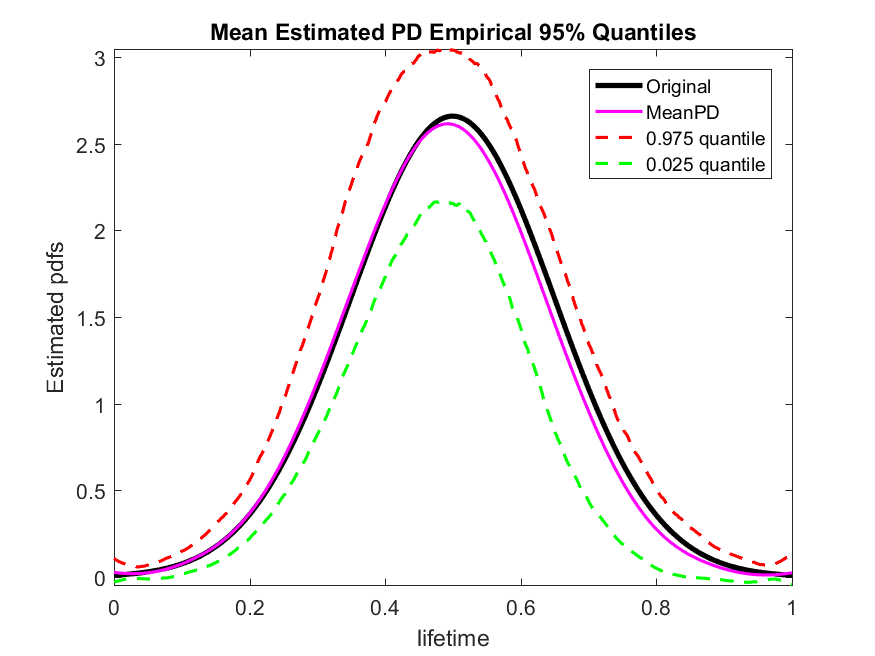}
       \caption{}
       \label{fig:normalQ1000Efro}
    \end{subfigure}
\caption{Results for 95\% empirical quantiles and average estimate for Normal distribution using Symmlet5.(a)-(d) correspond to the partial data approach (for $N=100,200,500,1000$, respectively).}
\label{fig:qnormal}
\end{figure}

\begin{figure}[!htb]
   \centering
   \begin{subfigure}[b]{0.25\textwidth}
       \includegraphics[width=\textwidth]{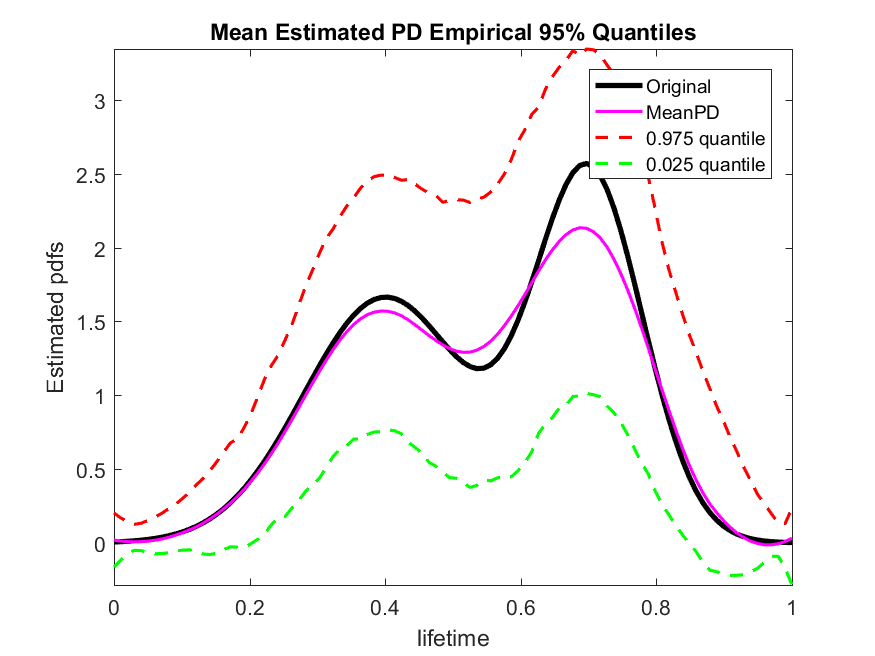}
       \caption{}
       \label{fig:bimodalQ100Efro}
   \end{subfigure}\hfill
   \begin{subfigure}[b]{0.25\textwidth}
       \includegraphics[width=\textwidth]{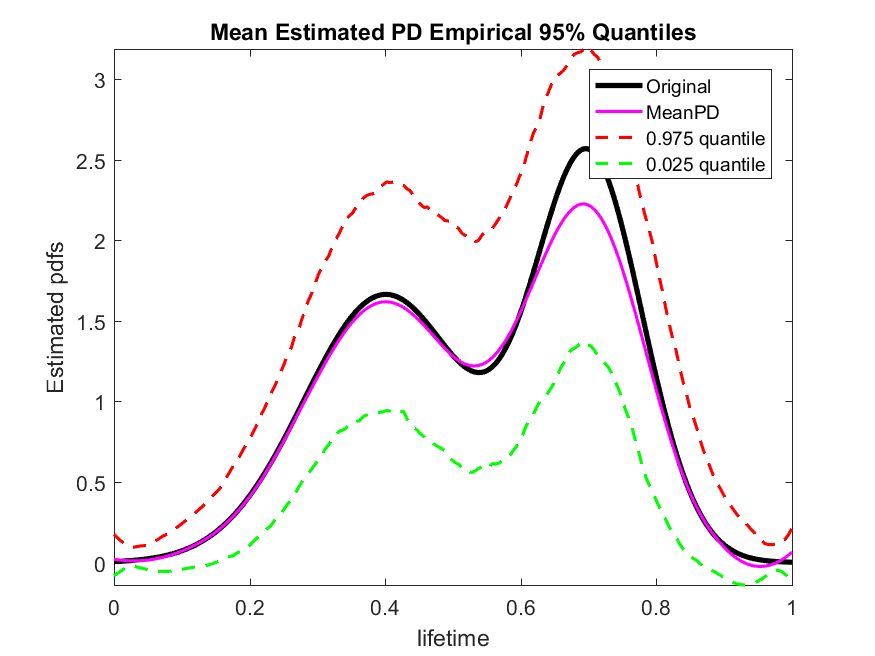}
       \caption{}
       \label{fig:bimodalQ200Efro}
    \end{subfigure}\hfill
    \begin{subfigure}[b]{0.25\textwidth}
       \includegraphics[width=\textwidth]{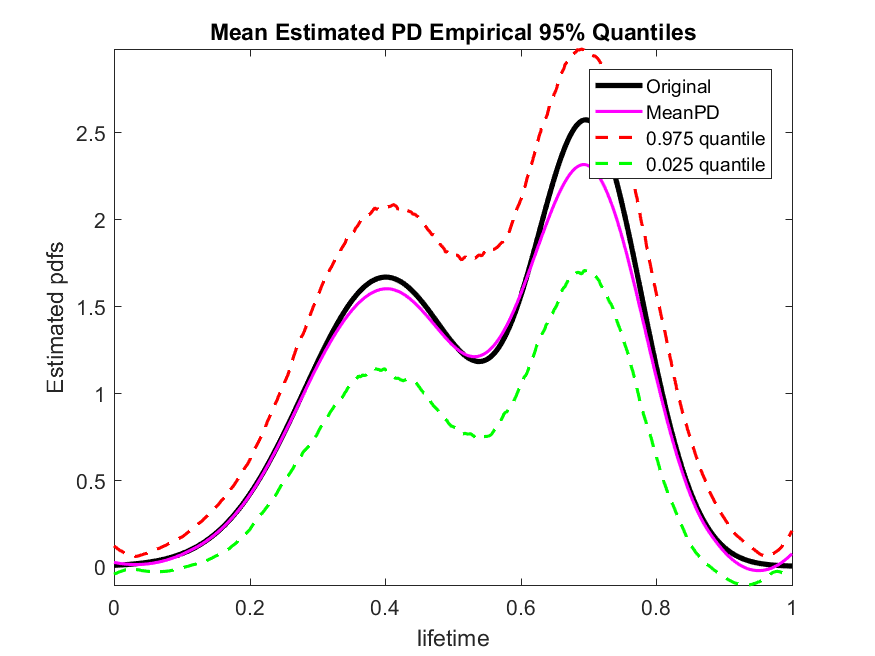}
       \caption{}
       \label{fig:bimodalQ500Efro}
    \end{subfigure}\hfill
    \begin{subfigure}[b]{0.25\textwidth}
       \includegraphics[width=\textwidth]{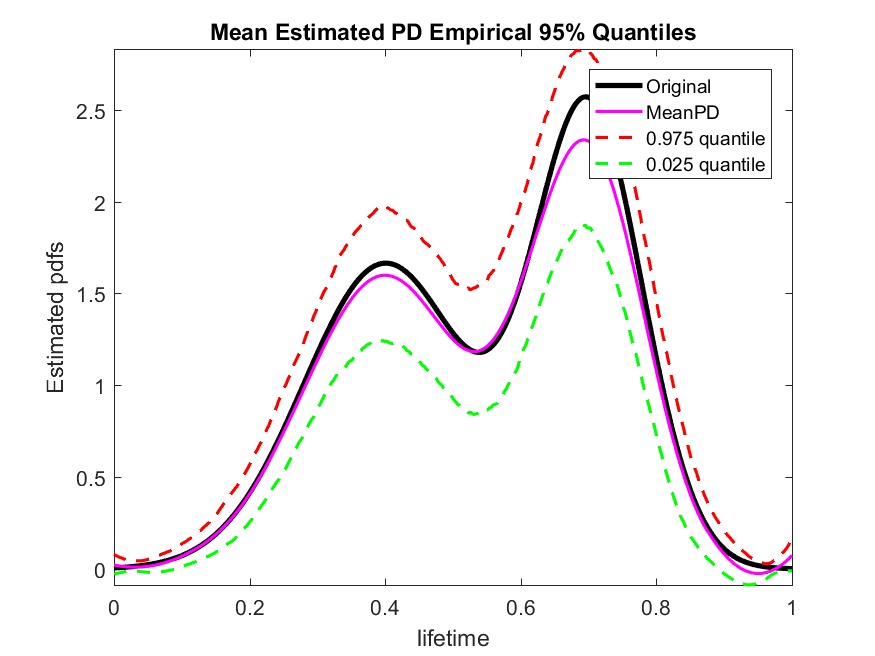}
       \caption{}
       \label{fig:bimodalQ1000Efro}
    \end{subfigure}
\caption{Results for 95\% empirical quantiles and average estimate for Bimodal distribution using Symmlet5.(a)-(d) correspond to the partial data approach (for $N=100,200,500,1000$, respectively).}
\label{fig:BimodalQ}
\end{figure}

\begin{figure}[!htb]
   \centering
   \begin{subfigure}[b]{0.25\textwidth}
       \includegraphics[width=\textwidth]{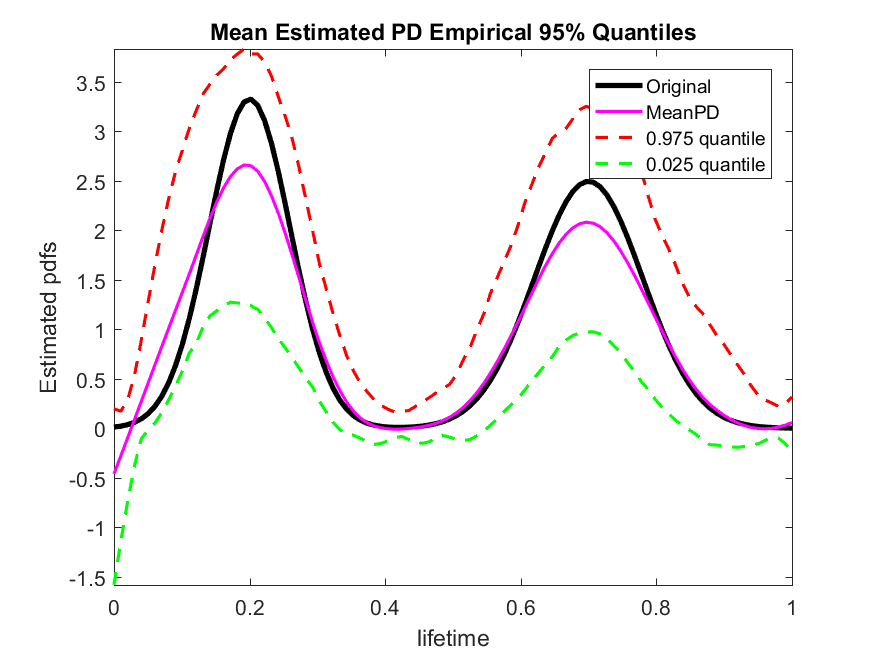}
       \caption{}
       \label{fig:strataQ100Efro}
   \end{subfigure}\hfill
   \begin{subfigure}[b]{0.25\textwidth}
       \includegraphics[width=\textwidth]{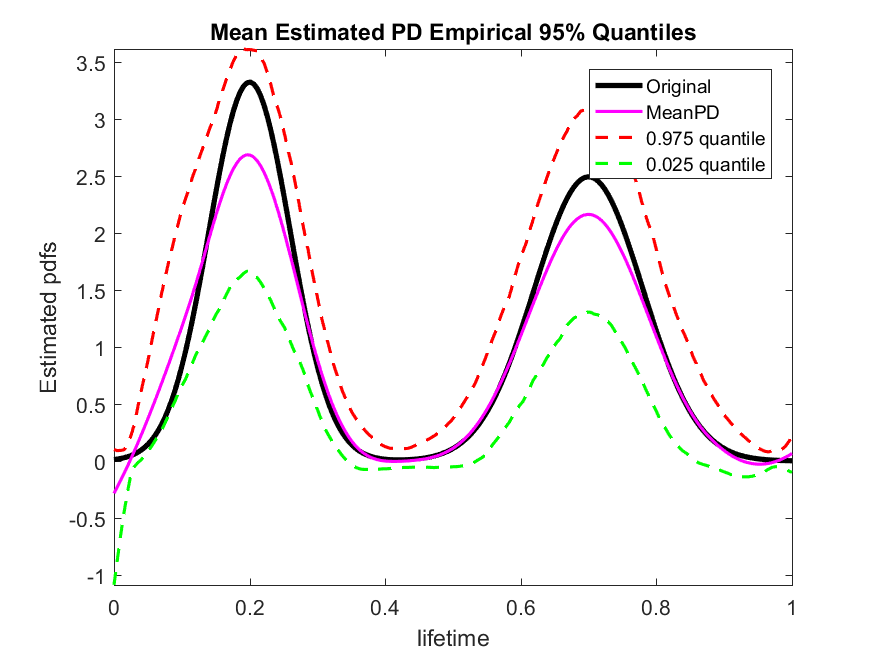}
       \caption{}
       \label{fig:strataQ200Efro}
    \end{subfigure}\hfill
    \begin{subfigure}[b]{0.25\textwidth}
       \includegraphics[width=\textwidth]{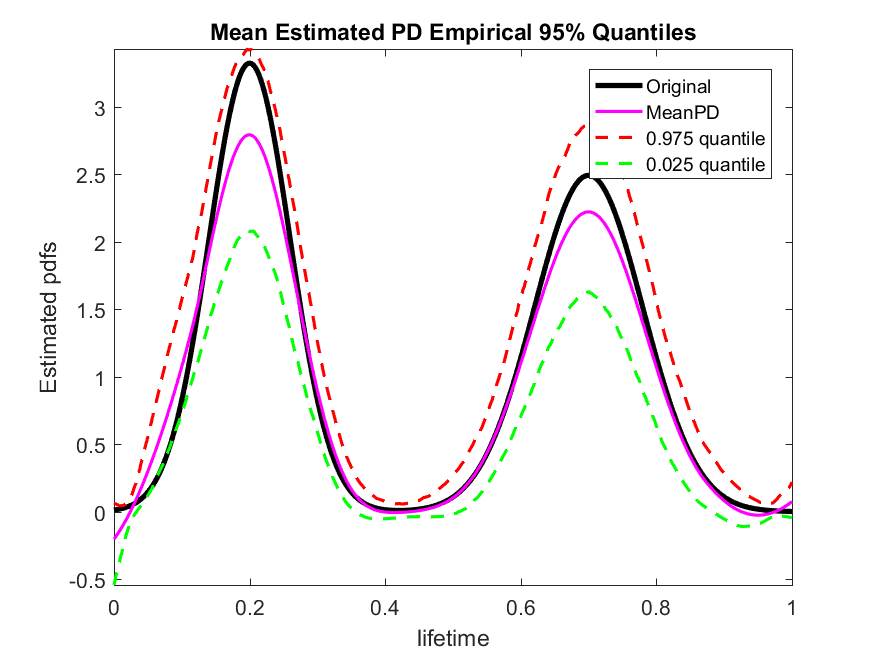}
       \caption{}
       \label{fig:strataQ500Efro}
    \end{subfigure}\hfill
    \begin{subfigure}[b]{0.25\textwidth}
       \includegraphics[width=\textwidth]{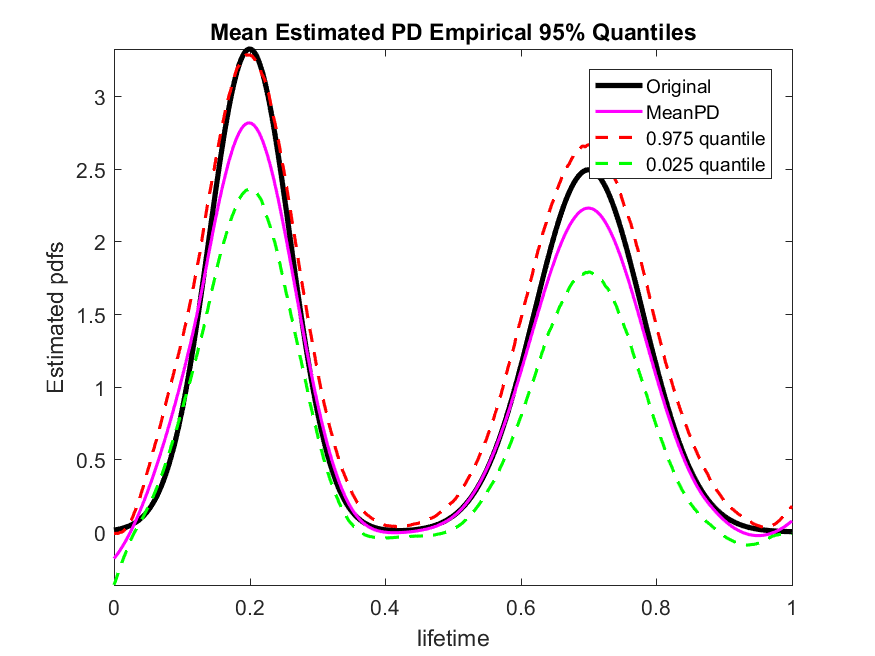}
       \caption{}
       \label{fig:strataQ1000Efro}
    \end{subfigure}
\caption{Results for 95\% empirical quantiles and average estimate for Strata distribution using Symmlet5.(a)-(d) correspond to the partial data approach (for $N=100,200,500,1000$, respectively).}
\label{fig:StrataQ}
\end{figure}

\begin{figure}[!htb]
   \centering
   \begin{subfigure}[b]{0.25\textwidth}
       \includegraphics[width=\textwidth]{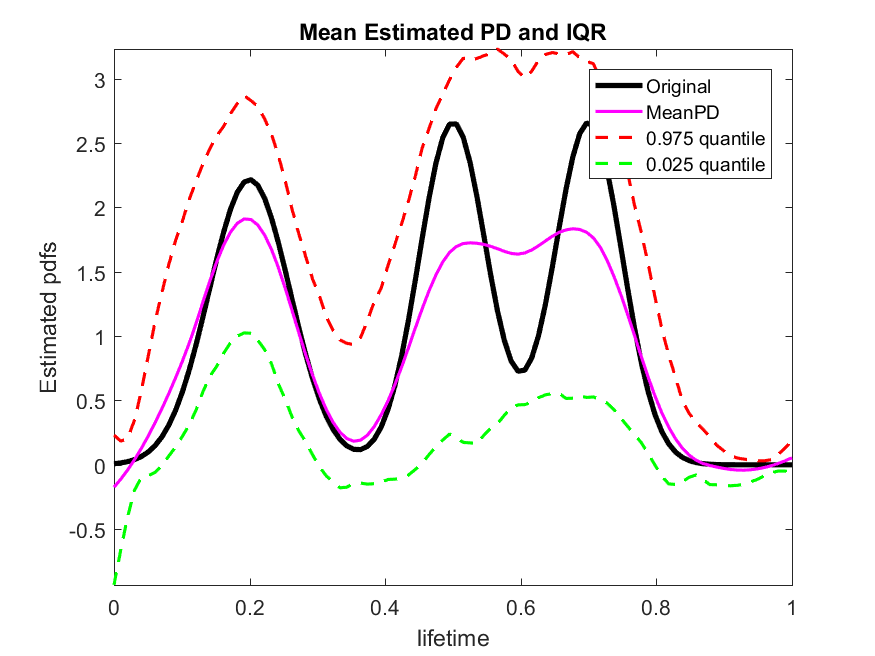}
       \caption{}
       \label{fig:multiQ100Efro}
   \end{subfigure}\hfill
   \begin{subfigure}[b]{0.25\textwidth}
       \includegraphics[width=\textwidth]{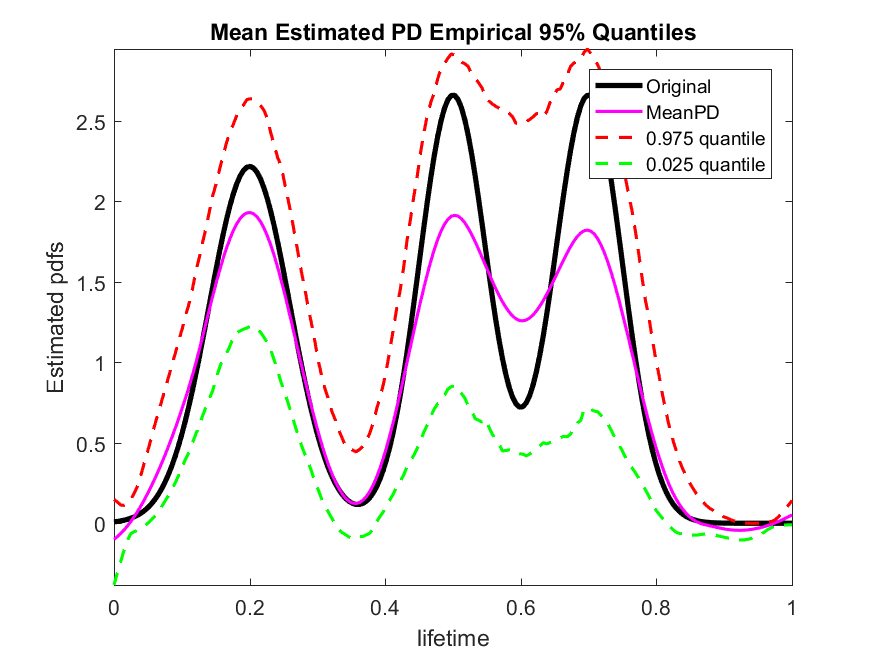}
       \caption{}
       \label{fig:multiQ200Efro}
    \end{subfigure}\hfill
    \begin{subfigure}[b]{0.25\textwidth}
       \includegraphics[width=\textwidth]{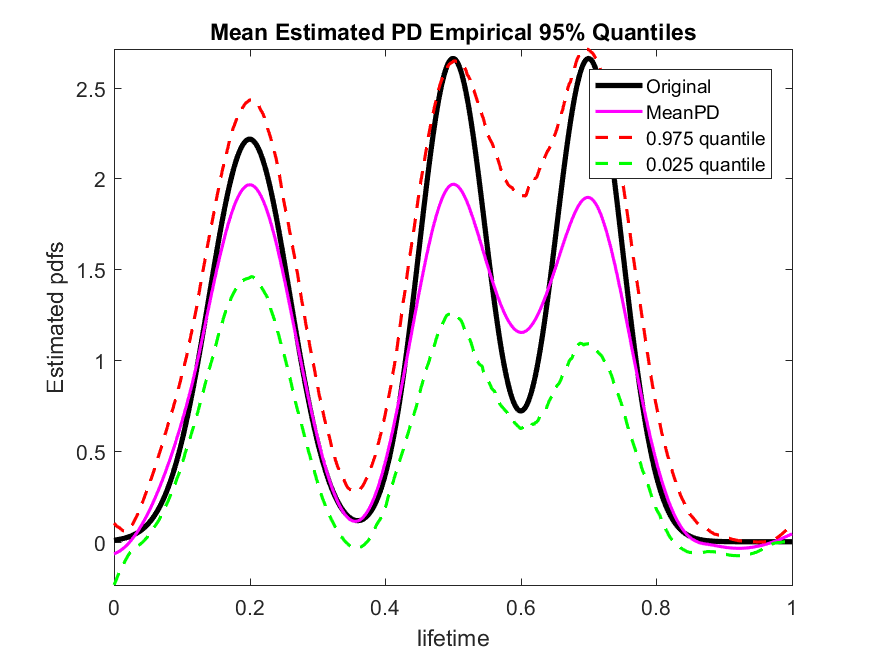}
       \caption{}
       \label{fig:multiQ500Efro}
    \end{subfigure}\hfill
    \begin{subfigure}[b]{0.25\textwidth}
       \includegraphics[width=\textwidth]{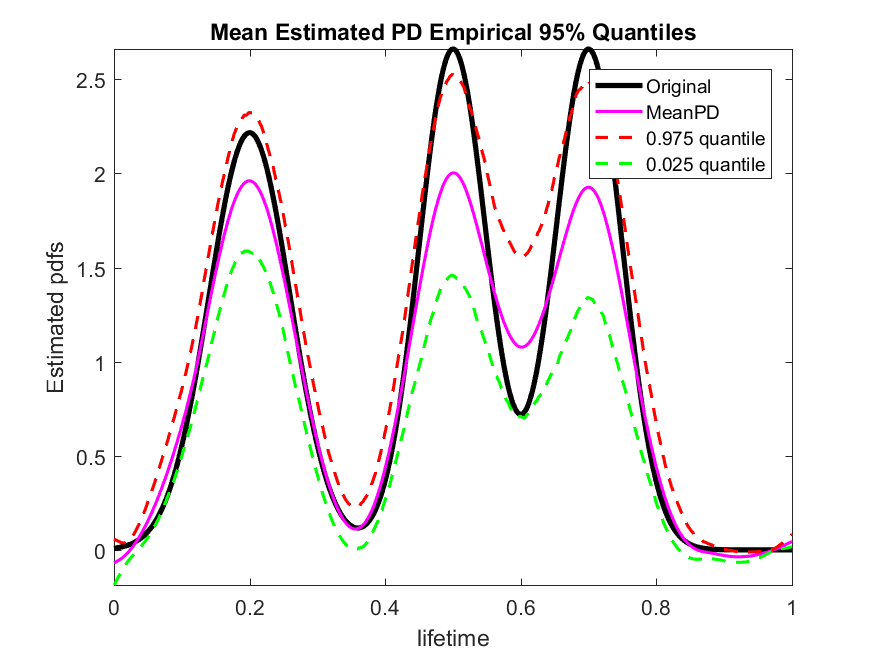}
       \caption{}
       \label{fig:multiQ1000Efro}
    \end{subfigure}
\caption{Results for 95\% empirical quantiles and average estimate for Multimodal distribution using Symmlet5.(a)-(d) correspond to the partial data approach (for $N=100,200,500,1000$, respectively).}
\label{fig:MultiQ}
\end{figure}

\begin{figure}[!htb]
   \centering
   \begin{subfigure}[b]{0.45\textwidth}
       \includegraphics[width=\textwidth]{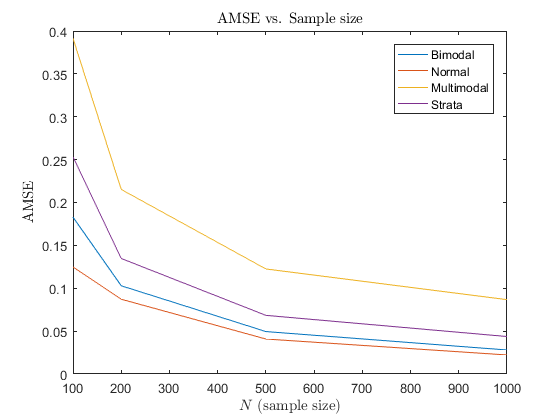}
       \caption{}
       \label{fig:AMSE}
   \end{subfigure}\hfill
   \begin{subfigure}[b]{0.45\textwidth}
       \includegraphics[width=\textwidth]{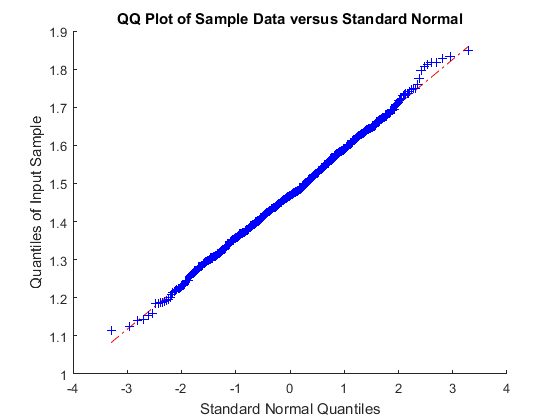}
       \caption{}
       \label{fig:QQ_Bimodal_X700_N2000}
    \end{subfigure}
\caption{(a) AMSE for baseline distributions. (b) Q-Q Plot for the density estimates for Bimodal Distribution, $N=1000$, $x=0.7$.}
\label{fig:AMSEqq}
\end{figure}

In this section, we summarize the results obtained for each baseline distribution. In particular, the following results are provided:

\begin{enumerate}[(a)]
  \item Tables \ref{tab:deltaPD} to \ref{tab:multimodalPD} present details for AMSE results obtained for each baseline distribution used in the study.
  \item In figures \ref{fig:deltaAll} - \ref{fig:multilAll}, dashed lines (red and blue) correspond to the average estimates for $\hat{f}^{PD}(x)$, computed at each data point $x$ from all $B=1000$ replications. The black line indicates the actual density function and the light blue and blue continuous lines represents the best estimates among all replications (i.e. the one with the smallest AMSE).
  \item In figures \ref{fig:qdelta} - \ref{fig:MultiQ}, dashed lines (red and green) correspond to the empirical 95\% quantiles computed at each data point $x$ from all $B=1000$ replications, for $\hat{f}^{b}(x)$ and $\hat{f}^{PD}(x)$ respectively. The blue and magenta lines show the average density estimates for the complete and partial data approach, respectively. The black line indicates the actual density function.
  \item Figure \ref{fig:AMSE} shows the AMSE vs. sample size plot.
  \item Figure \ref{fig:QQ_Bimodal_X700_N2000} exemplifies the asymptotic normality behavior of the density estimates, as proposed in \ref{limitDist}.
\end{enumerate}

\begin{table}[!htb]
\parbox{.50\linewidth}{
\centering
\scalebox{0.73}{
\begin{tabular}[h]{|c|c|c|c|c|}
\hline
PD Estimator & $N=100$ & $N=200$ & $N=500$ & $N=1000$  \\
\hline
\hline
Mean AMSE & 2.5954  &  0.3674   &  0.1856   &  0.2216    \\
\hline
St.Dev. AMSE  & 0.0986  &  0.1680   &  0.1301   &  0.1009   \\
\hline
Min AMSE  & 2.5149  &  0.2010   &  0.0112   &  0.0216 \\
\hline
Max AMSE  & 3.5061  &  1.3967   &  0.8243   &  0.6893 \\
\hline
\end{tabular}}
\caption{AMSE results for Delta distribution with Partial data estimator.} \label{tab:deltaPD}
}
\hfill
\parbox{.50\linewidth}{
\centering
\scalebox{0.73}{
\begin{tabular}[h]{|c|c|c|c|c|}
\hline
PD Estimator & $N=100$ & $N=200$ & $N=500$ & $N=1000$  \\
\hline
\hline
Mean AMSE & 0.1219  &  0.0821   &  0.0385   &  0.0214    \\
\hline
St.Dev. AMSE  & 0.0858  &  0.0524   &  0.0230   &  0.0129   \\
\hline
Min AMSE  & 0.0036  &  0.0086   &  0.0037   &  0.0031 \\
\hline
Max AMSE  & 0.5426  &  0.5058   &  0.1764   &  0.0872 \\
\hline
\end{tabular}}
\caption{AMSE results for Normal distribution with Partial data estimator.}
\label{tab:normalPD}
}
\end{table}


\begin{table}[!htb]

\parbox{.50\linewidth}{
\centering
\scalebox{0.73}{
\begin{tabular}[h]{|c|c|c|c|c|}
\hline
PD Estimator & $N=100$ & $N=200$ & $N=500$ & $N=1000$  \\
\hline
\hline
Mean AMSE & 0.1764  &  0.1041   &  0.0494   &  0.0296    \\
\hline
St.Dev. AMSE  & 0.1110  &  0.0620   &  0.0275   &  0.0175   \\
\hline
Min AMSE  & 0.0175  &  0.0123   &  0.0041   &  0.0030 \\
\hline
Max AMSE  & 0.9177  &  0.4933   &  0.1850   &  0.1323 \\
\hline
\end{tabular}}
\caption{AMSE results for Bimodal distribution with Partial data estimator.}
\label{tab:bimodalPD}
}
\hfill
\parbox{.50\linewidth}{
\centering
\scalebox{0.73}{
\begin{tabular}[h]{|c|c|c|c|c|}
\hline
PD Estimator & $N=100$ & $N=200$ & $N=500$ & $N=1000$  \\
\hline
\hline
Mean AMSE & 0.2468  &  0.1422   &  0.0731   &  0.0491    \\
\hline
St.Dev. AMSE  & 0.1485  &  0.0854   &  0.0420   &  0.0243   \\
\hline
Min AMSE  & 0.0225  &  0.0130   &  0.0078   &  0.0102 \\
\hline
Max AMSE  & 1.0432  &  0.6857   &  0.3657   &  0.1783 \\
\hline
\end{tabular}}
\caption{AMSE results for Strata distribution with Partial data estimator.}
\label{tab:strataPD}
}
\end{table}

\begin{table}[!htb]
\centering
\scalebox{0.73}{
\begin{tabular}[h]{|c|c|c|c|c|}
\hline
PD Estimator & $N=100$ & $N=200$ & $N=500$ & $N=1000$  \\
\hline
\hline
Mean AMSE & 0.3838  &  0.2183   &  0.1321   &  0.2216    \\
\hline
St.Dev. AMSE  & 0.1595  &  0.1108   &  0.0652   &  0.2193   \\
\hline
Min AMSE  & 0.0619  &  0.0289   &  0.0171   &  0.2193 \\
\hline
Max AMSE  & 1.0382  &  0.5863   &  0.4589   &  0.2193 \\
\hline
\end{tabular}}
\caption{AMSE results for Multimodal distribution with Partial data estimator.}
\label{tab:multimodalPD}
\end{table}


\subsection{Remarks and comments.}

\begin{enumerate}[(i)]
  \item From the resulting figures, it is possible to observe that the proposed estimator is able to recover the underlying density in the presence of right-censored observations. Also, estimates (Best and Mean) with respect to the sample size, suggests a bias effect in the vicinity of the underlying distribution modes.
  \item In terms of the sensibility of the estimator's performance to the scaling functions, we observed during our experiments that results obtained using Symmlets, Coiflets and Daubechies wavelets are similar.
  \item From the quantiles plots, the empirical quantiles of the estimated densities contain the actual values of the target density in most of its support. Moreover, for all baseline distributions except for the Multimodal, this is the case. On the contrary, the regions where the 95\% empirical quantiles do not contain the true density value are observed to occur in the vicinities of the distribution modes. This could be caused by the choice of the multiresolution index $J$, the post-processing smoothing procedure and/or by the censoring effect.
  \item As the sample size increases, it was observed that the interval $|\hat{f_{N}}_{0.975}(x)-\hat{f_{N}}_{0.025}|$ monotonically decreases in coherence with the theoretical convergence results shown in section \ref{StuteError}.
  \item From the AMSE plot (\ref{fig:AMSE}), it is possible to observe that all baseline distributions present a similar error decay behavior. Moreover, results contained in tables \ref{tab:deltaPD} to \ref{tab:multimodalPD}, imply that as $N$ grows, the standard deviation and range of AMSE decays in accordance with the convergence rates proposed for both estimators.
  \item Figure \ref{fig:QQ_Bimodal_X700_N2000}, suggest normality of the estimated density values, which is coherent with results presented in section \ref{limitDist}. This property of the estimators allows the construction of confidence intervals and the application of standard statistical inference tools that could be useful in practical situations. However, to make this applicable, the Variance of $\hat{f}^{PD}(x)$ in accordance with (\ref{eq:158}) needs to be estimated.
  \item  In most of presented figures it is possible to observe that at the extremes of the support sometimes the estimated density values are slightly negative. This effect is consistent with the boundary effect noted in \cite{Antoniadis1997} by Antoniadis. As was mentioned in the introduction, a possible remedial measure could be application the approach proposed by \cite{Pinheiro1997}. Another possibility is using  $\hat{f}_{+}(x)=\max\left\{0,\hat{f}^{PD}(x) \right\}$, as proposed in \cite{Antoniadis1999}.
\end{enumerate}

\section{Real Data application and comparison with other Estimators.}\label{RealData}

In this section we consider the implementation of the proposed estimator on the datasets utilized by Antoniadis et al. in \cite{Antoniadis1999}. To compare our approach with other popular estimators, we will also use the non-parametric Kernel density estimator with optimal bandwidth and the smoothed histogram using local polynomials based on the actual samples.

\medskip

\begin{figure}[!htb]
   \centering
   \begin{subfigure}[b]{0.45\textwidth}
       \includegraphics[width=\textwidth]{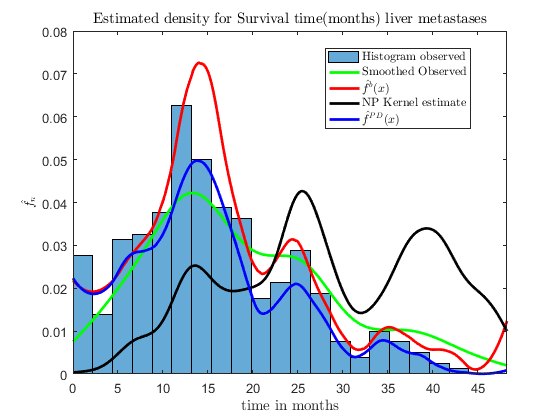}
       \caption{}
       \label{fig:divorce}
   \end{subfigure}\hfill
   \begin{subfigure}[b]{0.45\textwidth}
       \includegraphics[width=\textwidth]{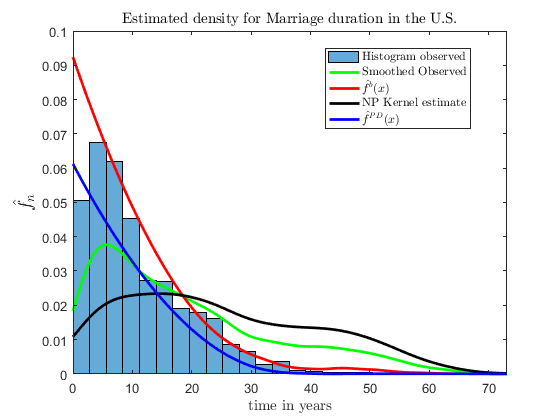}
       \caption{}
       \label{fig:marriage}
    \end{subfigure}
\caption{Results for the application of the data driven estimators in real datasets. (a) corresponds to Liver metastases data and (b) to marriage duration in the U.S.}
\label{fig:realdata}
\end{figure}

The first application considers the data studied by Haupt and Mansmann (1995)\footnote{The data set is available at CART for Survival Data. Statlib Archive \url{http://lib.stat.cmu.edu/S/survcart}.}. In their research, they analized the survival times for patients with liver metastases from a colorectal tumour without other distant metastases. In their data, they have a total of 622 patients from which 43.64\% of the samples are censored. The obtained results are given in Fig.\ref{fig:realdata} (a).

\medskip

Our next practical application, considers the study of marriage dissolution based on a longitudinal survey conducted in the U.S.\footnote{Data set available at \url{http://data.princeton.edu/wws509/datasets} and was adapted from an example in the software aML (See  Lillard and Panis (2000), aML Multilevel Multiprocess Statistical Software, Release 1.0, EconWare, LA, California.)} The unit of observation is the couple and the event of interest is the time from marriage to divorce. Interviewed and widowhood are considered as censoring events. Couples with different educational levels and ethnicity were considered. The original data considered 3371 couples with 30.61\% of samples being censored. The obtained results are given in Fig.\ref{fig:realdata} (b).

\medskip

From figure \ref{fig:realdata} (a), it can be observed that the complete data estimator (in red) shows boundary effects, since after $45$ months, according to the data there are almost no patients alive. However, both complete data and partial data estimators are able to catch the individual modes shown by the histogram without over smoothing as compared to the smoothed histogram (in green). Also, the estimators are able to keep the proportions between the histogram modes as compared to the Kernel density estimator with universal bandwidth (in black).

\medskip

From figure \ref{fig:realdata} (b), it is possible to observe the fairly exponential behavior of the density estimates. Both the complete data and the partial data are able to follow the rate of decay of the Histogram envelope and do not overestimate the density values in the right tails, which is consistent with the data (from data, it is highly unlikely that a certain couple would last married longer than 45 years); both local polynomial and kernel density estimator fail to account for that fact, while assigning significant density to times above 40 years.

\section{Conclusions and Discussion.}\label{Conclusions}
This paper introduced an empirical wavelet-based method to estimate the density in the case of randomly censored data. We proposed estimators based on the partial and complete sample, showing statistical properties of bias, consistency and limiting distribution. Also, we derived convergence rates for the expected $\mathbb{L}_{2}$ error using $J=\lfloor(\log_{2}(N)-\log_{2}(\log(N)))\rfloor$ for the multiresolution index.

\medskip
Both estimators were implemented and tested using different baseline distributions via a theoretical simulation study, showing good performance in the presence of significantly censored data. This simulation study shows that in theory, the estimator attains the large sample behavior that was proposed: it is asymptotically unbiased and mean-square consistent.

\medskip
Regarding the effect of censoring in the estimates, we observed that our method is robust enough to handle censoring proportions of nearly 50\% while achieving acceptable estimation results. Moreover, in the case of no censoring, the method converges to the usual orthogonal wavelet-series estimator (See remarks in section \ref{StuteBias}).

\medskip
From a real data application viewpoint, the proposed method was capable to uncover modes that were be hard to detect by other methods in the used datasets, avoiding the problem of modes over-smoothing that methods such as non-parametric kernels exhibited. Also, the estimator was capable of capturing exponential rates of decay of the underlying density, preventing the overestimation of density values in regions of the support with near-zero empirical mass.

\medskip

Based on the results seen in the simulation study and the real data testing, we can argue that our estimator yields interesting interpretations and results; it has good asymptotic properties and is relatively easy to implement. Also, it offers a useful and competitive alternative for the problem of density estimation with censored data, with respect to multimodal identification and exponential decay adjustment.

\medskip

Finally, some of the drawbacks that were observed throughout this paper were the possibility of obtaining negative values for the density estimates (highly likely at the tails) and also boundary problems resulting from the periodic wavelet extension approach. Also, another important remark worth noting is the fact that it is possible that the estimated density does no integrate to 1. Nonetheless, for most of these problems there are possible solutions such as the ones proposed in \cite{Antoniadis1997} and \cite{Pinheiro1997}.

\newpage

\bibliography{Refpaper1}

\begin{thebibliography}{10}

\bibitem{Deroye1985}
L.~Deroye and L.~Gy\"orfi.
\newblock {\em Nonparametric Density Estimation}.
\newblock John Wiley \& Sons, 1985.

\bibitem{Parzen1962}
E.~Parzen.
\newblock On estimation of a probability density function and mode.
\newblock {\em The Annals of Statistics}, 33:1065--1073, 1962.

\bibitem{Rossenblat1956}
M.~Rossenblat.
\newblock Remarks on some nonparametric estimates of a density function.
\newblock {\em The Annals of Mathematical Statistics}, 27:832--837, 1956.

\bibitem{Antoniadis1997}
Antoniadis A.
\newblock Wavelets in statistics: A review.
\newblock Technical report, University of Joseph Fourier, Laboratorie IMAG-LMC,
  38041 Grenoble Cedex 9, France, 1997.

\bibitem{Efromovich1999}
Sam Efromovich.
\newblock {\em Nonparametric Curve Estimation, Methods, Theory and
  Applications}.
\newblock Springer Series in Statistics. Springer, first edition, 1999.

\bibitem{Antoniadis1999}
Antoniadis A., Gregoire G., and G.~Nason.
\newblock Density and hazard rate estimation for right-censored data by using
  wavelets methods.
\newblock {\em J. Roy. Statist. Soc.}, 61:63--84, 1999.

\bibitem{Daubechies1992}
Ingrid Daubechies.
\newblock Ten lectures on wavelets.
\newblock {\em CBMS-NSF regional conferences series in applied mathematics},
  1992.

\bibitem{Vidakovic1999}
Brani Vidakovic.
\newblock {\em Statistical Modeling by Wavelets}.
\newblock Wiley, New York, 1999.

\bibitem{Pinheiro1997}
A.~Pinheiro and B.~Vidakovic.
\newblock Estimating the square root of a density via compactly supported
  wavelets.
\newblock {\em Computational Statistic and Data Analysis}, 25:399--415, 1997.

\bibitem{Donoho1996}
Donoho D., Johnstone I.M., Kerkyacharian G., and Picard D.
\newblock Density estimation by wavelets thresholding.
\newblock {\em The Annals of Statistics}, 2:508--539, 1996.

\bibitem{Vanucci1998}
M.~Vanucci.
\newblock Nonparametric density estimation using wavelets.
\newblock Discussion paper 95-26, Department of Statistics, Texas A\&M
  University, Duke University, U.S.A., 1998.

\bibitem{Li2003}
Li~L.
\newblock Non-linear wavelet-based density estimator under random censorship.
\newblock {\em Journal of Statistical planning and Inference}, 117:35--58,
  2003.

\bibitem{Li2007}
Li~L.
\newblock On the minimax optimality of wavelet estimators with censored data.
\newblock {\em Journal of Statistical planning and Inference}, 137:1138--1150,
  2007.

\bibitem{Zou2017}
Yu-Ye Zou and Han-Ying Liang.
\newblock Wavelet estimation of density for censored data with censoring
  indicator missing at random.
\newblock {\em A Journal of Theoretical and Applied Statistics}, 2017.

\bibitem{Restrepo1996}
J.M. Restrepo, G.~Leaf, and G.~Schlossnagle.
\newblock Periodized daubechies wavelets.
\newblock Technical report, Mathematics and Computer Science Division, Argonne,
  National Laboratory, Argonne, IL 60439, U.S.A., 1996.

\bibitem{Donoho1993}
D.~Donoho.
\newblock Nonlinear wavelet methods for recovery of signals, densities and
  spectra from indirect and noisy data.
\newblock {\em Proceedings of Symposia in Applied Mathematics}, 47:173--205,
  1993.

\bibitem{Stute1995}
W.~Stute.
\newblock The central limit theorem under random censorship.
\newblock {\em The Annals of Statistics}, 23:422--439, 1995.

\bibitem{Stute1994}
W.~Stute.
\newblock Strong and weak representation of cumulative hazard function and
  kaplan-meier estimator on increasing sets.
\newblock {\em Journal of Statistical Planning and Inference}, 42:315--329,
  1994.

\bibitem{DasGupta2008}
Anirban DasGupta.
\newblock {\em Asymptotic Theory of Probability and Statistics}.
\newblock Springer, 2008.

\bibitem{Hardle1998}
Wolfgang H\"ardle, Gerard Kerkyacharian, Dominique Picard, and Alexander
  Tsybakov.
\newblock {\em Wavelets, Approximation, and Statistical Applications}, volume
  129 of {\em Lecture Notes in Statistics}.
\newblock Springer-Verlag New York, 1 edition, 1998.

\end{thebibliography}
\bibliographystyle{unsrt}

\newpage

\appendix

\section{Derivation of the unbiased partial-data estimator.} \label{der:unbiased}

In this section we provide the derivation for the partial-data estimator proposed in \ref{partial}. From (\ref{eq:33}) and (\ref{eq:34}), it follows:
\begin{equation}\label{eq:35}
\mathbb{E}(\hat{f}_{J}(x))=\sum_{k=0}^{2^{J}-1}\mathbb{E}\left[ \tilde{c_{Jk}}\right]\cdot \phi^{per}_{J,k}(x)\,.
\end{equation}
Using (\ref{eq:32}), the expectation in the left hand side (lhs) of (\ref{eq:35}) is given by:

\begin{equation}\label{eq:36}
\mathbb{E}\left[ \tilde{c_{Jk}}\right]=\mathbb{E}\left[ \frac{1}{N}\sum_{i=1}^{N}\frac{\phi^{per}_{J,k}(Y_{(i)})}{1-\hat{G}(Y_{(i)})}\right]
-\mathbb{E}\left[ \frac{1}{N}\sum_{i=1}^{N}\frac{\mathbbm{1}_{(\delta_{i}=0)}(1-\hat{F}(Y_{(i)}))}{1-\hat{G}(Y_{(i)})}\phi^{per}_{J,k}(Y_{(i)}) \right]\,.
\end{equation}
Assuming iid samples  and $G(y)$ known, the first expectation on the rhs of (\ref{eq:36}) can be obtained as:
\begin{equation}\label{eq:38}
\mathbb{E}\left[ \frac{1}{N}\sum_{i=1}^{N}\frac{\phi^{per}_{J,k}(Y_{(i)})}{1-\hat{G}(Y_{(i)})}\right]=\mathbb{E}_{Y}\left[\frac{\phi^{per}_{J,k}(Y)}{1-G(Y)} \right]\,.
\end{equation}
Similarly, provided iid samples, and both $F(y)$ and $G(y)$ known, the expectation of the second term in the rhs of (\ref{eq:36}) can be obtained as:

\begin{equation}\label{eq:39}
\mathbb{E}\left[ \frac{1}{N}\sum_{i=1}^{N}\frac{\mathbbm{1}_{(\delta_{i}=0)}(1-\hat{F}(Y_{(i)}))}{1-\hat{G}(Y_{(i)})}\phi^{per}_{J,k}(Y_{(i)})\right]=
\mathbb{E}_{Y,\delta=0}\left[\frac{(1-F(Y))\phi^{per}_{J,k}(Y)}{1-G(Y)} \right]\,.
\end{equation}
Since $f_{Y,\delta}(y,\delta=0)=g(y)(1-F(y))$, it follows:

\begin{equation}\label{eq:40}
\resizebox{.9 \textwidth}{!}
{ $\mathbb{E}_{Y,\delta=0}\left[\frac{(1-F(Y))\phi^{per}_{J,k}(Y)}{1-G(Y)} \right]=\mathbb{E}_{T}\left[\frac{(1-F(T))\phi^{per}_{J,k}(T)}{1-G(T)} \right]-
\mathbb{E}_{T}\left[\frac{F(T)(1-F(T))\phi^{per}_{J,k}(T)}{1-G(T)} \right]$ }\,.
\end{equation}
Finally, combining (\ref{eq:38}) and (\ref{eq:40}), it follows:
\begin{equation}\label{eq:41}
\resizebox{.9 \textwidth}{!}
{ $\mathbb{E}\left[ \tilde{c_{Jk}}\right]=\mathbb{E}_{Y}\left[\frac{\phi^{per}_{J,k}(Y)}{1-G(Y)} \right]-\mathbb{E}_{T}\left[\frac{(1-F(T))\phi^{per}_{J,k}(T)}{1-G(T)} \right]+
\mathbb{E}_{T}\left[\frac{F(T)(1-F(T))\phi^{per}_{J,k}(T)}{1-G(T)} \right]$ }\,.
\end{equation}
Using (\ref{eq:23}) and (\ref{eq:41}), (\ref{eq:41}) takes the form:
\begin{equation}\label{eq:42}
\mathbb{E}\left[ \tilde{c_{Jk}}\right]=c_{Jk}+\mathbb{E}_{T}\left[\frac{F(T)(1-F(T))\phi^{per}_{J,k}(T)}{1-G(T)} \right]\,,
\end{equation}
which further implies that for (\ref{eq:35}), it follows:
\begin{equation}\label{eq:43a}
\mathbb{E}(\hat{f}_{J}(x))=f_{J}(x)+\sum_{k=0}^{2^{J}-1}\mathbb{E}_{T}\left[\frac{F(T)(1-F(T))\phi^{per}_{J,k}(T)}{1-G(T)} \right]\phi^{per}_{J,k}(x)\,.
\end{equation}
To facilitate notation, define $b_{J,k}=\mathbb{E}_{T}\left[\frac{F(T)(1-F(T))\phi^{per}_{J,k}(T)}{1-G(T)} \right]$. Thus, (\ref{eq:35}) can be represented as:
\begin{equation}\label{eq:43}
\mathbb{E}(\hat{f}_{J}(x))=f_{J}(x)+\sum_{k=0}^{2^{J}-1}b_{J,k}\cdot\phi^{per}_{J,k}(x)\,.
\end{equation}
Using the same approach as in (\ref{eq:25}), $b_{J,k}$ (i.e. the wavelet coefficient that define the bias of $\hat{f}_{J}(x)$ can be estimated from the sample as follows:
\begin{equation}\label{eq:44}
\tilde{b}_{J,k}=\frac{1}{N}\sum_{i=1}^{N}\mathbbm{1}_{(\delta_{i}=0)}\frac{\hat{F}(Y_{i})(1-\hat{F}(Y_{i}))\phi^{per}_{J,k}(Y_{i})}{1-\hat{G}(Y_{i})}\,.
\end{equation}
Therefore, the biased-corrected version of the estimator can be represented as:
\begin{equation}\label{eq:45}
\hat{f}_{J}^{*}(x)=\hat{f}_{J}(x)-\sum_{k=0}^{2^{J}-1}\tilde{b}_{J,k}\cdot\phi^{per}_{J,k}(x)\,,
\end{equation}
\begin{equation}\label{eq:50}
\hat{f}_{J}^{*}(x)=\sum_{k=0}^{2^{J}-1}\tilde{c}^{*}_{J,k}\cdot\phi^{per}_{J,k}(x)\,,
\end{equation}
where:
\begin{equation}\label{eq:51}
\resizebox{.9 \textwidth}{!}
{ $\tilde{c}^{*}_{J,k}=\tilde{c}_{J,k}-\tilde{b}_{J,k}=\frac{1}{N}\sum_{i=1}^{N}(\frac{1}{1-\hat{G}(Y_{(i)})} - \frac{\mathbbm{1}_{(\delta_{(i)}=0)}(1-\hat{F}(Y_{(i)}))}{1-\hat{G}(Y_{(i)})}-\frac{\mathbbm{1}_{(\delta_{(i)}=0)}\hat{F}(Y_{(i)})(1-\hat{F}(Y_{(i)}))}{1-\hat{G}(Y_{(i)})})\cdot \phi^{per}_{J,k}(Y_{(i)})$}\,.
\end{equation}
Note that (\ref{eq:51}) can be further simplified into:
\begin{equation}\label{eq:52}
\tilde{c}^{*}_{J,k}=\frac{1}{N}\sum_{i=1}^{N}(\frac{1-\mathbbm{1}_{(\delta_{(i)}=0)}(1-\hat{F}(Y_{(i)}))(1+\hat{F}(Y_{(i)}))}{1-\hat{G}(Y_{(i)})})\phi^{per}_{J,k}(Y_{(i)})\,.
\end{equation}
Computing the expectation of the bias-correction coefficient $\tilde{b}_{Jk}$, it follows:
\begin{equation}\label{eq:53}
\mathbb{E}_{Y}\left[\tilde{b}_{Jk}\right]=b_{Jk}-\mathbb{E}_{T}\left[\frac{F(T)^{2}(1-F(T))}{1-G(T)}\phi^{per}_{Jk}(T) \right]\,.
\end{equation}
Therefore, the bias of $\tilde{b}_{Jk}$ can be corrected by defining $\tilde{b}^{*}_{Jk}=\tilde{b}_{Jk}+\mathbb{E}_{T}\left[\frac{F(T)^{2}(1-F(T))}{1-G(T)}\phi^{per}_{Jk}(T) \right]$.
Using the empirical argument as in (\ref{eq:44}), $\tilde{b}^{*}_{Jk}$ can be estimated by:
\begin{equation}\label{eq:54}
\tilde{b}^{*}_{Jk}=\tilde{b}_{Jk}+\frac{1}{N}\sum_{i=1}^{N}\frac{\mathbbm{1}_{(\delta_{(i)}=0)}F(Y_{i})^{2}(1-F(Y_{i}))}{1-G(Y_{i})}\phi^{per}_{Jk}(Y_{i})\,.
\end{equation}
This implies that the updated bias-corrected estimator of $b_{Jk}$ can be represented as:
\begin{equation}\label{eq:55}
\tilde{b}^{*}_{Jk}=\frac{1}{N}\sum_{i=1}^{N}\frac{\mathbbm{1}_{(\delta_{(i)}=0)}F(Y_{i})(1-F(Y_{i}))(1+F(Y_{i}))}{1-G(Y_{i})}\phi^{per}_{Jk}(Y_{i})\,.
\end{equation}
Taking the expectation of $\tilde{b}^{*}_{Jk}$, it follows:
\begin{equation}\label{eq:56}
\mathbb{E}_{Y}\left[\tilde{b}^{*}_{Jk} \right]=b_{Jk}-\mathbb{E}_{T}\left[\frac{F(T)^{3}(1-F(T))}{1-G(T)}\phi^{per}_{Jk}(T) \right]\,.
\end{equation}

Following the same methodology used to derive (\ref{eq:55}), an updated bias-corrected estimate of $\tilde{b}^{*}_{Jk}$, denoted by $\tilde{b}^{**}_{Jk}$ can be represented as:
\begin{equation}\label{eq:57}
\tilde{b}^{**}_{Jk}= \frac{1}{N}\sum_{i=1}^{N}\frac{\mathbbm{1}_{(\delta_{(i)}=0)}F(Y_{i})(1-F(Y_{i}))(1+F(Y_{i})+F(Y_{i})^{2}))}{1-G(Y_{i})}\phi^{per}_{Jk}(Y_{i})\,.
\end{equation}
Taking the expectation of $\tilde{b}^{**}_{Jk}$, it follows:
\begin{equation}\label{eq:58}
\mathbb{E}_{Y}\left[\tilde{b}^{**}_{Jk} \right]=b_{Jk}-\mathbb{E}_{T}\left[\frac{F(T)^{4}(1-F(T))}{1-G(T)}\phi^{per}_{Jk}(T) \right]\,.
\end{equation}
This implies that the bias-corrected estimate of $b_{Jk}$ represented as $\tilde{b}^{***}_{Jk}=\tilde{b}^{**}_{Jk}+\mathbb{E}_{T}\left[\frac{F(T)^{4}(1-F(T))}{1-G(T)}\phi^{per}_{Jk}(T) \right]$ can be iteratively updated. Thus, following the same process as before, it follows:
\begin{equation}\label{eq:59}
\tilde{b}^{***}_{Jk}=\frac{1}{N}\sum_{i=1}^{N}\frac{\mathbbm{1}_{(\delta_{(i)}=0)}F(Y_{i})(1-F(Y_{i}))(1+F(Y_{i})+F(Y_{i})^{2}+F(Y_{i})^{3}))}{1-G(Y_{i})}\phi^{per}_{Jk}(Y_{i})\,.
\end{equation}
From the last set of equations, it follows that this process can be repeated sequentially, infinitely many times. This implies that:
\begin{equation}\label{eq:60}
\tilde{\tilde{b}}_{Jk}=\frac{1}{N}\sum_{i=1}^{N}\frac{\mathbbm{1}_{(\delta_{(i)}=0)}F(Y_{i})(1-F(Y_{i}))\sum_{l=0}^{\infty}F(Y_{i})^{l}}{1-G(Y_{i})}\phi^{per}_{Jk}(Y_{i})\,,
\end{equation}
provided $0<F(Y)<1$. Therefore, it follows that $\sum_{l=0}^{\infty}F(Y_{i})^{l}$ is a convergent series. In fact, it is a geometric power series that satisfies:
\begin{equation}\label{eq:61}
\sum_{l=0}^{\infty}F(Y_{i})^{l}=\frac{1}{1-F(Y_{i})}\,.
\end{equation}
Therefore, this implies that (\ref{eq:60}) takes the form:
\begin{equation}\label{eq:62}
\tilde{\tilde{b}}_{Jk}=\frac{1}{N}\sum_{i=1}^{N}\frac{\mathbbm{1}_{(\delta_{(i)}=0)}F(Y_{i})}{1-G(Y_{i})}\phi^{per}_{Jk}(Y_{i})\,.
\end{equation}
Clearly, $\tilde{\tilde{b}}_{Jk}$ is an unbiased estimate of $b_{Jk}$. Therefore, we conclude that the unbiased estimate of the $c_{Jk}$ coefficient, denoted by $\tilde{\tilde{c}}_{Jk}$ is given by:
\begin{equation}\label{eq:63}
\tilde{\tilde{c}}_{Jk}=\tilde{c}_{Jk}-\tilde{\tilde{b}}_{Jk}=\frac{1}{N}\sum_{i=1}^{N}\frac{\mathbbm{1}_{(\delta_{(i)}=1)}}{1-G(Y_{i})}\phi^{per}_{Jk}(Y_{i})\,,
\end{equation}
thus, it is possible to define the partial-data density estimator $\hat{f}^{PD}(x)$ as:
\begin{equation}\label{eq:34c}
\hat{f}^{PD}(x)=\sum_{k=0}^{2^{J}-1}\tilde{c}_{Jk}\cdot \phi^{per}_{J,k}(x)\,,
\end{equation}
where:

\begin{equation}\label{eq:34d2}
\tilde{c}_{Jk}=\frac{1}{N}\sum_{i=1}^{N}\frac{\mathbbm{1}_{(\delta_{(i)}=1)}}{1-\hat{G}(Y_{i})}\phi^{per}_{Jk}(Y_{i})\,,
\end{equation}
which is unbiased for $f_{J}(x)$, provided $G(y)$ is known and $0<F(Y)<1$.

\newpage
\section{Proof of Proposition 1} \label{proof:prop1}

Assume the following conditions are satisfied:
\begin{enumerate}[(i)]
\item The scaling function $\phi$ that generates the orthonormal set $\left\{\phi_{Jk}^{per}, 0\leq k \leq 2^{J}\right\} $ has compact support and satisfies $||\theta_{\phi}(x)||_{\infty}=C<\infty$, for $\theta_{\phi}(x):=\sum_{r\in\mathbb{Z}}|\phi(x-r)|$.
\item $\exists$ $F\in \mathbb{L}_{2}(\mathbb{R})$ such that $|K(x,y)|\leq F(x-y)$, for all $x,y\, \in \mathbb{R}$, where $K(x,y)=\sum_{k\in\mathbb{Z}}\phi(x-k)\phi(y-k)$.
\item For $s=m+1$, $m\geq1$, integer, $\int |x|^{s}F(x)dx<\infty$.
\item $\int (y-x)^{l}K(x,y)dy=\delta_{0,l}$ for $l=0,...,s$.
\item The density $f$ belongs to the $s$-sobolev space $W_{2}^{s}([0,1])$, $s>1$ defined as:
\begin{equation}
\nonumber
W_{2}^{s}([0,1])=\left\{f\,|\,f\in\mathbb{L}_{2}([0,1]),\, \exists \, f^{(1)},...,f^{(s)}\,\text{s.t.}\,f^{(l)}\in \mathbb{L}_{2}([0,1]),\,l=1,...,s \right\}.
\end{equation}
\end{enumerate}

\medskip

Then, it follows:
\begin{equation}\label{eq:72a1}\mathop{\sup}\limits_{f\in W_{2}^{s}([0,1])}\mathbb{E}\left[||\hat{f}^{PD}(x)-f(x) ||_{2}^{2} \right]\leq C_{1}\frac{2^{J}}{N}+C_{2}2^{-2sJ}\,,\text{and} \end{equation}
for $J=\lfloor \log_{2}(N)-\log_{2}(\log(N)) \rfloor$:
\begin{equation}\label{eq:72b1}
\sigma^{2}_{J}(x)=\mathcal{O}(\log(N)^{-1})\,,
\end{equation}
\begin{equation}\label{eq:46a1}
\mathbb{E}\left[\parallel f(x)-\hat{f}^{PD}(x) \parallel_{2}^{2}\right] \, \leq \, \mathcal{O}(N^{-s}\log(N)^{s})
\end{equation}
\medskip
for $C_{1}>0\,,\,C_{2}>0$ independent of $J$ and $N$, provided $\exists$ $ \alpha_{1}$ $\mid$ $0<\alpha_{1}<\infty$, $C_{T} \in (0,1) $ such that $(1-G(y)) \geq C_{T}e^{-\alpha_{1}y}$ for $y \in [0,1)$,
and $0 \leq F(y) \leq 1$ $\forall y \in [0,1]$.

\subsection*{Proof}

Note that $\hat{f}^{PD}(x)$ can be expressed as follows:
\begin{equation}\label{eq:proofProp1.1}
\hat{f}^{PD}(x)=\frac{1}{N}\sum_{i=1}^{N}w_{i}K_{J}(Y_{i},x)\,,
\end{equation}
where $w_{i}=\frac{\delta_{i}}{1-G(Y_{i})}$, and $K_{J}(x,Y_{i})=2^{J}\sum_{k\in\mathbb{Z}}\phi(2^{J}x-k)\phi(2^{J}y-k)$, for $i=1,...,N$.
\medskip
Since it is assumed that $\exists$ $ \alpha_{1}$ $\mid$ $0<\alpha_{1}<\infty$, $C_{T} \in (0,1) $ such that $(1-G(y)) \geq C_{T}e^{-\alpha_{1}y}$ for $y \in [0,1)$, this implies that $0\leq w_{i}\leq \frac{e^{\alpha_{1}}}{C_{T}}$, for $i=1,...,N$.

\medskip

Also, it is possible to bound the $\mathbb{L}_{2}$ risk of the estimator $\hat{f}^{PD}(x) $ as follows:

\begin{equation}\label{eq:proofProp1.2}
\mathbb{E}\left[||\hat{f}^{PD}(x)-f(x) ||_{2}^{2} \right]\leq 2\left\{\mathbb{E}\left[||\hat{f}^{PD}(x)-\mathbb{E}[\hat{f}^{PD}(x) ] ||_{2}^{2} \right]+||\mathbb{E}[\hat{f}^{PD}(x) ]-f(x) ||_{2}^{2} \right\}\,,
\end{equation}
where the first term in the rhs of (\ref{eq:proofProp1.1}) corresponds to $Var(\hat{f}^{PD}(x))$ and the second, to $bias(\hat{f}^{PD}(x))$.

\subsubsection*{Bound for $\mathbb{E}\left[||\hat{f}^{PD}(x)-\mathbb{E}[\hat{f}^{PD}(x) ] ||_{2}^{2} \right]$}
From (\ref{eq:proofProp1.1}), it follows:
\begin{equation}\nonumber
\hat{f}^{PD}(x)-\mathbb{E}[\hat{f}^{PD}(x)]=\frac{1}{N}\sum_{i=1}^{N}\left(w_{i}K_{J}(x,Y_{i})-\mathbb{E}[w_{i}K_{J}(x,Y_{i})]\right)\,.
\end{equation}
Define $Z_{i}(x)=w_{i}K_{J}(x,Y_{i})-\mathbb{E}[w_{i}K_{J}(x,Y_{i})]$ and $\tilde{Z}_{i}(x)=K_{J}(x,Y_{i})-\mathbb{E}[K_{J}(x,Y_{i})]$. Clearly, $\mathbb{E}[Z_{i}(x)]=\mathbb{E}[\tilde{Z}_{i}(x)]=0$. This implies:
\begin{equation}\nonumber
|\hat{f}^{PD}(x)-\mathbb{E}[\hat{f}^{PD}(x) | \leq \frac{e^{\alpha_{1}}}{C_{T}}\frac{1}{N}\left|\sum_{i=1}^{N}\tilde{Z}_{i}(x) \right|\,,
\end{equation}
since $0\leq w_{i}\leq \frac{e^{\alpha_{1}}}{C_{T}}$, for $i=1,...,N$. Therefore, it follows:

\begin{eqnarray}
\nonumber
|\hat{f}^{PD}(x)-\mathbb{E}[\hat{f}^{PD}(x) |^{2} &\leq& \frac{e^{2\alpha_{1}}}{C_{T}^{2}}\frac{1}{N^{2}}\left|\sum_{i=1}^{N}\tilde{Z}_{i}(x) \right|^{2} \\
\nonumber
\mathbb{E}\left[\int_{0}^{1}|\hat{f}^{PD}(x)-\mathbb{E}[\hat{f}^{PD}(x) |^{2}dx \right] & \leq & \frac{e^{2\alpha_{1}}}{C_{T}^{2}}\frac{1}{N^{2}}\mathbb{E}\left[\int_{0}^{1}\left|\sum_{i=1}^{N}\tilde{Z}_{i}(x) \right|^{2}dx \right]\,.
\end{eqnarray}
From conditions (i) and (ii), Fubini's thorem implies:
\begin{eqnarray}
\nonumber
\mathbb{E}\left[\int_{0}^{1}|\hat{f}^{PD}(x)-\mathbb{E}[\hat{f}^{PD}(x) |^{2}dx \right] & \leq & \frac{e^{2\alpha_{1}}}{C_{T}^{2}}\frac{1}{N^{2}}\int_{0}^{1}\mathbb{E}\left[\left|\sum_{i=1}^{N}\tilde{Z}_{i}(x) \right|^{2} \right]dx \\
\label{eq:proofProp1.4}
& \leq & \frac{e^{2\alpha_{1}}}{C_{T}^{2}}\frac{1}{N}\int_{0}^{1}\mathbb{E}[\tilde{Z}_{1}(x)^{2} ]dx\,,
\end{eqnarray}
where (\ref{eq:proofProp1.4}) follows from the fact that $\tilde{Z}_{i}(x)$ are iid, with $\mathbb{E}[ \tilde{Z}_{i}(x)]=0$, and $\mathbb{E}[\tilde{Z}_{i}(x)^{2}]<\infty$. This, together with the application of Rosenthal's inequality implies $\mathbb{E}\left[\left|\sum_{i=1}^{N}\tilde{Z}_{i}(x) \right|^{2}\right] \leq \sum_{i=1}^{N}\mathbb{E}[\tilde{Z}_{i}(x)^{2} ]=N\,\mathbb{E}[\tilde{Z}_{1}(x)^{2} ]$.

\medskip

Since $\mathbb{E}[\tilde{Z}_{1}(x)^{2}]=\mathbb{E}[K_{J}(x,Y_{1})^{2} ]-\left(K_{J}f_{Y}(x) \right)^{2}\leq \mathbb{E}[K_{J}(x,Y_{1})^{2} ]$, where $K_{J}f_{Y}(x)=\int_{0}^{1}K_{J}(x,u)f_{Y}(u)du$, and the fact that $|K_{J}(x,y) |=2^{J}|K(2^{J}x,2^{J}y)|$, it follows from (\ref{eq:proofProp1.4}) and condition (ii):
\begin{equation}
\nonumber
\mathbb{E}\left[||\hat{f}^{PD}(x)-\mathbb{E}[\hat{f}^{PD}(x)] ||_{2}^{2} \right] \leq \frac{e^{2\alpha_{1}}}{C_{T}^{2}}\frac{1}{N}\int_{0}^{1}\mathbb{E}[K_{J}(x,Y_{1})^{2} ]dx
\end{equation}

\begin{eqnarray}
\nonumber
\int_{0}^{1}\mathbb{E}[K_{J}(x,Y_{1})^{2} ]dx &\leq & 2^{J}\int_{0}^{1}\left[\int_{-2^{J}y}^{2^{J}(1-y)}F^{2}(v)dv \right]f_{Y}(y)dy \\
\label{eq:proofProp1.5}
&\leq & 2^{J}||F||_{2}^{2}\,.
\end{eqnarray}
Therefore, substituting (\ref{eq:proofProp1.5}) into (\ref{eq:proofProp1.4}), it follows:
\begin{equation}\label{eq:proofProp1.6}
\mathbb{E}\left[||\hat{f}^{PD}(x)-\mathbb{E}[\hat{f}^{PD}(x)] ||_{2}^{2} \right] \leq \frac{||F||_{2}^{2}e^{2\alpha_{1}}}{C_{T}^{2}}\frac{2^{J}}{N}\,.
\end{equation}

\subsubsection*{Bound for $||\mathbb{E}[\hat{f}^{PD}(x) ]-f(x) ||_{2}^{2}$}
According to corollary 8.2 \cite{Hardle1998}, if $f\in W_{2}^{s}([0,1])$ then $||K_{J}f-f||_{2}^{2}=\mathcal{O}\left(2^{-2Js}\right)$. Furthermore, assume conditions (i)-(iv) are satisfied. Since $\mathbb{E}[\hat{f}^{PD}(x)]=K_{J}f(x)$, it follows:

\begin{equation}\label{eq:proofProp1.7}
||\mathbb{E}[\hat{f}^{PD}(x) ]-f(x) ||_{2}^{2} \leq C_{2}\,2^{-2Js}\,.
\end{equation}
Finally, putting together (\ref{eq:proofProp1.6}) and (\ref{eq:proofProp1.7}), it follows:
\begin{equation}\label{eq:proofProp1.8}
\mathop{\sup}\limits_{f\in W_{2}^{s}([0,1])}\mathbb{E}\left[||\hat{f}^{PD}(x)-f(x) ||_{2}^{2} \right]\leq C_{1}\frac{2^{J}}{N}+C_{2}2^{-2sJ}\,,
\end{equation}
as desired, for $C_{1}=\frac{||F||_{2}^{2}e^{2\alpha_{1}}}{C_{T}^{2}}$ and $C_{2}>0$, independent of $N$ and $J$.

\medskip
From (\ref{eq:proofProp1.8}), by choosing $J=\lfloor \log_{2}(N)-\log_{2}(\log(N)) \rfloor$, it follows that $\sigma^{2}_{J}(x)=\mathcal{O}(\log(N)^{-1})$. Furthermore, this also implies that $\mathop{\sup}\limits_{f\in W_{2}^{s}([0,1])}\mathbb{E}\left[||\hat{f}^{PD}(x)-f(x) ||_{2}^{2} \right] = \mathcal{O}(N^{-s}\log(N)^{2}$, which completes the proof.

\subsubsection*{Remarks}
Note that from (\ref{eq:proofProp1.8}), it is possible to choose the multiresolution level $J$ such that the upper bound for the $\mathbb{L}_{2}$ risk is minimized. In this context, it is possible to show that $J^{*}(N)=\frac{1}{2s+1}\log_{2}\left(\frac{2s\,C_{2}}{C_{1}} \right)+\frac{1}{2s+1}\log_{2}(N)$ achieves that result. Moreover, under this choice of $J$, it follows:

\begin{equation}\nonumber
\mathop{\sup}\limits_{f\in W_{2}^{s}([0,1])}\mathbb{E}\left[||\hat{f}^{PD}(x)-f(x) ||_{2}^{2} \right]\leq \tilde{C}N^{-\frac{2s}{2s+1}}\,.
\end{equation}

\newpage
\section{Proof of Proposition 2} \label{proof:prop2}
Under the assumptions and definitions stated in \ref{prop1} and \ref{StuteBias}, and choosing $J=\lfloor \log_{2}(N)-\log_{2}(\log(N)) \rfloor$, it follows:
\begin{eqnarray}
\mathop{\sup}\limits_{f\in W_{2}^{s}([0,1])}\mathbb{E}\left[\parallel f(x)-\hat{f}^{PD}(x) \parallel_{2}^{2}\right] &=& \mathcal{O}(N^{-s}\log(N)^{s})\,. \label{eq:proofProp2.1}
\end{eqnarray}
\subsection*{Proof}
Assume conditions (i)-(iv) established in \ref{proof:prop1} are satisfied. Furthermore, assume $\exists\, \gamma >0$ and a constant $C\in(0,1)$ such that $1-\hat{G}(y)\geq Ce^{-\gamma y}$, for $y\in[0,1)$. Note that $\hat{f}^{PD}(x)$ can be expressed as follows:
\begin{equation}\label{eq:proofProp2.2}
\hat{f}^{PD}(x)=\frac{1}{N}\sum_{i=1}^{N}w_{i}K_{J}(Y_{i},x)\,,
\end{equation}
where $w_{i}=\frac{\delta_{i}}{1-\hat{G}(Y_{i})}$, and $K_{J}(x,Y_{i})=2^{J}\sum_{k\in\mathbb{Z}}\phi(2^{J}x-k)\phi(2^{J}y-k)$, for $i=1,...,N$. Since it is assumed that $\exists\, \gamma >0$ and a constant $C\in(0,1)$ such that $1-\hat{G}(y)\geq Ce^{-\gamma y}$, for $y\in[0,1)$, this implies that $0\leq w_{i}\leq \frac{e^{\gamma}}{C}$, for $i=1,...,N$. Thus, following the same methodology as in \ref{proof:prop1}, it follows that by choosing $J=\lfloor \log_{2}(N)-\log_{2}(\log(N)) \rfloor$:
\begin{eqnarray}
\mathop{\sup}\limits_{f\in W_{2}^{s}([0,1])}\mathbb{E}\left[\parallel f(x)-\hat{f}^{PD}(x) \parallel_{2}^{2}\right] &=& \mathcal{O}(N^{-s}\log(N)^{s})\,. \label{eq:proofProp2.3}
\end{eqnarray}
\subsubsection*{Remarks}
\begin{enumerate}[(i)]
\item Observe that by following the same methodology as in \ref{proof:prop1}, it is possible to obtain:
\begin{equation}\nonumber
\mathop{\sup}\limits_{f\in W_{2}^{s}([0,1])}\mathbb{E}\left[||\hat{f}^{PD}(x)-f(x) ||_{2}^{2} \right]\leq C_{1}\frac{2^{J}}{N}+C_{2}2^{-2sJ}\,,
\end{equation}
for $C_{1}=\frac{||F||_{2}^{2}e^{2\gamma}}{C^{2}}$ and $C_{2}>0$, independent of $N$ and $J$.
\item The last result implies that by choosing $J^{*}(N)=\frac{1}{2s+1}\log_{2}\left(\frac{2s\,C_{2}}{C_{1}} \right)+\frac{1}{2s+1}\log_{2}(N)$, the $\mathbb{L}_{2}$ risk of the estimator $\hat{f}^{PD}(x)$ when $G$ is unknown is also mean square consistent, and achieves a convergence rate of the order $\sim N^{-\frac{2s}{2s+1}}$.
\end{enumerate}

\newpage
\section{Proof of Proposition 3} \label{proof:prop3}

From (\ref{eq:138}), and for $N$ large it follows that the rhs of (\ref{eq:139}) corresponds to the sum of normally distributed random variables $\sim N(0,\sigma_{Jk}^{2}) $ which is indeed a normally distributed random variable. To obtain its variance, it can be used the fact that $Cov\left(\sqrt{N}(\tilde{c}_{Jk}-c_{Jk})\,,\sqrt{N}(\tilde{c}_{Jl}+c_{Jl}) \right) = N\,\mathbb{E}\left[(\tilde{c}_{Jk}-c_{Jk})(\tilde{c}_{Jl}-c_{Jl}) \right]$. Thus, (\ref{eq:121}) implies:
\begin{equation} \label{eq:145}
\mathbb{E}\left[N(\tilde{c}_{Jk}-c_{Jk})(\tilde{c}_{Jl}-c_{Jl}) \right]=N\,\left(\mathbb{E}\left[\tilde{c}_{Jk}\tilde{c}_{Jl} \right]-c_{Jk}c_{Jl} \right)-(c_{Jk}-c_{Jl})\mathcal{O}(\log(N))\,.
\end{equation}
Using (\ref{eq:113}), it follows:
\begin{equation}\label{eq:146}
\tilde{\tilde{c}}_{Jk}\tilde{\tilde{c}}_{Jl} =A_{1}+A_{2}+A_{3}+A_{4}+A_{5}+A_{6}+A_{7}+A_{8}+A_{9}\,,
\end{equation}
where:
\begin{eqnarray}
  A_{1} &=& \frac{1}{N^{2}} \sum_{i=1}^{N}\sum_{j=1}^{N}\frac{\delta_{i}\delta_{j}\phi_{Jk}^{per}(Y_{i})\phi_{Jl}^{per}(Y_{j})}{(1-G_{T}(Y_{i}) )(1-G_{T}(Y_{j}) )}  \label{eq:147a}\\
  A_{2} &=& \frac{1}{N^{2}} \sum_{i=1}^{N}\sum_{j=1}^{N}\frac{\delta_{i}\phi_{Jk}^{per}(Y_{i})U_{jl}}{1-G_{T}(Y_{i})} \label{eq:147b}\\
  A_{3} &=& \frac{1}{N}R_{Nl} \sum_{i=1}^{N}\frac{\delta_{i}\phi_{Jk}^{per}(Y_{i})}{1-G_{T}(Y_{i})} \label{eq:147c}\\
  A_{4} &=& \frac{1}{N^{2}} \sum_{i=1}^{N}\sum_{j=1}^{N}\frac{\delta_{j}\phi_{Jl}^{per}(Y_{j})U_{ik}}{1-G_{T}(Y_{j})} \label{eq:147d}\\
  A_{5} &=& \frac{1}{N^{2}} \sum_{i=1}^{N}\sum_{j=1}^{N}U_{ik}U_{jl} \label{eq:147e}\\
  A_{6} &=& \frac{1}{N}R_{Nl}\sum_{i=1}^{N}U_{ik} \label{eq:147f}\\
  A_{7} &=& \frac{1}{N}R_{Nk} \sum_{i=1}^{N}\frac{\delta_{j}\phi_{Jl}^{per}(Y_{j})}{1-G_{T}(Y_{j})} \label{eq:147g}\\
  A_{8} &=& \frac{1}{N}R_{Nk}\sum_{i=1}^{N}U_{il} \label{eq:147h}\\
  A_{9} &=& R_{Nk}R_{Nl} \label{eq:147i}\,.
\end{eqnarray}
From the last set of equations, it is possible to observe that the following pairs have the same structure (i.e. they are symmetric counter parts of each other) $(A_{2},A_{4})$, $(A_{3},A_{7})$ and $(A_{6},A_{8})$.
\medskip

Now, assuming that $\mathbb{E}\left[ \frac{\delta^{2}\phi_{Jk}^{per}(Y)\phi_{Jl}^{per}(Y)}{(1-G(Y))^{2}}\right]$ is finite (provided (\ref{eq:140}), (\ref{eq:141}), and the assumptions stated above) for $A_{1}$, it follows:
\begin{eqnarray}
\nonumber 
  \mathbb{E}\left[A_{1}\right] &=& \frac{1}{N^{2}}\mathbb{E}\left[ \sum_{i=1}^{N}\sum_{j=1}^{N}\frac{\delta_{i}\phi_{Jk}^{per}(Y_{i})U_{jl}}{1-G_{T}(Y_{i})}\right] \\
   &=& \frac{1}{N}\mathbb{E}\left[ \frac{\delta^{2}\phi_{Jk}^{per}(Y)\phi_{Jl}^{per}(Y)}{(1-G(Y))^{2}}\right]+\frac{N-1}{N}c_{Jk}c_{Jl} \label{eq:146b}\,.
\end{eqnarray}
Consider possible upper bounds for $\gamma_{1,Jk}(x)$ and $\gamma_{2,Jk}(x)$. Using the corresponding definitions stated in \ref{StuteBias}, it follows:
\begin{eqnarray}
\nonumber 
  \gamma_{1,Jk}(x) & = & \frac{1}{(1-F_{X}(x) ) (1-G_{T}(x) )} \int_{x}^{1}\phi_{Jk}^{per}(u)f_{X}(u)du \\
   & \leq & \frac{\parallel f_{X} \parallel_{\infty}\,M\,2^{-\frac{J}{2}}}{c(1-G_{T}(x))^{\beta+1}} \\
   & \leq & \frac{e^{\frac{\alpha_{1}(\beta+1)}{2}}\parallel f_{X} \parallel_{\infty}\,M\,2^{-\frac{J}{2}}}{c\,C_{T}^{\frac{\beta+1}{2}}}\,. \label{eq:147}
\end{eqnarray}
Similarly, for $\gamma_{2,Jk}(x)$, it follows:
\begin{eqnarray}
\nonumber
\gamma_{2,Jk}(x) & \leq & \int_{0}^{1}\frac{|\phi_{Jk}^{per}(u)|f_{X}(u)du}{(1-F_{X}(u))(1-G_{T}(u))} \\
\nonumber
& \leq & \int_{0}^{1}\frac{|\phi_{Jk}^{per}(u)|f_{X}(u)du}{c(1-G_{T}(u))^{\beta+1}} \\
& \leq & \frac{e^{\frac{\alpha_{1}(\beta+1)}{2}}\parallel f_{X} \parallel_{\infty}\,M\,2^{-\frac{J}{2}}}{c\,C_{T}^{\frac{\beta+1}{2}}}\,. \label{eq:150}
\end{eqnarray}
Therefore, the last result implies that for $k,l = 0,...,2^{J}-1$ and $\tilde{i} \in \left\{0,1\right\}$:
\begin{eqnarray}
\nonumber
\gamma_{\tilde{i},Jk}(x)\gamma_{\tilde{i},Jl}(x) & \leq & \frac{\parallel f_{X} \parallel_{\infty}^{2}\,M^{2}\,2^{-J}}{c^{2}(1-G_{T}(x))^{2(\beta+1)}} \\
\nonumber
  & \leq & \frac{e^{\alpha_{1}(\beta+1)}\parallel f_{X} \parallel_{\infty}^{2}\,M^{2}\,2^{-J}}{c^{2}C_{T}^{\beta+1}} \\
  & \leq & \mathcal{O}(N^{-1}\log(N))\,. \label{eq:148}
\end{eqnarray}
Using the last result,it follows:
\begin{eqnarray}
\nonumber
\mathbb{E}\left[(1-\delta)\gamma_{1,Jk}(Y)\gamma_{2,Jl}(Y) \right] & \leq & \frac{e^{\alpha_{1}(\beta+1)}\parallel f_{X} \parallel_{\infty}^{2}\,M^{2}\,2^{-J}}{c^{2}C_{T}^{\beta+1}}\int_{0}^{1}(1-G(u))f_{X}(u)du  \\
 & \leq & \frac{e^{\alpha_{1}(\beta+1)}\parallel f_{X} \parallel_{\infty}^{2}\,M^{2}\,2^{-J}}{c^{2}C_{T}^{\beta+1}}\,. \label{eq:149}
\end{eqnarray}
Clearly, from the last result the same upper bound holds for $\mathbb{E}\left[(1-\delta)^{2}\gamma_{1,Jk}(Y)\gamma_{1,Jl}(Y) \right]$ and  $\mathbb{E}\left[\gamma_{2,Jk}(Y)\gamma_{2,Jl}(Y) \right]$.

\medskip

Now, for the pair $(A_{2},A_{4})$, it follows:
\begin{eqnarray}
\nonumber 
  \mathbb{E}\left[A_{2}\right] &=& \frac{1}{N^{2}}\mathbb{E}\left[ \sum_{i=1}^{N}\sum_{j=1}^{N}\frac{\delta_{i}\phi_{Jk}^{per}(Y_{i})U_{jl}}{1-G_{T}(Y_{i})}\right]\\
  \nonumber
   &=& -\frac{1}{N}\mathbb{E}\left[\frac{\delta\phi_{Jk}^{per}(Y)\gamma_{2,Jk}(Y)}{1-G_{T}(Y)} \right] \\
   \nonumber
  & \leq & \frac{1}{N}\frac{e^{\frac{\alpha_{1}(\beta+1)}{2}}\parallel f_{X} \parallel_{\infty}\,M\,2^{-\frac{J}{2}}}{c\,C_{T}^{\frac{\beta+1}{2}}}\int_{0}^{1}| \phi_{Jk}^{per}(u)|c(1-G_{T}(u))^{\beta-1}g_{T}(u)du \\
  \nonumber
  & \leq & \frac{1}{N}\frac{e^{\frac{\alpha_{1}(\beta+1)}{2}}\parallel f_{X} \parallel_{\infty}\parallel g_{T} \parallel_{\infty}\,M^{2}\,2^{-J}}{C_{T}^{\frac{\beta+1}{2}}} \\
  & \leq & \mathcal{O}(N^{-2}\log(N))\,, \label{eq:150}
\end{eqnarray}
In the case of the pair $(A_{3},A_{7})$ we have:
\begin{eqnarray}
\nonumber 
   \mathbb{E}\left[A_{3}\right] &=& \frac{1}{N}\mathbb{E}\left[R_{Nl} \sum_{i=1}^{N}\frac{\delta_{i}\phi_{Jk}^{per}(Y_{i})}{1-G_{T}(Y_{i})} \right] \\
   & \leq & \mathcal{O}(N^{-1}\log(N))c_{Jk} \label{eq:151}
\end{eqnarray}

For the term $A_{5}$ we have the following:

\begin{eqnarray}
\nonumber 
  \mathbb{E}\left[A_{5}\right] &=& \frac{1}{N^{2}}\mathbb{E}\left[\sum_{i=1}^{N}\sum_{j=1}^{N}U_{ik}U_{jl} \right] \\
\nonumber
 &=& \frac{1}{N}\mathbb{E}\left[U_{k}U_{l} \right]
\end{eqnarray}

Therefore, using the definition of $U_{k}$:

\begin{equation}
\resizebox{.9 \textwidth}{!}
{$
\mathbb{E}\left[A_{5}\right] = \frac{1}{N}\mathbb{E}\left[ (1-\delta)^{2}\gamma_{1,Jk}(Y)\gamma_{1,Jl}(Y) - (1-\delta)\gamma_{1,Jk}(Y)\gamma_{2,Jl}(Y)-(1-\delta)\gamma_{1,Jl}(Y)\gamma_{2,Jk}(Y)+\gamma_{2,Jk}(Y)\gamma_{2,Jl}(Y)\right] $} \nonumber
\end{equation}

From the last result and (\ref{eq:148}), it is clear that:

\begin{equation}
\mathbb{E}\left[A_{5}\right] \leq  \mathcal{O}(N^{-2}\log(N)) \label{eq:152}
\end{equation}

Now, for the pair $(A_{6},A_{8})$ it is clear from the zero mean condition of $U_{k}$ and the fact that $R_{N}=\mathcal{O}(N^{-1}\log(N))$ that:

\begin{eqnarray}
\mathbb{E}\left[A_{6}\right] & \leq & \mathcal{O}(N^{-2}\log(N)) \label{eq:153} \\
\mathbb{E}\left[A_{9}\right] & \leq & \mathcal{O}(N^{-2}\log(N)^{2}) \label{eq:154}
\end{eqnarray}

Putting together (\ref{eq:146b})-(\ref{eq:154}) in (\ref{eq:146}) we get:

\begin{equation}\label{eq:155}
\resizebox{.9 \textwidth}{!}
{$
\mathbb{E}\left[\tilde{\tilde{c}}_{Jk}\tilde{\tilde{c}}_{Jl}\right] \leq \frac{1}{N}\mathbb{E}\left[ \frac{\delta^{2}\phi_{Jk}^{per}(Y)\phi_{Jl}^{per}(Y)}{(1-G(Y))^{2}}\right]+\frac{N-1}{N}c_{Jk}c_{Jl} + \mathcal{O}(N^{-2}\log(N))+\mathcal{O}(N^{-2}\log(N)^{2})+\mathcal{O}(N^{-1}\log(N))(c_{Jk}+c_{Jl}) $}
\end{equation}

Therefore, (\ref{eq:145}) becomes:

\begin{equation}\label{eq:156}
\mathbb{E}\left[N(\tilde{\tilde{c}}_{Jk}-c_{Jk})(\tilde{\tilde{c}}_{Jl}-c_{Jl}) \right] \leq \mathbb{E}\left[ \frac{\delta^{2}\phi_{Jk}^{per}(Y)\phi_{Jl}^{per}(Y)}{(1-G(Y))^{2}}\right]-c_{Jk}c_{Jl}+\mathcal{O}(N^{-1}\log(N)^{2})
\end{equation}

Therefore, for $N$ large the last result suggests that:

\begin{equation}\label{eq:157}
Cov\left(\sqrt{N}(\tilde{\tilde{c}}_{Jk}-c_{Jk})\,,\sqrt{N}(\tilde{\tilde{c}}_{Jl}+c_{Jl}) \right) \approx \mathbb{E}\left[ \frac{\delta^{2}\phi_{Jk}^{per}(Y)\phi_{Jl}^{per}(Y)}{(1-G(Y))^{2}}-c_{Jk}c_{Jl}\right]
\end{equation}

Finally, in light of the last result and the properties of the Normal Distribution, result (\ref{eq:144}) follows. Therefore,

\begin{equation}\label{eq:158}
\resizebox{.93 \textwidth}{!}
{$
\hat{f}^{PD}(x) \mathop{\sim}\limits^{app.}N\left(f(x)\,,\frac{1}{N}\sum_{k=0}^{2^{J}-1}\sigma_{Jk}^{2}(\phi_{Jk}^{per}(x))^{2}+\frac{2}{N}\sum_{k<l}\mathbb{E}\left[ \frac{\delta^{2}\phi_{Jk}^{per}(Y)\phi_{Jl}^{per}(Y)}{(1-G(Y))^{2}}-c_{Jk}c_{Jl} \right]\phi_{Jk}^{per}(x)\phi_{Jl}^{per}(x)\right) $}
\end{equation}

\end{document}